\documentclass[12pt]{article}
\usepackage{cite,epsfig,amsmath,amssymb,axodraw,color,verbatim,colordvi}
 
\newcommand{\beq}{\begin{equation}}
\newcommand{\eeq}{\end{equation}}
\newcommand{\bea}{\begin{eqnarray}}
\newcommand{\eea}{\end{eqnarray}}
\newcommand{\Fig}[1]{Fig.\,\ref{#1}}
\newcommand{\fig}[1]{\ref{#1}}
\newcommand{\Figs}[1]{Figs.\,\ref{#1}}
\newcommand{\Figsand}[2]{Figs.\,\ref{#1}\ and \ref{#2}}
\newcommand{\Eq}[1]{Eq.\,(\ref{#1})}
\newcommand{\eq}[1]{(\ref{#1})}
\newcommand{\Eqs}[1]{Eqs.\,(\ref{#1})}
\newcommand{\Eqsand}[2]{Eqs.\,(\ref{#1}) and (\ref{#2})}
\newcommand{\Eqsto}[2]{Eqs.\,(\ref{#1}) to (\ref{#2})}
\newcommand{\Tab}[1]{Tab.\,\ref{#1}}

\newcommand{\Tabsand}[2]{Tabs.\,\ref{#1} and \ref{#2}}
\newcommand{\Sec}[1]{Sec.\,\ref{#1}}
\newcommand{\sect}[1]{\ref{#1}}
\newcommand{\Secs}[1]{Secs.\,\ref{#1}}
\newcommand{\Secsand}[2]{Secs.\,\ref{#1} and \ref{#2}}
\newcommand{\Secsto}[2]{Secs.\,\ref{#1} to \ref{#2}}
\newcommand{\App}[1]{App.\,\ref{#1}}
\newcommand{\f}{\frac}
\newcommand{\non}{\nonumber}
\newcommand{\gs}{g}
\newcommand{\as}{\alpha_s}
\newcommand{\aem}{\alpha}
\newcommand{\aemMSbar}{\alpha_{\overline{\rm MS}}}
\newcommand{\sws}{\sin^2 \theta_{\scriptscriptstyle W}}
\newcommand{\swq}{\sin^4 \theta_{\scriptscriptstyle W}}
\newcommand{\swsMSbar}{\sin^2 \hat{\theta}_{\scriptscriptstyle W}}
\newcommand{\MW}{M_{\scriptscriptstyle W}}
\newcommand{\MZ}{M_{\scriptscriptstyle Z}}
\newcommand{\mc}{m_c}
\newcommand{\mb}{m_b}
\newcommand{\mt}{m_t}
\newcommand{\mtpole}{M_t}
\newcommand{\mf}{m_f}
\newcommand{\mtau}{m_\tau}
\newcommand{\muh}{\mu_{\scriptscriptstyle W}}
\newcommand{\mul}{\mu}
\newcommand{\muf}{\mu_\nf}
\newcommand{\muc}{\mu_c}
\newcommand{\mub}{\mu_b}
\newcommand{\mut}{\mu_t}
\newcommand{\cf}{C_{\scriptstyle F}}
\newcommand{\ca}{C_{\scriptstyle A}}

\newcommand{\nf}{f}
\newcommand{\betazero}{\beta_0}
\newcommand{\betaone}{\beta_1}
\newcommand{\betatwo}{\beta_2}
\newcommand{\betak}{\beta_k}
\newcommand{\gammamczero}{\gamma_{m}^{(0)}}
\newcommand{\gammamcone}{\gamma_{m}^{(1)}}
\newcommand{\gammamctwo}{\gamma_{m}^{(2)}}
\newcommand{\gammamck}{\gamma_{m}^{(k)}}
\newcommand{\Heff}{{\cal H}_{\rm eff}}

\newcommand{\GF}{G_F}
\newcommand{\BR}{{\cal B}}

\newcommand{\GeV}{{\rm \ GeV}}
\newcommand{\MSbar}{\overline{\rm MS}}

\newcommand{\re}{{\rm Re}}
\newcommand{\im}{{\rm Im}}
\newcommand{\ord}{{\cal O}}
\newcommand{\eps}{\epsilon}
\newcommand{\sL}{{\scalebox{0.6}{$L$}}}

\newcommand{\Ktopinunu}{K^+ \to \pi^+ \nu \bar{\nu}}
\newcommand{\stodnunu}{s \to d \nu \bar{\nu}}

\newcommand{\KLtopinunu}{K_L \to \pi^0 \nu \bar{\nu}}

\newcommand{\Ktopinunus}{K \to \pi \nu \bar{\nu}}
\newcommand{\Ktopienu}{K^+ \to \pi^0 e^+ \nu}

\newcommand{\cl}{{\rm Cl}_2}
\newcommand{\li}{{\rm Li}_2}
\newcommand{\Qpaper}{Q}
\newcommand{\Qmisiak}{Q'}
\newcommand{\Epaper}{E}
\newcommand{\Emisiak}{E'}
\newcommand{\QQ}{\Qpaper \hspace{-0.0mm} \Qpaper}
\newcommand{\EE}{\Epaper \hspace{-0.0mm} \Epaper}
\newcommand{\QE}{\Qpaper \hspace{-0.2mm} \Epaper}

\newcommand{\gammadown}[1]{\gamma_{#1}}
\newcommand{\gammaup}[1]{\gamma^{#1}}

\newcommand{\zetathree}{\zeta (3)}
\newcommand{\zetax}{\zeta (x)}
\newcommand{\bare}{{0}}
\newcommand{\PcX}{P_c (X)}
\newcommand{\dPcu}{\delta P_{c, u}}
\newcommand{\PcXLO}{P_c^{(0)} (X)}
\newcommand{\PcXNLO}{P_c^{(1)} (X)}
\newcommand{\PcXNNLO}{P_c^{(2)} (X)}
\newcommand{\PcXk}{P_c^{(k)} (X)}

\newcommand{\Qmuh}{(\muh)}
\newcommand{\Qmub}{(\mub)}
\newcommand{\nfive}{5}
\newcommand{\nfour}{4}

\newcommand{\ckmfitter}{CKMfitter Group~}
\newcommand{\rfit}{{\it R}fit~}
\newcommand{\utfit}{{\bf{U}}\kern-.24em{\bf{T}}\kern-.21em{\it{fit}}\@
Collaboration~}
\newcommand{\ckm}{CKMfitter}
\newcommand{\ut}{{\bf{U}}\kern-.24em{\bf{T}}\kern-.21em{\it{fit}}\@}

\let\oldmarginpar\marginpar
\renewcommand\marginpar[1]{\-\oldmarginpar[\raggedleft\scriptsize\sf
#1]{\raggedright\scriptsize\sf #1}} 

\newcommand{\figcsmatch}{Fig.~9}

\newcommand{\pcseriesequ}{Eq.~(13)}
\newcommand{\bilocinsequ}{Eq.~(19)}
\newcommand{\qzequ}{Eq.~(37)}
\newcommand{\blrgeequ}{(47)}
\newcommand{\gamavequ}{Eq.~(51)}
\newcommand{\qacsdefequ}{Eqs.~(52)}
\newcommand{\axevequ}{Eq.~(53)}
\newcommand{\zaaequ}{Eq.~(55)}
\newcommand{\rmatequ}{Eqs.~(83) and (84)}
\newcommand{\pensolequ}{Eq.~(86)}
\newcommand{\zinput}{Eqs.~(54), (55), (62) and (63)}
\newcommand{\zsec}{Sec.~6.1}
\newcommand{\sectwo}{Sec.~3}

\newcommand{\ov}{\overline}

\begin{document}
\allowdisplaybreaks
\thispagestyle{empty}
\rightline{TUM-HEP-600/05}
\rightline{IPPP/05/74}
\rightline{DCPT/05/148}
\rightline{TTP06-06}
\rightline{FERMILAB-PUB-05-512-T}
\rightline{ZU-TH 23/05}
\rightline{hep-ph/0603079}
\rightline{}
\rightline{\today}
\vspace*{1.2truecm}
\bigskip
\bigskip

\centerline{\LARGE \bf Charm Quark Contribution to {\boldmath
$\Ktopinunu$}} 
\vspace{0.2cm}
\centerline{\LARGE \bf at Next-to-Next-to-Leading Order} 

\vskip1truecm
\centerline{\large\bf Andrzej~J.~Buras$^a$, Martin~Gorbahn$^{b,c}$,}
\vspace{0.1cm}
\centerline{\large\bf Ulrich~Haisch$^{d,e}$ and Ulrich~Nierste$^{c,d}$}
\bigskip
\begin{center}{
{\em $^a$ Physik Department, Technische Universit\"at M\"unchen, \\  
D-85748 Garching, Germany} \\
\vspace{.3cm}
{\em $^b$ IPPP, Physics Department, University of Durham, \\ DH1 3LE,
Durham, UK} \\
\vspace{.3cm}
{\em $^c$ Institut f\"ur Theoretische Teilchenphysik, Universit\"at 
Karlsruhe, \\ D-76128 Karlsruhe, Germany} \\  
\vspace{.3cm}
{\em $^d$ Theoretical Physics Department, Fermilab, \\ Batavia, IL
60510, USA} \\
\vspace{.3cm}
{\em $^e$ Institut f\"ur Theoretische Physik, Universit\"at Z\"urich, 
\\ CH-8057 Z\"urich, Switzerland}
}\end{center}

\newpage 
\thispagestyle{empty}

\centerline{\bf Abstract}
\vspace*{0.5cm}

\noindent We calculate the complete next-to-next-to-leading order QCD
corrections to the charm contribution of the rare decay
$\Ktopinunu$. We encounter several new features, which were absent in
lower orders. We discuss them in detail and present the results for
the two-loop matching conditions of the Wilson coefficients, the
three-loop anomalous dimensions, and the two-loop matrix elements of
the relevant operators that enter the next-to-next-to-leading order
renormalization group analysis of the $Z$-penguin and the electroweak
box contribution. The inclusion of the next-to-next-to-leading order
QCD corrections leads to a significant reduction of the theoretical
uncertainty from $\pm  9.8 \%$ down to $\pm 2.4 \%$ in the relevant
parameter $\PcX$, implying the leftover scale uncertainties in
$\BR (\Ktopinunu)$ and in the determination of $| V_{td} |$, $\sin 2  
\beta$, and $\gamma$ from the $\Ktopinunus$ system to be $\pm 1.3 \%$,
$\pm 1.0 \%$, $\pm 0.006$, and $\pm 1.2^\circ$, respectively. For the  
charm quark $\MSbar$ mass $\mc (\mc) = \left ( 1.30 \pm 0.05 \right )
\!\! \GeV$ and $| V_{us} | = 0.2248$ the next-to-leading order value
$\PcX = 0.37 \pm 0.06$ is modified to $\PcX = 0.38 \pm 0.04$ at the 
next-to-next-to-leading order level with the latter error fully
dominated by the uncertainty in $\mc (\mc)$. We present tables for
$\PcX$ as a function of $\mc (\mc)$ and $\as (\MZ)$ and a very
accurate analytic formula that summarizes these two dependences as
well as the dominant theoretical uncertainties. Adding the recently 
calculated long-distance contributions we find $\BR (\Ktopinunu) =
\left ( 8.0 \pm 1.1 \right ) \times 10^{-11}$ with the present
uncertainties in $\mc (\mc)$ and the Cabibbo-Kobayashi-Maskawa
elements being the dominant individual sources in the quoted error. We
also emphasize that improved calculations of the long-distance
contributions to $\Ktopinunu$ and of the isospin breaking corrections
in the evaluation of the weak current matrix elements from $\Ktopienu$
would be valuable in order to increase the potential of the two golden
$\Ktopinunus$ decays in the search for new physics.  

\vspace*{1.0cm}

\newpage  

\section{Introduction} 
\label{sec:intro}

The rare decay $\Ktopinunu$ plays together with $\KLtopinunu$ an
outstanding role in the field of flavor changing neutral current
(FCNC) processes both in the standard model (SM) \cite{UT} and in
all of its extensions \cite{Gino, Buras:2004uu}. The main reason for
this is its theoretical cleanness and its large sensitivity to
short-distance QCD effects that can be systematically calculated using
an effective field theory framework. The hadronic matrix element of
this decay can be extracted, including isospin breaking corrections
\cite{Marciano:1996wy}, from the accurately measured leading
semileptonic decay $\Ktopienu$, and the remaining long-distance
contributions \cite{LD} turn out to be small \cite{Isidori:2005xm},
and in principle calculable by means of lattice QCD
\cite{Isidori:2005tv}.       

Consequently the SM decay rate of $\Ktopinunu$ can be expressed almost
entirely in terms of the Cabibbo-Kobayashi-Maskawa (CKM) \cite{CKM}
parameters, the top and the charm quark mass, and the strong coupling
constant $\as (\MZ)$ that enters the QCD corrections calculated within
renormalization group (RG) improved perturbation theory. Beyond the SM
additional parameters like new couplings and masses of new heavy
particles will be present in the decay rate, but from the point of
view of hadronic effects, the theoretical cleanness of the
prediction will not be affected by these non-standard contributions.  

In view of this, the theoretical uncertainties in the decay rate of
$\Ktopinunu$ are at leading order essentially only of perturbative
origin and in order to be able to test the SM and its extensions to a
high degree of precision it is important to evaluate the first
non-trivial and higher order QCD corrections to this decay mode.   

To be specific, the low-energy effective Hamiltonian for the
$\Ktopinunus$ system can be written in the SM as follows
\cite{Buchalla:1993wq, Buchalla:1998ba}   
\beq \label{eq:Heff}
\Heff = \f{4 \GF}{\sqrt{2}} \f{\aem}{2 \pi \sws} \sum_{\ell = e, \mu,
\tau} \left ( \lambda_c X^\ell (x_c) + \lambda_t X(x_t) \right ) (
\bar{s}_\sL \gamma_\mu d_\sL ) ( \bar {\nu_\ell}_\sL \gamma^{\mu}
{\nu_\ell}_\sL ) \, .      
\eeq
Here $\GF$, $\aem$, and $\sws$ denote the Fermi constant, the
electromagnetic coupling, and the weak mixing angle, respectively. The
sum over $\ell$ extends over all lepton flavors, $\lambda_i =
V_{is}^\ast V_{id}$ are the relevant CKM factors and $f_\sL$ are
left-handed fermion fields. The dependence on the charged lepton mass
is negligible for the top quark contribution. In the charm quark
sector this is the case only for the electron and the muon but not for
the tau lepton.  

The function $X (x_t)$ in \Eq{eq:Heff} depends on the top quark
$\MSbar$ \cite{Bardeen:1978yd} mass through $x_t = \mt^2
(\mut)/\MW^2$. It originates from $Z$-penguin and electroweak box
diagrams with an internal top quark. Sample diagrams are shown in
\Fig{fig:penguinandbox}. As the relevant operator has a vanishing
anomalous dimension and the energy scales involved are of the order of
the electroweak scale or higher, the function $X (x_t)$ can be
calculated within ordinary perturbation theory. It is known through
next-to-leading order (NLO) \cite{Buchalla:1998ba, X,
Misiak:1999yg}. The inclusion of these $\ord (\as)$ corrections
allowed to reduce the $\pm 6 \%$ uncertainty due to the top quark
matching scale $\mut = \ord (\mt)$ present in the leading order (LO)
formula down to $\pm 1 \%$. Consequently the reached theoretical
accuracy on the top quark contribution to $\Ktopinunu$ and in the
amplitude of $\KLtopinunu$, where only $X (x_t)$ enters, is
satisfactory.      

The function $X^\ell (x_c)$ in \Eq{eq:Heff} relevant only for
$\Ktopinunu$ depends on the charm quark $\MSbar$ mass through $x_c
= \mc^2 (\muc)/\MW^2$. As now both high- and low-energy scales,
namely $\muh = \ord (\MW)$ and $\muc = \ord (\mc)$ are involved, a
complete RG analysis of this term is required. In this manner, large
logarithms $\ln (\muc^2/\muh^2)$ are resummed to all orders in
$\as$. At LO such an analysis has been performed in \cite{LO}. The
large scale uncertainty due to $\muc$ of $\pm 26 \%$ in this result 
was a strong motivation for the NLO analysis of this contribution
\cite{Buchalla:1993wq, Buchalla:1998ba}.     

{%
\begin{figure}[!t]
\begin{center}
$\begin{array}{c@{\hspace{15mm}}c}
\scalebox{0.65}{
\begin{picture}(150,134) (30,-45)
\SetWidth{0.5}
\SetColor{Black}
\Photon(75,74)(135,74){6}{5}
\ArrowLine(135,74)(180,89)
\ArrowLine(105,14)(135,74)
\Photon(105,14)(105,-31){6}{4}
\Vertex(75,74){2.83}
\Vertex(135,74){2.83}
\Vertex(105,14){2.83}
\ArrowLine(75,74)(105,14)
\ArrowLine(30,89)(75,74)
\Text(125,-8)[]{\Large{\Black{$Z$}}}
\Text(151,42)[]{\Large{\Black{$u, c, t$}}}
\Text(58,42)[]{\Large{\Black{$u, c, t$}}}
\Text(105,94)[]{\Large{\Black{$W$}}}
\Text(157,96)[]{\Large{\Black{$d$}}}
\Text(54,94)[]{\Large{\Black{$s$}}}
\end{picture}
} & 
\scalebox{0.65}{
\begin{picture}(193,157) (17,-26)
\SetWidth{0.5}
\SetColor{Black}
\ArrowLine(210,112)(165,97)
\ArrowLine(165,7)(210,-8)
\Vertex(75,7){2.83}
\Vertex(75,97){2.83}
\Vertex(165,97){2.83}
\Vertex(165,7){2.83}
\Text(120,117)[]{\Large{\Black{$W$}}}
\Text(186,118)[]{\Large{\Black{$\nu$}}}
\Text(200,52)[]{\Large{\Black{$e,\mu, \tau$}}}
\Text(120,-13)[]{\Large{\Black{$W$}}}
\Text(185,-18)[b]{\Large{\Black{$\nu$}}}
\Photon(75,97)(165,97){6}{8}
\Photon(75,7)(165,7){6}{8}
\ArrowLine(165,97)(165,7)
\ArrowLine(30,112)(75,97)
\Text(54,118)[]{\Large{\Black{$s$}}}
\ArrowLine(75,97)(75,7)
\ArrowLine(75,7)(30,-8)
\Text(42,52)[]{\Large{\Black{$u, c, t$}}}
\Text(52,-14)[]{\Large{\Black{$d$}}}
\end{picture}
}
\end{array}$
\vspace{2mm}
\caption{Examples of $Z$-penguin and electroweak box diagrams that
contribute both to $\Ktopinunu$ and $\KLtopinunu$. Here and in the
following we do not display the neutrino line attached to the
$Z$-boson.}   
\label{fig:penguinandbox}
\end{center}
\end{figure}
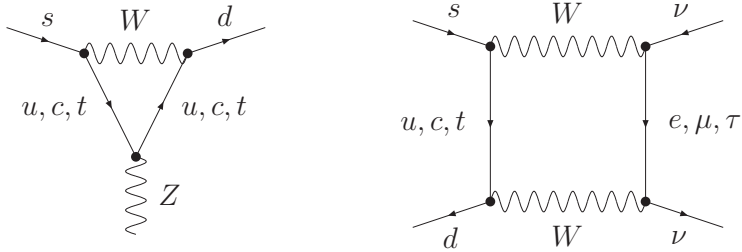
}%

Defining the phenomenologically useful parameter 
\beq \label{eq:defPcX}
\PcX = \f{1}{\lambda^4} \left ( \f{2}{3} X^e (x_c) + \f{1}{3} X^\tau
(x_c) \right ) \, , 
\eeq 
with $\lambda = | V_{us} |$, one finds for $\lambda = 0.2248$ at NLO
\cite{Buras:2005gr}\footnote{The numerical results presented here
differ somewhat from \cite{Buras:2005gr}. For the present numerical
evaluation the program used in \cite{Buras:2005gr} has been changed
slightly in order to implement the various theoretical errors in an
improved fashion.}   
\beq \label{eq:PcXNLO}
\PcX = 0.369 \pm 0.036_{\rm theory} \pm 0.033_{\mc} \pm 0.009_{\as} \,
,
\eeq
where the parametric errors correspond to the ranges of the charm
quark $\MSbar$ mass $\mc (\mc)$ and the strong coupling constant $\as
(\MZ)$ given in \Tab{tab:input}. The theoretical error summarizes 
uncertainties due to various scales and different methods of computing
$\as (\muc)$ from $\as (\MZ)$. Details on how the quoted errors have
been obtained will be given in \Sec{sec:numerics}.       

Provided $\PcX$ is known with a sufficient precision, a measurement
of $\Ktopinunu$, either alone or together with one of $\KLtopinunu$,
allows for precise determinations of the CKM parameters \cite{UT}. The
comparison of this standard unitarity triangle (UT) with the one from
$B$-physics offers a stringent and unique test of the SM. In
particular for the branching ratios $\BR (\Ktopinunu)$ and $\BR
(\KLtopinunu)$ close to their SM predictions, one finds that a given
uncertainty $\sigma (\PcX)$ translates into \cite{Buras:2005gr}     
\begin{align} \label{eq:ckmerrors}
\f{\sigma \left ( | V_{td} | \right)}{| V_{td} |} & = \pm 0.41 \, 
\f{\sigma \left (\PcX \right )}{\PcX} \, , \non \\[1mm]
\f{\sigma \left ( \sin 2 \beta \right)}{\sin 2 \beta} & = \pm 0.34 \,
\f{\sigma \left (\PcX \right )}{\PcX} \, , \non \\[1mm] 
\f{\sigma \left ( \gamma \right)}{\gamma} & = \pm 0.83 \, \f{\sigma
\left (\PcX \right )}{\PcX} \, ,  
\end{align}
with similar formulas given in \cite{Buras:2004uu}. Here $V_{td}$ is
the element of the CKM matrix and $\beta$ and $\gamma$ are the angles
in the standard UT. As the uncertainties in \Eq{eq:PcXNLO} coming from
the charm quark mass and the CKM parameters should be decreased in the
coming years it is also desirable to reduce the theoretical
uncertainty in $\PcX$.  

To this end, we extend here the NLO analysis of $\PcX$ presented in
\cite{Buchalla:1993wq, Buchalla:1998ba} to the next-to-next-to-leading
order (NNLO) \cite{Buras:2005gr}. We encounter several new features,
which were absent in lower orders. First, closed quark loops in gluon
propagators occur, resulting in a novel dependence of $\PcX$ on the top
quark mass and in non-trivial matching corrections at the bottom quark
threshold scale $\mub = \ord (\mb)$. Second, the contributions from
the vector component of the $Z$-boson coupling are non-trivial at NNLO
and are only found to vanish in the sum of several contributions,
which involve a flavor off-diagonal wave function renormalization. 
Third, the presence of anomalous triangle diagrams involving a top quark
loop, two gluons, and a $Z$-boson makes it necessary to introduce a
Chern-Simons operator \cite{Chern:1974ft, Larin:1993tq} in order to
obtain the correct anomalous Ward identity of the axial-vector current
\cite{ABJ}. The inclusion of such a Chern-Simons term is also required
to compensate for the anomalous contributions from triangle diagrams
with a bottom quark loop. Since all these effects arise first at NNLO,
they are not included in the theoretical uncertainty quoted in
\Eq{eq:PcXNLO}, which has been estimated from the variation of scales
and different methods of evaluating $\as (\muc)$ from $\as (\MZ)$. The
only way to control their size is to compute them explicitly, which is a
further strong motivation for our NNLO calculation.

Our paper is organized as follows. In \Sec{sec:master} we give
formulas for $\PcX$ and $\BR (\Ktopinunu)$ at NNLO in a form suitable
for phenomenological applications. In particular we present tables
that show $\PcX$ for different values of $\as (\MZ)$, $\muc$, and $\mc
(\mc)$ and we give a simple analytic formula for $\PcX$ that
approximates the exact numerical result with high
accuracy. \Sec{sec:guide} is meant to be a guide to the subsequent  
\Secsto{sec:CC}{sec:B} that describe our calculation in
detail. These sections are naturally rather technical and might be
skipped by readers mainly interested in phenomenological applications
of our result. \Sec{sec:final} contains another accurate approximate
formula for $\PcX$ that summarizes the dominant parametric and
theoretical uncertainties. In \Sec{sec:numerics} we present the
numerical analysis of the NNLO formulas. In particular we analyze
various scale uncertainties that are drastically reduced by going from
NLO to NNLO. We present the result for $\BR (\Ktopinunu)$ and $\BR
(\KLtopinunu)$ and we investigate the parametric and theoretical
uncertainties in the determination of the CKM parameters with the
latter being significantly reduced through our calculation. In the
course of this section we also present results provided by the
\ckmfitter \cite{Charles:2004jd} and the \utfit
\cite{Ciuchini:2000de}. We conclude in \Sec{sec:summary}. Some
technical details as well as additional material has been relegated to
the appendices.  

\section{Master Formulas at NNLO}
\label{sec:master}

\subsection{Preliminaries}
\label{subsec:pre}

In this section we will present the formula for $\BR (\Ktopinunu)$
based on the low-energy effective Hamiltonian given in \Eq{eq:Heff}
extended to include the recently calculated contributions of
dimension-eight four fermion operators generated at the charm quark
scale $\muc$, and of genuine long-distance contributions which can be
described within the framework of chiral perturbation theory
\cite{Isidori:2005xm}. These contributions can be effectively included
by adding  
\beq \label{eq:dPcu}
\dPcu = 0.04 \pm 0.02 \, ,
\eeq
to the relevant parameter $\PcX$. The quoted error in $\dPcu$ can in
principle be reduced by means of lattice QCD \cite{Isidori:2005tv}.   

For completeness we will also recall the formula for $\BR
(\KLtopinunu)$ which we will use in \Sec{sec:numerics} to obtain its
updated SM value in view of the recently modified value of the top
quark mass \cite{Group:2005cc}.   

\subsection{Branching Ratio for {\boldmath $\Ktopinunu$}}
\label{subsec:BRKp}

After summation over the three neutrino flavors the resulting
branching ratio for $\Ktopinunu$ can be written as
\cite{Isidori:2005xm, Buchalla:1993wq, Buchalla:1998ba}  
\beq \label{eq:BR}
\BR ( \Ktopinunu) = \kappa_+ \left [ \left ( \f{\im
\lambda_t}{\lambda^5} X (x_t) \right )^2 + \left ( \f{\re 
\lambda_t}{\lambda^5} X (x_t) + \f{\re \lambda_c}{\lambda} \left (\PcX
+ \dPcu \right ) \right )^2 \right ] \, , 
\eeq
with
\beq \label{eq:kappap}
\kappa_+ = r_{K^+} \f{3 \aem^2 \, \BR (\Ktopienu)}{2 \pi^2 \swq} 
\lambda^8 = \left ( 5.04 \pm 0.17 \right ) \times 10^{-11} \left (
\f{\lambda}{0.2248} \right )^8 \, . 
\eeq
Here the parameter $r_{K^+} = 0.901 \pm 0.027$ summarizes isospin
breaking corrections in relating $\Ktopinunu$ to $\Ktopienu$
\cite{Marciano:1996wy}. The apparent strong dependence of $\BR
(\Ktopinunu)$ on $\lambda$ is spurious as both $\PcX$ and $\dPcu$ are
proportional to $1/\lambda^4$. In quoting the value for $\PcX$ and
$\BR (\Ktopinunu)$ we will set $\lambda = 0.2248$. $\aem$ and $\sws$
entering $\BR (\Ktopinunu)$ are naturally evaluated at the electroweak
scale \cite{Bobeth:2003at}. Then the leading term in the heavy top
expansion of the electroweak two-loop corrections to $X (x_t)$ amounts
to typically $-1 \%$ for the $\MSbar$ definition of $\aem$ and $\sws$
\cite{Buchalla:1997kz}. In obtaining the numerical value of
\Eq{eq:kappap} we have employed $\aem = \aemMSbar (\MZ) = 1/127.9$,
$\sws = \swsMSbar = 0.231$, and $\BR (\Ktopienu) = \left ( 4.93 \pm
0.07 \right ) \times 10^{-2}$ \cite{PDG}. We remark that in writing
$\BR (\Ktopinunu)$ in the form of \Eq{eq:BR} we have omitted a term
proportional to $\left ( X^e (x_c) - X^\tau (x_c) \right )^2$. Its
effect on $\BR (\Ktopinunu)$ is around $+0.2 \%$.

The function $X (x_t)$ entering \Eqs{eq:Heff}, \eq{eq:BR} and
\eq{eq:BRL} is given in NLO accuracy by 
\beq \label{eq:X} 
X (x_t) = X_0 (x_t) + \f{\as (\mut)}{4 \pi} X_1 (x_t) = \eta_X X_0
(x_t (\mt)) \, , 
\eeq
with 
\beq \label{eq:etaX}
\eta_X = 0.986 \pm 0.009 \, .
\eeq
The contribution stemming from $Z$-penguin and electroweak box
diagrams without QCD corrections reads \cite{x} 
\beq \label{eq:X0} 
X_0 (x_t) = -\f{2 x_t + x_t^2}{8 (1 - x_t)} - \f{6 x_t - 3 x_t^2}{8 (1
- x_t)^2} \ln x_t \, ,  
\eeq
while the QCD corrections to it take the following form
\cite{Buchalla:1998ba, X, Misiak:1999yg}  
\beq \label{eq:X1}
\begin{split}
X_1 (x_t) & = -\f{29 x_t - x_t^2 - 4 x_t^3}{3 (1 - x_t)^2} - \f{x_t
+ 9 x_t^2 - x_t^3 - x_t^4}{(1 - x_t)^3} \ln x_t \\
& + \f{8 x_t + 4 x_t^2 + x_t^3 - x_t^4}{2 (1 - x_t)^3} \ln^2 x_t -
\f{4 x_t - x_t^3}{(1 - x_t)^2} \li (1 - x_t) + 8 x_t \f{\partial X_0
(x_t)}{\partial x_t} \ln \f{\mut^2}{\MW^2} \, .   
\end{split}
\eeq
Here $\li (x) = -\int^x_0 dt \, \ln (1 - t)/t$. The explicit
$\mut$-dependence of the last term in \Eq{eq:X1} cancels to the
considered order the $\mut$-dependence of the leading term $X_0 (x_t
(\mut))$. The factor $\eta_X$ summarizes the NLO corrections. Its
error has been obtained by varying $\mut$ in the range $60 \! \GeV \le
\mut \le 240 \! \GeV$ on the left-hand side of \Eq{eq:X} while keeping
$\mut$ fixed at $\mt = \mt (\mt)$ on the right-hand side of the same
equation. The leftover $\mut$-dependence in $X (x_t)$ is slightly
below $\pm 1 \%$.        

The uncertainty in $X (x_t)$ is then dominated by the experimental
error in the mass of the top quark. Converting the top quark pole mass
of $\mtpole = \left ( 172.7 \pm 2.9 \right ) \!\! \GeV$
\cite{Group:2005cc} at three loops to $\mt (\mtpole)$
\cite{Melnikov:2000qh} and relating $\mt (\mtpole)$ to $\mt (\mt) =
\left ( 163.0 \pm 2.8 \right ) \!\! \GeV$ using one-loop accuracy, we 
find   
\beq \label{eq:Xnum}
X (x_t) = 1.464 \pm 0.041 \, .
\eeq
The given uncertainty combines linearly an error of $\pm 0.028$ due to
the error of $\mt (\mt)$ and an error of $\pm 0.013$ obtained by
varying $\mut$ in the range given above.                  

{%
\renewcommand{\arraystretch}{1.25}
\begin{table}[!t]
\begin{center}
\begin{tabular}{|c|c|c|c|c|c|}
\hline  
\multicolumn{1}{|c|}{} & \multicolumn{5}{c|}{$\PcX$} \\
\hline 
$\as (\MZ) \hspace{1.5mm} \backslash \hspace{1.5mm} \muc
\hspace{1.5mm} [\!\!\GeV]$ & 1.0 & 1.5 & 2.0 & 2.5 & 3.0 \\
\hline 
0.115 & 0.393 & 0.397 & 0.395 & 0.392 & 0.388 \\
0.116 & 0.389 & 0.394 & 0.391 & 0.388 & 0.383 \\
0.117 & 0.384 & 0.390 & 0.387 & 0.383 & 0.379 \\
0.118 & 0.380 & 0.386 & 0.383 & 0.379 & 0.374 \\
0.119 & 0.375 & 0.381 & 0.379 & 0.374 & 0.369 \\
0.120 & 0.370 & 0.377 & 0.374 & 0.369 & 0.364 \\
0.121 & 0.365 & 0.372 & 0.369 & 0.364 & 0.359 \\
0.122 & 0.359 & 0.368 & 0.364 & 0.359 & 0.354 \\
0.123 & 0.353 & 0.363 & 0.359 & 0.354 & 0.348 \\
\hline 
\end{tabular} 
\vspace{2mm}
\caption{The parameter $\PcX$ in NNLO approximation for various values
of $\as (\MZ)$ and $\muc$. In quoting the numerical values for $\PcX$
we have set $\lambda = 0.2248$, $\mt (\mt) = 163.0 \! \GeV$, $\mc
(\mc) = 1.30 \! \GeV$, $\muh = 80.0 \! \GeV$, and $\mub = 5.0 \!
\GeV$.}                 
\label{tab:pcasmuc}
\end{center}
\end{table}
}%

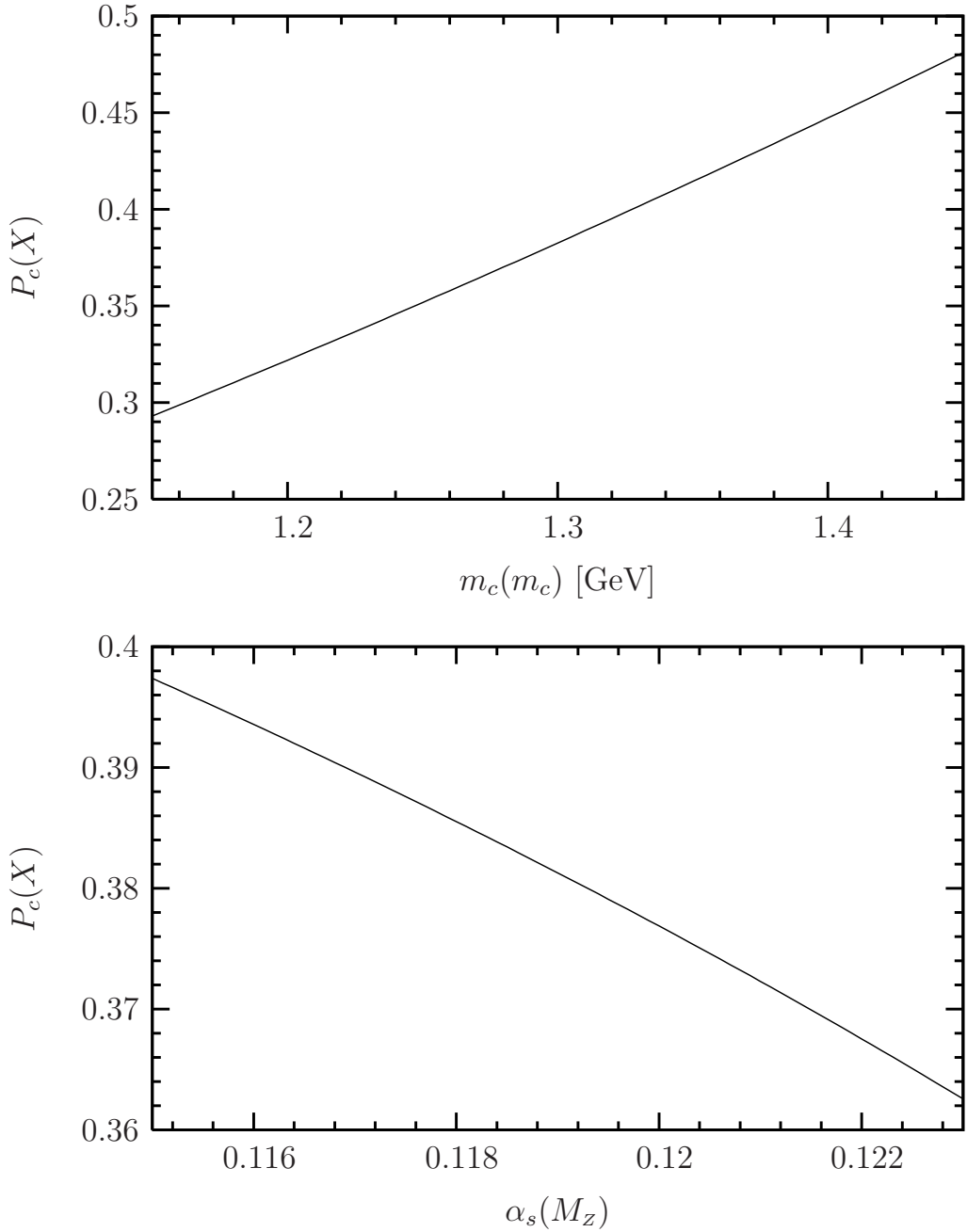
\begin{figure}[!p]
\begin{center}
\scalebox{1.1}{
\begingroup%
  \makeatletter%
  \newcommand{\GNUPLOTspecial}{%
    \@sanitize\catcode`\%=14\relax\special}%
  \setlength{\unitlength}{0.1bp}%
\begin{picture}(3600,2160)(0,0)%
\special{psfile=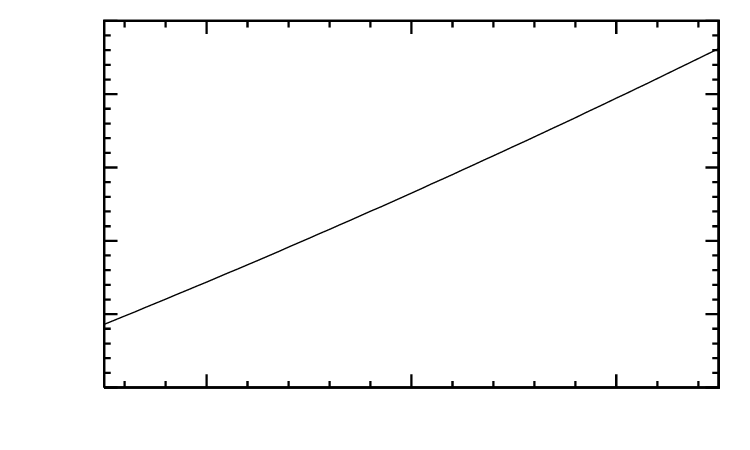 llx=0 lly=0 urx=360 ury=216 rwi=3600}
\put(1975,0){\makebox(0,0){\scalebox{1.0}{$\mc(\mc)$ [GeV]}}}%
\put(100,1180){%
\special{ps: gsave currentpoint currentpoint translate
270 rotate neg exch neg exch translate}%
\makebox(0,0)[b]{\shortstack{\scalebox{1.0}{$\PcX$}}}%
\special{ps: currentpoint grestore moveto}%
}%
\put(2958,200){\makebox(0,0){ 1.4}}%
\put(1975,200){\makebox(0,0){ 1.3}}%
\put(992,200){\makebox(0,0){ 1.2}}%
\put(450,2060){\makebox(0,0)[r]{ 0.5}}%
\put(450,1708){\makebox(0,0)[r]{ 0.45}}%
\put(450,1356){\makebox(0,0)[r]{ 0.4}}%
\put(450,1004){\makebox(0,0)[r]{ 0.35}}%
\put(450,652){\makebox(0,0)[r]{ 0.3}}%
\put(450,300){\makebox(0,0)[r]{ 0.25}}%
\end{picture}%
\endgroup
 }\\[5mm]
\scalebox{1.1}{
\begingroup%
  \makeatletter%
  \newcommand{\GNUPLOTspecial}{%
    \@sanitize\catcode`\%=14\relax\special}%
  \setlength{\unitlength}{0.1bp}%
\begin{picture}(3600,2160)(0,0)%
\special{psfile=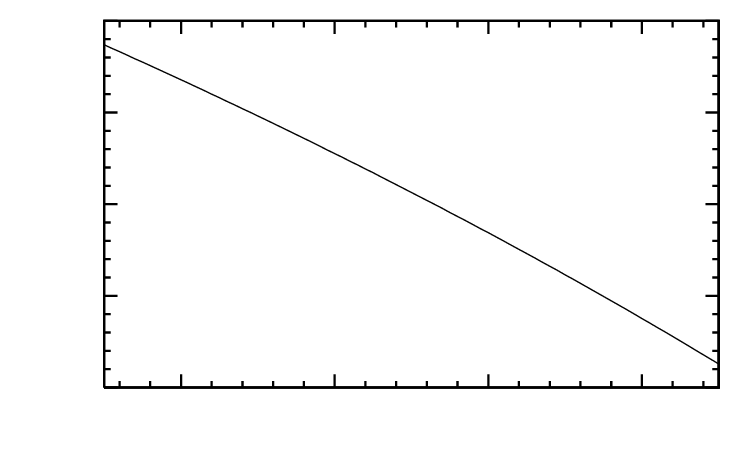 llx=0 lly=0 urx=360 ury=216 rwi=3600}
\put(1975,0){\makebox(0,0){\scalebox{1.0}{$\as(\MZ)$}}}%
\put(100,1180){%
\special{ps: gsave currentpoint currentpoint translate
270 rotate neg exch neg exch translate}%
\makebox(0,0)[b]{\shortstack{\scalebox{1.0}{$\PcX$}}}%
\special{ps: currentpoint grestore moveto}%
}%
\put(3081,200){\makebox(0,0){ 0.122}}%
\put(2344,200){\makebox(0,0){ 0.12}}%
\put(1606,200){\makebox(0,0){ 0.118}}%
\put(869,200){\makebox(0,0){ 0.116}}%
\put(450,2060){\makebox(0,0)[r]{ 0.4}}%
\put(450,1620){\makebox(0,0)[r]{ 0.39}}%
\put(450,1180){\makebox(0,0)[r]{ 0.38}}%
\put(450,740){\makebox(0,0)[r]{ 0.37}}%
\put(450,300){\makebox(0,0)[r]{ 0.36}}%
\end{picture}%
\endgroup
 }
\end{center}
\vspace{0mm}
\caption{The parameter $\PcX$ at NNLO as a function of $\mc (\mc)$
(upper plot) and $\as (\MZ)$ (lower plot). In obtaining the numerical
values for $\PcX$ we have set $\lambda = 0.2248$, $\mt (\mt) = 163.0
\! \GeV$, $\muh = 80.0 \! \GeV$, $\mub = 5.0 \! \GeV$, $\muc = 1.50
\! \GeV$, $\as (\MZ) = 0.1187$ (upper plot), and $\mc (\mc) = 1.30 \!
\GeV$ (lower plot).}         
\label{fig:mcasplot}
\end{figure}

As opposed to $X (x_t)$ the charm quark contribution, represented by
the parameter $\PcX$ in \Eq{eq:defPcX}, involves several different
scales like $\muh$, $\mub$, and $\muc$. In order to control the size of
the perturbative corrections to $\PcX$ the large logarithms associated
with these scales have to be resummed to all orders in $\as$ using RG
techniques. Keeping terms to first order in $\as$, the perturbative
expansion of $\PcX$ has the following general structure \beq
\label{eq:PcXPT} \PcX = \frac{4 \pi}{\as (\muc)} \PcXLO + \PcXNLO +
\f{\as (\muc)}{4 \pi} \PcXNNLO \, , \eeq where we have suppressed the
dependence of the expansion coefficients $\PcXk$ on the involved
physical and unphysical mass scales for simplicity. The leading term
$\PcXLO$ has been worked out in \cite{LO} while the NLO correction
$\PcXNLO$ has been calculated in \cite{Buchalla:1993wq,
  Buchalla:1998ba}.

The main goal of the present paper is the calculation of the NNLO term
$\PcXNNLO$. As indicated by the theoretical error in \Eq{eq:PcXNLO}, 
the sum of the first two terms in \Eq{eq:PcXPT} still exhibits sizable
unphysical scale dependences, in particular the one on
$\muc$. Besides, the NLO value of $\PcX$ depends in a non-negligible
way on the method used to compute $\as (\muc)$ from $\as (\MZ)$
\cite{Buras:2005gr}. The observed numerical difference is due to
higher order terms in $\as$ and has to be regarded as part of the
theoretical error. This source of uncertainty has not been taken into
account in previous NLO analyses of the charm quark contribution
\cite{Buras:2004uu, Buchalla:1993wq, Buchalla:1998ba}.     

{%
\renewcommand{\arraystretch}{1.25}
\begin{table}[!t]
\begin{center}
\begin{tabular}{|c|c|c|c|c|c|c|c|}
\hline  
\multicolumn{1}{|c|}{} & \multicolumn{7}{c|}{$\PcX$} \\
\hline 
$\as (\MZ) \hspace{1.5mm} \backslash \hspace{1.5mm} \mc (\mc)
\hspace{1.5mm} [\!\!\GeV]$ & 1.15 & 1.20 & 1.25 & 1.30 & 1.35 & 1.40 &
1.45 \\
\hline 
0.115 & 0.307 & 0.336 & 0.366 & 0.397 & 0.430 & 0.463 & 0.497 \\
0.116 & 0.303 & 0.332 & 0.362 & 0.394 & 0.426 & 0.459 & 0.493 \\
0.117 & 0.300 & 0.329 & 0.359 & 0.390 & 0.422 & 0.455 & 0.489 \\
0.118 & 0.296 & 0.325 & 0.355 & 0.386 & 0.417 & 0.450 & 0.484 \\
0.119 & 0.292 & 0.321 & 0.350 & 0.381 & 0.413 & 0.446 & 0.480 \\
0.120 & 0.288 & 0.316 & 0.346 & 0.377 & 0.409 & 0.441 & 0.475 \\
0.121 & 0.283 & 0.312 & 0.342 & 0.372 & 0.404 & 0.437 & 0.470 \\
0.122 & 0.279 & 0.307 & 0.337 & 0.368 & 0.399 & 0.432 & 0.465 \\
0.123 & 0.274 & 0.303 & 0.332 & 0.363 & 0.394 & 0.426 & 0.460 \\
\hline 
\end{tabular} 
\vspace{2mm}
\caption{The parameter $\PcX$ in NNLO approximation for various values
of $\as (\MZ)$ and $\mc (\mc)$. In quoting the numerical values for
$\PcX$ we have set $\lambda = 0.2248$, $\mt (\mt) = 163.0 \! \GeV$,
$\muh = 80.0 \! \GeV$, $\mub = 5.0 \! \GeV$, and $\muc = 1.50 \!
\GeV$.}                  
\label{tab:pcasmc}
\end{center}
\end{table}
}%

As we will demonstrate in \Sec{sec:numerics}, the inclusion of
$\PcXNNLO$ removes essentially the entire sensitivity of $\PcX$ on
$\muc$ and on higher order terms in $\as$ that effect the evaluation
of $\as (\muc)$ from $\as (\MZ)$. As a result, the final theoretical
error in $\PcX$ is reduced from $\pm 9.8 \%$ at NLO down to $\pm 2.4
\%$ at NNLO. After our calculation the theoretical accuracy on the 
charm quark contribution to $\Ktopinunu$ is thus also satisfactory.    

The analytic formula for the sum of the first two terms in
\Eq{eq:PcXPT} can be found in \cite{Buchalla:1993wq,
Buchalla:1998ba}. The formula for $\PcXNNLO$ is given in
\Secsto{sec:Z}{sec:final}. 
Setting $\lambda = 0.2248$, $\mt (\mt) = 163.0 \! \GeV$,
$\muh = 80.0 \! \GeV$, $\mub = 5.0 \! \GeV$, and $\muc = 1.50 \! \GeV$,
we derive an approximate formula for $\PcX$ as a function of $\mc
(\mc)$ and $\as (\MZ)$. It reads   
\beq \label{eq:mformula}
\PcX = 0.379 \left ( \f{\mc (\mc)}{1.30 \! \GeV} \right )^{2.155}
\left ( \f{\as (\MZ)}{0.1187} \right )^{-1.417} \, ,
\eeq
and approximates the exact NNLO result with an accuracy of better
than $\pm 1.0 \%$ in the ranges $1.15 \! \GeV \le \mc (\mc) \le 1.45 \!
\GeV$ and $0.1150 \le \as (\MZ) \le 0.1230$. 

The dependence of $\PcX$ on $\mc (\mc)$ and $\as (\MZ)$ can be seen in
\Fig{fig:mcasplot}, while in \Tabsand{tab:pcasmuc}{tab:pcasmc} we
present the exact values for $\PcX$ for different values of $\as
(\MZ)$, $\muc$, and $\mc (\mc)$. We observe that the $\muc$-dependence
is almost negligible and the dependence on $\as (\MZ)$ is small. On
the other hand the dependence on $\mc (\mc)$ is sizable. A reduction
of the error in $\mc (\mc)$, which is dominated by theoretical
uncertainties, is thus an important goal in connection with
$\Ktopinunu$.      

\begin{figure}[!p]
\begin{center}
\scalebox{1.1}{
\begingroup%
  \makeatletter%
  \newcommand{\GNUPLOTspecial}{%
    \@sanitize\catcode`\%=14\relax\special}%
  \setlength{\unitlength}{0.1bp}%
\begin{picture}(3600,2160)(0,0)%
\special{psfile=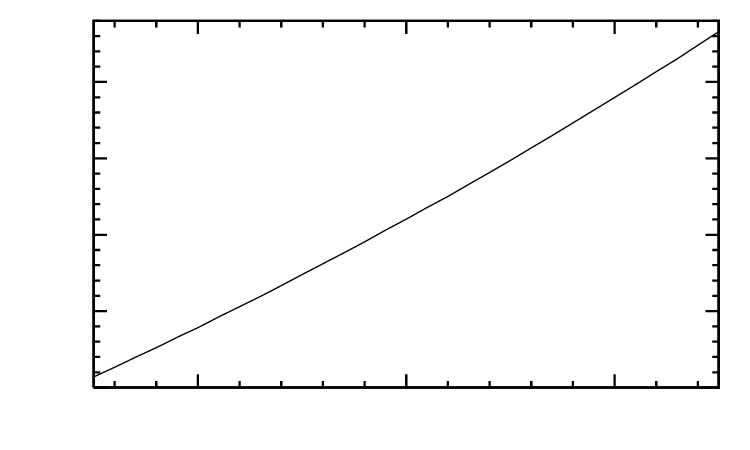 llx=0 lly=0 urx=360 ury=216 rwi=3600}
\put(1950,0){\makebox(0,0){\scalebox{1.0}{$\mc(\mc)$ [GeV]}}}%
\put(100,1180){%
\special{ps: gsave currentpoint currentpoint translate
270 rotate neg exch neg exch translate}%
\makebox(0,0)[b]{\shortstack{\scalebox{1.0}{$\BR(\Ktopinunu)~~[10^{-11}]$}}}%
\special{ps: currentpoint grestore moveto}%
}%
\put(2950,200){\makebox(0,0){ 1.4}}%
\put(1950,200){\makebox(0,0){ 1.3}}%
\put(950,200){\makebox(0,0){ 1.2}}%
\put(400,1767){\makebox(0,0)[r]{ 9}}%
\put(400,1400){\makebox(0,0)[r]{ 8.5}}%
\put(400,1033){\makebox(0,0)[r]{ 8}}%
\put(400,667){\makebox(0,0)[r]{ 7.5}}%
\put(400,300){\makebox(0,0)[r]{ 7}}%
\end{picture}%
\endgroup
 }\\[5mm]
\scalebox{1.1}{
\begingroup%
  \makeatletter%
  \newcommand{\GNUPLOTspecial}{%
    \@sanitize\catcode`\%=14\relax\special}%
  \setlength{\unitlength}{0.1bp}%
\begin{picture}(3600,2160)(0,0)%
\special{psfile=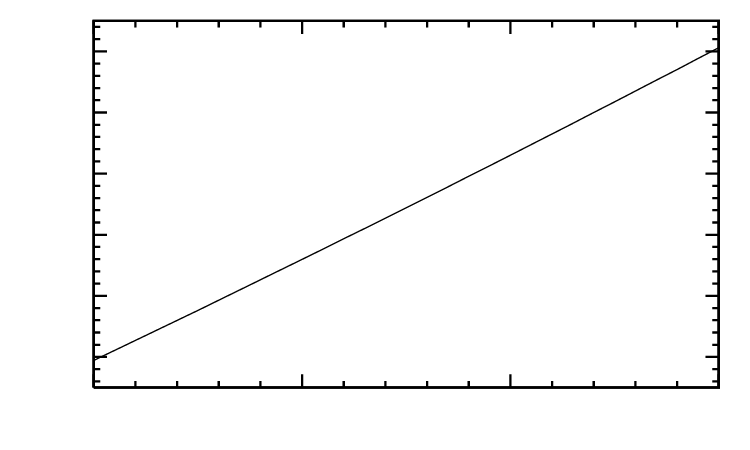 llx=0 lly=0 urx=360 ury=216 rwi=3600}
\put(1950,0){\makebox(0,0){\scalebox{1.0}{$\mtpole$ [GeV]}}}%
\put(100,1180){%
\special{ps: gsave currentpoint currentpoint translate
270 rotate neg exch neg exch translate}%
\makebox(0,0)[b]{\shortstack{\scalebox{1.0}{$\BR(\Ktopinunu)~~[\GeV]$}}}%
\special{ps: currentpoint grestore moveto}%
}%
\put(3450,200){\makebox(0,0){ 180}}%
\put(2450,200){\makebox(0,0){ 175}}%
\put(1450,200){\makebox(0,0){ 170}}%
\put(450,200){\makebox(0,0){ 165}}%
\put(400,1913){\makebox(0,0)[r]{ 8.6}}%
\put(400,1620){\makebox(0,0)[r]{ 8.4}}%
\put(400,1327){\makebox(0,0)[r]{ 8.2}}%
\put(400,1033){\makebox(0,0)[r]{ 8}}%
\put(400,740){\makebox(0,0)[r]{ 7.8}}%
\put(400,447){\makebox(0,0)[r]{ 7.6}}%
\end{picture}%
\endgroup
 }
\end{center}
\vspace{0mm}
\caption{The branching ratio $\BR (\Ktopinunu)$ at NNLO as a function
of $\mc (\mc)$ (upper plot) and $\mtpole$ (lower plot). In obtaining
the numerical values for $\BR (\Ktopinunu)$ we have used $\muh = 80.0
\! \GeV$, $\mub = 5.0 \! \GeV$, and $\muc = 1.50 \! \GeV$, 
and set all input parameters to their central values as given in
\Tab{tab:input}.}          
\label{fig:BRmcmtplot}
\end{figure}

Employing the central values of the input parameters summarized in
\Tab{tab:input}, the dependence of $\BR (\Ktopinunu)$ on $\mc (\mc)$
is described by      
\beq \label{eq:mcformula}
\BR (\Ktopinunu) = 8.15 \left ( \f{\mc (\mc)}{1.30 \! \GeV} \right
)^{1.19} \times 10^{-11} \, ,
\eeq
which approximates the exact NNLO result with an accuracy of almost
$\pm 0.6 \%$ in the range $1.15 \! \GeV \le \mc (\mc) \le 1.45 \!
\GeV$. Similarly, the dependence on $\mtpole$ is given by  
\beq \label{eq:mtformula}
\BR (\Ktopinunu) = 8.11 \left ( \f{\mtpole}{172.7 \! \GeV} \right
)^{1.46} \times 10^{-11} \, ,
\eeq
which approximates the exact NNLO result with an accuracy of better
than $\pm 0.1 \%$ in the range $165.0 \! \GeV \le \mtpole \le 180.0 \!
\GeV$. The dependences in \Eqsand{eq:mcformula}{eq:mtformula} are
exhibited in \Fig{fig:BRmcmtplot}. We remark that the present analysis
of the UT is practically independent of the exact value of the top
quark mass. In obtaining both \Eqsand{eq:mcformula}{eq:mtformula} we
have therefore set for simplicity $\im \lambda_t$, $\re \lambda_t$, and
$\re \lambda_c$ to their central values as given in
\Tab{tab:input}. Furthermore we have used the numerical value of
$\PcX$ evaluated at $\muh = 80.0 \! \GeV$, $\mub = 5.0 \! \GeV$, and
$\muc = 1.5 \! \GeV$. A detailed numerical analysis of various
uncertainties in $\BR (\Ktopinunu)$ will be presented in
\Sec{sec:numerics}.     

\subsection{Branching Ratio for {\boldmath $\KLtopinunu$}}
\label{subsec:BRL}

In the case of $\KLtopinunu$ the charm quark contribution and the
long-distance effects are negligible so that the relevant branching
ratio is given simply as follows \cite{Buchalla:1993wq,
Buchalla:1998ba}   
\beq \label{eq:BRL}
\BR (\KLtopinunu) = \kappa_L \left ( \f{\im \lambda_t}{\lambda^5} X(
x_t) \right )^2 \, ,
\eeq
where
\beq \label{eq:kappaL}
\kappa_L = \kappa_+ \f{r_{K_L}}{r_{K^+}} \f{\tau(K_L)}{\tau(K^+)}=  
(2.20 \pm 0.07) \times 10^{-10} \left ( \f{\lambda}{0.2248} \right )^8
\, .
\eeq
Here we have summed over the three neutrino flavors and used
$\tau(K_L)/\tau(K^+) = 4.16 \pm 0.04$ \cite{PDG}. $r_{K_L} = 0.944 \pm
0.028$ is the isospin breaking correction from \cite{Marciano:1996wy}
with $\kappa_+$ given in \Eq{eq:kappap}. Due to the absence of $\PcX$
in \Eq{eq:BRL}, $\BR (\KLtopinunu)$ is plagued only by small
theoretical uncertainties coming from $\mut$ and $\kappa_L$. The total
parametric uncertainty stemming from $\mt (\mt)$ and $\im \lambda_t$
is on the other hand sizeable. The latter errors should be decreased
significantly in the coming years, so that a precise prediction for
$\BR (\KLtopinunu)$ should be possible in this decade.    

\subsection{Summary}
\label{subsec:summary}

As far as perturbative uncertainties are concerned, with the NNLO
correction to the charm quark contribution to $\BR (\Ktopinunu)$ at
hand, $\Ktopinunu$ has been put nearly on the same level as
$\KLtopinunu$. The leftover scale ambiguities are all small, so that
the reached theoretical accuracy is now satisfactory in both
decays. Similarly the errors due to uncertainties in $\as (\MZ)$ and
$\mt (\mt)$ are small. The future of precise predictions for $\BR
(\Ktopinunu)$ will depend primarily on the reduction of the errors in
$\mc (\mc)$ and in the CKM parameters, whereas $\BR (\KLtopinunu)$ is
practically only affected by the uncertainties in the CKM
elements. Non-negligible uncertainties arise in both cases also from
the theoretical error of the isospin breaking corrections and in the
case of $\BR (\Ktopinunu)$ from the uncertainty associated with the
long-distance corrections.        

On the other hand the determination of the CKM parameters and of the
UT from the $\Ktopinunus$ system will depend on the progress in the
determination of $\mc (\mc)$ and the measurements of both branching
ratios. Also a further reduction of the error in $| V_{cb} |$,
$r_{K^+}$, $r_{K_L}$, and $\dPcu$ would be very welcome in this
respect.   

\section{Guide to the NNLO Calculation}
\label{sec:guide}

In analogy to many other FCNC processes, perturbative QCD effects lead
to a sizable modification of the purely electroweak contribution to 
$\Ktopinunu$ by generating large logarithms of the form $L = \ln
(\muc^2/\muh^2)$. A suitable framework to achieve the necessary
resummation of the logarithmic enhanced corrections is the
construction of an effective field theory by integrating out the heavy
degrees of freedom. In this context short-distance QCD corrections can
be systematically calculated by solving the RG equation that governs
the scale dependence of the Wilson coefficient functions of the
relevant local operators built out of the light and massless SM
fields. 

As several new features, which were absent in LO and NLO, enter the RG
analysis of $\PcX$ at NNLO, the actual calculation is rather involved
and it is worthwhile to outline first its general structure. This will
be done in this section, whereas the details of the computation of the
different ingredients that are necessary to obtain the NNLO correction
to $\PcX$ will be described in \Secsto{sec:CC}{sec:B}.   

The key element of the RG analysis of the charm quark contribution to
the $\stodnunu$ transition is the mixing of the bilocal composite
operators 
\beq \label{eq:QPpm}
Q^P_{\pm} = -i \int \! d^4 x \hspace{1mm} T \big (
Q^c_\pm (x) Q_Z (0) - Q^u_\pm (x) Q_Z (0) \big ) \, , 
\eeq
and
\beq \label{eq:QB}
Q^B = -i \int \! d^4 x \hspace{1mm} T \big ( Q^c_3
(x) Q^c_4 (0) - Q^u_3 (x) Q^u_4 (0) \big ) \, , 
\eeq
where T denotes the usual time-ordering, into 
\beq \label{eq:Qv}
Q_\nu = \f{\mc^2}{\gs^2 \mul^{2 \eps}} \sum_{\ell = e, \mu, \tau}
\left (\bar s_\sL \gamma_\mu d_\sL \right ) \left ( \bar
{\nu_\ell}_\sL \gamma^\mu {\nu_\ell}_\sL \right ) \, .
\eeq
Here $\mc$ is the charm quark $\MSbar$ mass $\mc (\mul)$, and the
inverse powers of $\gs$ have been introduced for later convenience,
following \cite{Buchalla:1993wq}. One may arbitrarily shift such
factors from the Wilson coefficient into the definition of the
operator. The factor $\mul^{-2 \eps}$ stems from the relation
${\gs}_\bare = Z_\gs \gs \mul^\eps$, where $Z_\gs$ denotes the
renormalization constant of $\gs$, and the fact that $Q_\nu$ written
in terms of bare fields and parameters must be $\mul$-independent. All
bare quantities will carry the subscript $\bare$ hereafter. The
operators $Q_\pm^q$, $Q_3^q$, and $Q_4^q$ entering
\Eqsand{eq:QPpm}{eq:QB} will be defined as we proceed. 

Interestingly the charm quark contribution to the $\stodnunu$
amplitude involves large logarithms even in the absence of QCD
interactions, because $Q^P_{\pm}$ and $Q^B$ mix into $Q_\nu$ through
one-loop diagrams containing no gluon. The relevant Feynman graphs can
be seen in \Fig{fig:LL}. Factoring out $\GF$ and $\aem$ the charm
quark contribution to the $\stodnunu$ amplitude thus receives
corrections of $\ord (\as^n L^{n + 1})$ at LO, of $\ord (\as^n L^n)$
at NLO, and of $\ord (\as^n L^{n - 1})$ at NNLO. This structure of
large logarithms explains the peculiar expansion of $\PcX$ in
\Eq{eq:PcXPT} with the leading term being of $\ord (1/\as)$ rather
than $\ord (1)$. 

{%
\begin{figure}[!t]
\begin{center}
$\begin{array}{c@{\hspace{15mm}}c}
\scalebox{0.65}{
\begin{picture}(120,137) (45,-30)
\SetWidth{0.5}
\SetColor{Black}
\ArrowLine(45,92)(105,77)
\ArrowLine(105,77)(165,92)
\ArrowLine(105,2)(165,-13)
\ArrowArc(97.5,39.5)(38.24,-78.69,78.69)
\Text(75,99)[]{\Large{\Black{$\nu$}}}
\Text(135,99)[]{\Large{\Black{$\nu$}}}
\Text(160,38)[]{\Large{\Black{$u, c$}}}
\Text(75,-22)[]{\Large{\Black{$s$}}}
\Text(135,-20)[]{\Large{\Black{$d$}}}
\GBox(100,-3)(110,7){0.0}
\GBox(100,72)(110,82){0.0}
\ArrowLine(45,-13)(105,2)
\ArrowArc(112.5,39.5)(38.24,101.31,258.69)
\Text(50,38)[]{\Large{\Black{$u, c$}}}
\end{picture}
} & 
\scalebox{0.65}{
\begin{picture}(120,138) (45,-30)
\SetWidth{0.5}
\SetColor{Black}
\ArrowLine(45,93)(105,78)
\ArrowLine(165,93)(105,78)
\ArrowLine(105,3)(45,-12)
\ArrowLine(105,3)(165,-12)
\ArrowArcn(97.5,40.5)(38.24,78.69,-78.69)
\Text(75,100)[]{\Large{\Black{$s$}}}
\Text(135,99)[]{\Large{\Black{$\nu$}}}
\Text(170,39)[]{\Large{\Black{$e, \mu, \tau$}}}
\Text(75,-21)[]{\Large{\Black{$d$}}}
\Text(135,-22)[]{\Large{\Black{$\nu$}}}
\GBox(100,-2)(110,8){0.0}
\GBox(100,73)(110,83){0.0}
\ArrowArc(112.5,40.5)(38.24,101.31,258.69)
\Text(50,39)[]{\Large{\Black{$u, c$}}}
\end{picture}
}
\end{array}$
\vspace{2mm}
\caption{One-loop diagrams in the low-energy effective theory that
involve a large leading logarithms. The black squares represent the
insertion of the bilocal composite operators $Q^P_\pm$ (left) and
$Q^B$ (right). The charm quark line is open in the left diagram and in
all subsequent diagrams in which the charm quark couples to
$Q_\pm^c$.}           
\label{fig:LL}
\end{center}
\end{figure}
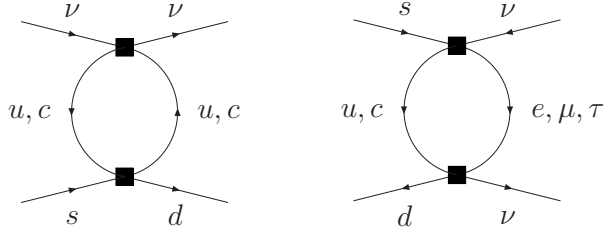
}%

Since there is no mixing between the bilocal composite operators
$Q_\pm^P$ and $Q^B$, the RG analysis of $\PcX$ naturally splits into
two parts: one for the $Z$-penguin contribution which involves
$Q_\pm^P$ and one for the electroweak box contribution that brings
$Q^B$ into play. As the structure of the $Z$-penguin is more
complicated than the one of the electroweak boxes it is useful to
discuss the former type of contributions first. This will be done in
\Sec{sec:Z}. After the detailed exposition of the $Z$-penguin
contribution it is straightforward to repeat the analysis in the
analogous, but somewhat simpler case of the electroweak boxes. The
essential steps of this calculation will be discussed in \Sec{sec:B}.

The main components of the calculations of \Secsand{sec:Z}{sec:B}
performed here at the NNLO level are: $i)$ the two-loop $\ord (\as^2)$
corrections to the initial conditions of the relevant Wilson
coefficients at $\muh$, $ii)$ the three-loop $\ord (\as^3)$ ADM
describing the mixing of the associated physical operators, $iii)$ the
two-loop $\ord (\as^2)$ threshold corrections to the Wilson
coefficients at $\mub$, and $iv)$ the two-loop matrix elements $\ord
(\as^2)$ of the relevant operators at $\muc$.  

The current-current operators $Q_\pm^q$ which enters the bilocal
composite operator $Q^P_\pm$ are familiar from the non-leptonic $|
\Delta S | = 1$ transitions. As their mixing under renormalization is
not affected by the presence of the other operators, it is convenient
to perform their RG analysis before discussing the $Z$-penguin
contribution itself. This calculation involves the first three
aforementioned steps as we will explain in \Sec{sec:CC}.

The second ingredient of the bilocal composite operator $Q^P_\pm$ is
the neutral-current operator $Q_Z$, which describes the interactions
of neutrinos and quarks mediated by $Z$-boson exchange. It is a linear
combination of the usual vector and axial-vector couplings of the
left-handed neutrino current to quarks, and a Chern-Simons operator
that describes the coupling of neutrinos to two and three gluons. The
inclusion of the Chern-Simons operator is essential to guarantee the
non-renormalization of $Q_Z$ to all orders in $\as$, which plays an
important role in the RG analysis of the $Z$-penguin sector. The
subleties arising in connection with $Q_Z$ will be reviewed in
\Sec{subsec:ZEX} before analyzing the $Z$-penguin contribution itself.

Finally, we will discuss the operators $Q_3^q$ and $Q_4^q$. These
operators are the building blocks of the bilocal composite operator
$Q^B$ and describe the interactions between leptons and quarks
mediated by $W$-exchange. They will be briefly discussed in
\Sec{subsec:WEX}.         

In summary, after the three preparatory sections, namely
\Secs{sec:CC}, \sect{subsec:ZEX}, and \sect{subsec:WEX}, that discuss
the dimension-six operators $Q_\pm^q$, $Q_Z$, $Q_3^q$, and $Q_4^q$, the
actual NNLO calculation relevant for the charm quark contribution to
$\Ktopinunu$ is presented in \Secsand{sec:Z}{sec:B} for the
$Z$-penguin and the electroweak box contributions, respectively. The
result of these efforts will be summarized in \Sec{sec:final}. 

\section{Current-Current Interactions} 
\label{sec:CC}

\subsection{Effective Hamiltonian}
\label{subsec:HeffCC}

As we are only interested in the charm quark contribution to the
$\stodnunu$ transition, we can drop the parts of the low-energy
effective Hamiltonian that are proportional to the CKM factor
$\lambda_t$. The unitarity of the CKM matrix then allows one to express
all the relevant contributions in terms of one independent CKM factor,
namely $\lambda_c$. $\PcX$ receives contributions from $Z$-penguin and
electroweak box diagrams with internal charm and up quarks. Examples
are depicted in \Fig{fig:penguinandbox}. For scales $\mul$ in the
range $\muc \le \mul \le \muh$ the four-quark interaction mediated by
$W$-boson exchange is described by the effective current-current
Hamiltonian        
\beq \label{eq:HCC}
\Heff^{CC} = \f{4 \GF}{\sqrt{2}} \lambda_c \sum_{i = \pm} C_i (\mul)
\left ( Q_i^c - Q_i^u \right ) \, ,    
\eeq
where 
\beq \label{eq:Qpm}
Q_\pm^q = \f{1}{2} \left ( ( \bar s_\sL^\alpha \gamma_{\mu}
q_\sL^\alpha ) ( \bar q_\sL^\beta \gamma^{\mu} d_\sL^\beta ) \pm (
\bar s_\sL^\alpha \gamma_{\mu} q_\sL^\beta ) ( \bar q_\sL^\beta
\gamma^{\mu} d_\sL^\alpha ) \right ) \, .
\eeq
Here $\alpha$ and $\beta$ are color indices. At LO the operators in
\Eq{eq:Qpm} renormalize multiplicatively. Beyond LO they mix into
so-called evanescent operators, which vanish algebraically in $n = 4$
dimensions \cite{Buras:1989xd, Dugan:1990df, Herrlich:1994kh,
collins}, but affect the values of the Wilson coefficients $C_\pm
(\mul)$. These operators can be chosen in such a way that the
renormalized matrix elements of $Q_\pm^q$ and its Fierz transform are
the same. For this choice $Q_\pm^q$ have well-defined and distinct
isospin quantum numbers and do not mix with each other at NLO and
beyond. The definitions of the evanescent operators required to
preserve the diagonal form of the NNLO anomalous dimension matrix
(ADM) in the $Q_\pm^q$ basis can be found in \App{app:evanescent}.      

\subsection{Initial Conditions}
\label{subsec:ICCC}

We now turn our attention to the calculation of the initial conditions
of $Q_\pm^q$. Dropping the unnecessary flavor index $q$ we write for
$i = \pm$ 
\beq \label{eq:Ci}
C_i (\muh) = C_i^{(0)} (\muh) + \f{\as (\muh)}{4 \pi} C_i^{(1)} 
(\muh) + \left ( \f{\as (\muh)}{4 \pi} \right )^2 C_i^{(2)} (\muh)
\, , 
\eeq
where $\as (\muh)$ denotes the strong coupling constant in the
$\MSbar$ scheme for five active quark flavors. The values of the
coefficients $C_\pm^{(k)} (\muh)$ are determined by matching Green's
functions in the full and the effective theory at $\muh$. In the NNLO
approximation this requires the calculation of
one-particle-irreducible two-loop diagrams. Sample SM graphs are 
displayed in \Fig{fig:ccmatching}. For what concerns the
regularization of infrared (IR) divergences we follow the procedure
outlined for example in \cite{IRdimensional, Bobeth:1999mk}, which
consists in using dimensional regularization for both IR and
ultraviolet (UV) divergences. While the former singularities are
removed by renormalization, the latter poles cancel out in the
difference between the full and the effective theory amplitudes. For
detailed descriptions of higher-order matching calculations of strong
and electroweak corrections applying the latter method we refer the
interested reader to \cite{Bobeth:1999mk, Gambino:2001au,
Misiak:2004ew}.           

{%
\begin{figure}[!t]
\begin{center}
$\begin{array}{c@{\hspace{10mm}}c}
\scalebox{0.65}{
\begin{picture}(180,144) (30,-31)
\SetWidth{0.5}
\SetColor{Black}
\Gluon(75,54)(75,84){6}{2.57}
\Gluon(75,-6)(75,24){6}{2.57}
\ArrowArc(75,39)(15,-90,90)
\ArrowArc(75,39)(15,90,270)
\ArrowLine(30,99)(75,84)
\ArrowLine(30,-21)(75,-6)
\ArrowLine(75,-6)(165,-6)
\ArrowLine(75,84)(165,84)
\Photon(165,84)(165,-6){6}{8}
\ArrowLine(165,84)(210,99)
\ArrowLine(165,-6)(210,-21)
\Vertex(75,-6){2.83}
\Vertex(75,24){2.83}
\Vertex(75,54){2.83}
\Vertex(75,84){2.83}
\Vertex(165,84){2.83}
\Vertex(165,-6){2.83}
\Text(55,105)[]{\Large{\Black{$s$}}}
\Text(120,99)[]{\Large{\Black{$s$}}}
\Text(186,105)[]{\Large{\Black{$q$}}}
\Text(190,39)[]{\Large{\Black{$W$}}}
\Text(48,39)[]{\Large{\Black{$t$}}}
\Text(102,39)[]{\Large{\Black{$t$}}}
\Text(57,69)[]{\Large{\Black{$g$}}}
\Text(120,-21)[]{\Large{\Black{$q$}}}
\Text(57,9)[]{\Large{\Black{$g$}}}
\Text(55,-31)[b]{\Large{\Black{$q$}}}
\Text(185,-31)[b]{\Large{\Black{$d$}}}
\end{picture}
} & 
\scalebox{0.65}{
\begin{picture}(180,144) (30,-31)
\SetWidth{0.5}
\SetColor{Black}
\ArrowLine(30,99)(75,84)
\ArrowLine(30,-21)(75,-6)
\ArrowLine(75,-6)(165,-6)
\Photon(165,84)(165,-6){6}{8}
\ArrowLine(165,84)(210,99)
\ArrowLine(165,-6)(210,-21)
\Vertex(75,-6){2.83}
\Vertex(75,84){2.83}
\Vertex(165,84){2.83}
\Vertex(165,-6){2.83}
\Text(55,105)[]{\Large{\Black{$s$}}}
\Text(98,99)[]{\Large{\Black{$s$}}}
\Text(186,105)[]{\Large{\Black{$q$}}}
\Text(190,39)[]{\Large{\Black{$W$}}}
\Text(120,-21)[]{\Large{\Black{$q$}}}
\Text(55,-31)[b]{\Large{\Black{$q$}}}
\Text(185,-31)[b]{\Large{\Black{$d$}}}
\ArrowLine(75,84)(120,84)
\ArrowLine(120,84)(165,84)
\Vertex(120,84){2.83}
\Gluon(75,-6)(75,39){6}{3.43}
\Gluon(75,39)(75,84){6}{3.43}
\Gluon(120,84)(75,39){6}{5.14}
\Vertex(75,39){2.83}
\Text(143,99)[]{\Large{\Black{$s$}}}
\Text(115,54)[]{\Large{\Black{$g$}}}
\Text(57,16)[]{\Large{\Black{$g$}}}
\Text(57,62)[]{\Large{\Black{$g$}}}
\end{picture}
}
\end{array}$
\vspace{2mm}
\caption{Sample diagrams for the $\ord (\as^2)$ corrections to the
initial values of the Wilson coefficients of the current-current
operators.}              
\label{fig:ccmatching}
\end{center}
\end{figure}
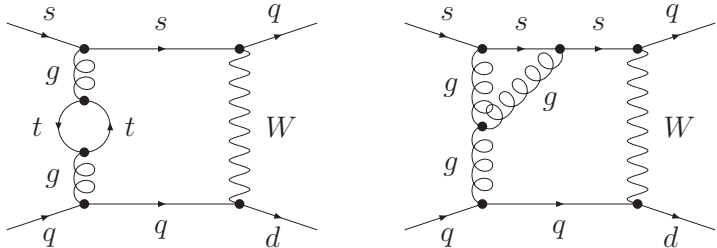
}%

Using naive dimensional regularization (NDR) \cite{Chanowitz:1979zu}
with a fully anticommuting $\gamma_5$, we obtain for the standard
choices of the Casimir invariants $\ca = 3$, $\cf = 4/3$, and five
active quark flavors  
\beq \label{eq:Cpm}
\begin{split}  
C_\pm^{(0)} (\muh) & = 1 \, , \\[4mm]
C_\pm^{(1)} (\muh) & = \pm \f{1}{2} \left ( 1 \mp \f{1}{3} \right)
\left ( 11 + 6 \ln \f{\muh^2}{\MW^2} \right ) \, , \\[1mm]     
C_\pm^{(2)} (\muh) & = -\f{1}{3600} \left ( 135677 \mp 124095 \right )
+ \f{1}{18} \left ( 7 \pm 51 \right ) \pi^2 \mp \f{1}{2} \left ( 1
\mp \f{1}{3} \right ) T (x_t) \\   
& -\f{5}{36} \left ( 11 \mp 249 \right ) \ln \f{\muh^2}{\MW^2} +
\f{1}{6} \left ( 7 \pm 51 \right ) \ln^2 \f{\muh^2}{\MW^2} \, . 
\end{split}
\eeq
The function $T (x_t)$ depends on the top quark $\MSbar$ mass via $x_t
= \mt^2 (\muh)/\MW^2$. It originates from SM diagrams like the one
shown on the left of \Fig{fig:ccmatching}. Subtracting the
corresponding terms in the gluon propagator in the momentum-space 
subtraction scheme at $q^2 = 0$, which ensures that $\as (\muh)$ is
continuous at the top quark threshold $\muh$,\footnote{This scheme
coincides with $\MSbar$ for $\mul \le \muh$. For details see
for example \cite{Misiak:2004ew, Bobeth:1999mk}.} we find         
\beq \label{eq:oneloopselfenergy}
\begin{split}
T (x_t) & = \f{112}{9} + 32 x_t + \left ( \f{20}{3} + 16 x_t \right )
\ln x_t \\  
& - \left ( 8 + 16 x_t \right ) \sqrt{4 x_t - 1} \, \cl \left ( 2
\arcsin \left ( \f{1}{2 \sqrt{x_t}} \right ) \right ) \, , 
\end{split}
\eeq
where $\cl (x) = {\rm Im} ( \li ( e^{i x} ) )$. As far as the one-loop
initial conditions, namely $C_\pm^{(1)} (\muh)$ are concerned, our
results agree with those of \cite{Buras:1989xd, Ciuchini:1993vr}. They
also agree with the results obtained in \cite{Altarelli:1980fi,
Chetyrkin:1997gb} after a transformation to our renormalization scheme
specified by \Eqs{eq:ccevanescent}. The general formalism of a change
of renormalization scheme discussed in detail in \cite{Gorbahn:2004my}
can also be used to verify that our result for the two-loop initial
conditions $C_\pm^{(2)} (\muh)$ coincides with the findings of
\cite{Bobeth:1999mk}. This is shown in \App{app:change}.      

\subsection{Anomalous Dimensions}
\label{subsec:ADMCC}

The Wilson coefficients are evolved from $\muh$ down to the relevant
low-energy scale $\mul$ with the help of the RG equation. In this way,
large logarithms of the form $\ln ( \mul^2/\muh^2 )$ are resummed to
all orders in $\as$. In mass-independent renormalization schemes like
$\MSbar$ the RG equation is given by  
\beq
\label{eq:renormalizationgroupequation} 
\mul \f{d}{d \mul} C_i (\mul) = \gamma_{j i} (\mul) C_j (\mul) \, ,  
\eeq 
where $\gamma_{j i} (\mul)$ is the entry of the ADM describing the
mixing of $Q_j$ into $Q_i$. In the case of $Q^q_\pm$ we will denote
the diagonal entries of $\hat{\gamma} (\mul)$ by $\gamma_\pm (\mul)$.

{%
\begin{figure}[!t]
\begin{center}
$\begin{array}{c@{\hspace{15mm}}c}
\scalebox{0.65}{
\begin{picture}(160,141) (25,-52)
\SetWidth{0.5}
\SetColor{Black}
\ArrowLine(45,81)(60,66)
\Vertex(60,66){2.83}
\CBox(110.66,15.34)(99.34,26.66){Black}{Black}
\Text(105,-4)[]{\Large{\Black{$Q^q_\pm$}}}
\ArrowLine(45,-39)(60,-24)
\ArrowLine(60,-24)(105,21)
\Vertex(60,-24){2.83}
\Text(40,21)[]{\Large{\Black{$g$}}}
\Vertex(127,44){2.83}
\Text(105,65)[]{\Large{\Black{$g$}}}
\ArrowLine(105,21)(128,44)
\Gluon(60,-24)(60,66){6}{8.57}
\Gluon(82,44)(127,44){6}{4.14}
\ArrowLine(60,66)(82,44)
\ArrowLine(82,44)(105,21)
\Vertex(82,44){2.83}
\ArrowLine(150,-24)(165,-39)
\Vertex(150,-24){2.83}
\ArrowLine(105,21)(150,-24)
\ArrowLine(128,44)(150,66)
\ArrowLine(150,66)(165,81)
\Vertex(150,66){2.83}
\Gluon(150,66)(150,-24){6}{8.57}
\Text(170,21)[]{\Large{\Black{$g$}}}
\Text(60,81)[]{\Large{\Black{$s$}}}
\Text(150,81)[]{\Large{\Black{$q$}}}
\Text(58,-44)[]{\Large{\Black{$q$}}}
\Text(150,-42)[]{\Large{\Black{$d$}}}
\end{picture}
} & 
\scalebox{0.65}{
\begin{picture}(165,141) (20,-52)
\SetWidth{0.5}
\SetColor{Black}
\ArrowLine(60,-24)(105,21)
\CBox(110.66,15.34)(99.34,26.66){Black}{Black}
\ArrowLine(45,81)(60,66)
\Vertex(60,66){2.83}
\Text(105,-4)[]{\Large{\Black{$Q^q_\pm$}}}
\ArrowLine(45,-39)(60,-24)
\Vertex(60,-24){2.83}
\Gluon(60,36)(60,66){6}{2.57}
\Gluon(60,-24)(60,6){6}{2.57}
\ArrowArc(60,21)(15,-90,90)
\ArrowArc(60,21)(15,90,270)
\Vertex(60,6){2.83}
\Vertex(60,36){2.83}
\Text(85,21)[]{\Large{\Black{$\nf$}}}
\Text(35,21)[]{\Large{\Black{$\nf$}}}
\Text(43,-9)[]{\Large{\Black{$g$}}}
\Text(43,51)[]{\Large{\Black{$g$}}}
\ArrowLine(60,66)(105,21)
\ArrowLine(105,21)(150,66)
\ArrowLine(105,21)(150,-24)
\ArrowLine(150,66)(165,81)
\ArrowLine(150,-24)(165,-39)
\Vertex(150,66){2.83}
\Vertex(150,-24){2.83}
\Gluon(150,66)(150,-24){6}{8.57}
\Text(170,21)[]{\Large{\Black{$g$}}}
\Text(58,-44)[]{\Large{\Black{$q$}}}
\Text(150,-42)[]{\Large{\Black{$d$}}}
\Text(60,81)[]{\Large{\Black{$s$}}}
\Text(150,81)[]{\Large{\Black{$q$}}}
\end{picture}
}
\end{array}$
\vspace{2mm}
\caption{Sample diagrams for the $\ord (\as^3)$ mixing among the
current-current operators. The fermion loop in the right Feynman graph
contains $\nf$ active quark flavors.}       
\label{fig:ccmixing}
\end{center}
\end{figure}
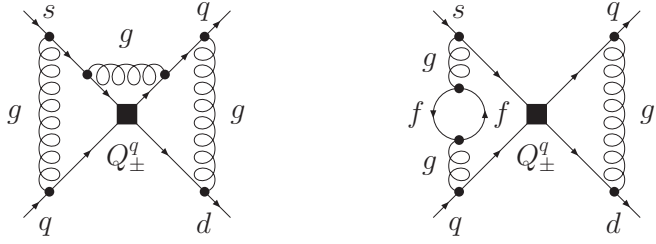
}%

In the NNLO approximation the ADM has the following perturbative
expansion 
\beq \label{eq:admexpansion}
\hat{\gamma} (\mul) = \f{\as (\mul)}{4 \pi} \hat{\gamma}^{(0)} + \left ( 
\f{\as (\mul)}{4 \pi} \right )^2 \hat{\gamma}^{(1)} + \left ( \f{\as
(\mul)}{4 \pi} \right )^3 \hat{\gamma}^{(2)} \, ,  
\eeq
where the coefficients $\hat{\gamma}^{(k)}$ can be extracted from the
one-, two-, and three-loop QCD renormalization constants in the
effective theory. The renormalization matrices are found by
calculating amputated Green's functions with single insertions of
$Q^q_\pm$ up to three loops. Sample diagrams are shown in
\Fig{fig:ccmixing}. The corresponding amplitudes are evaluated using
the method that has been described in \cite{Misiak:1994zw,
Chetyrkin:1997fm, Gambino:2003zm}. We keep the gauge parameter
arbitrary and find it to cancel from $\hat{\gamma}^{(k)}$, which
provides a powerful check of our calculation. To distinguish between
IR and UV divergences, we introduce a common mass $M$ for all fields
and expand all loop integrals in inverse powers of $M$. This makes the
calculation of the UV divergences possible at three loops, as $M$
becomes the only relevant internal scale, and three-loop tadpole
integrals with a single non-zero mass are known  \cite{Chetyrkin:1997fm,
Broadhurst:1998rz}. Comprehensive discussions of the technical
details of the renormalization of the effective theory and the actual
calculation of the operator mixing are given in \cite{Misiak:1994zw,
Gambino:2003zm}.  

While $\gamma^{(0)}_\pm$ is renormalization-scheme independent,
$\gamma^{(1)}_\pm$ and $\gamma^{(2)}_\pm$ are not. In the NDR scheme
supplemented by the definition of evanescent operators given in
\Eqs{eq:ccevanescent}, we obtain for $\ca = 3$, $\cf = 4/3$, and an 
arbitrary number of active quark flavors $\nf$ 
\beq \label{eq:gpm} 
\begin{split}
\gamma_\pm^{(0)} & = \pm 6 \left ( 1 \mp \f{1}{3} \right ) \, ,
\\[1mm] 
\gamma_\pm^{(1)} & = \left ( -\f{21}{2} \pm \f{2}{3} \nf \right )
\left ( 1 \mp \f{1}{3} \right ) \, , \\[1mm] 
\gamma_\pm^{(2)} & = \f{1}{300} \left ( 349049 \pm 201485 \right ) - 
\f{1}{1350} \left ( 115577 \mp 9795 \right ) \nf \\[1mm]
& \mp \f{130}{27} \left ( 1 \mp \f{1}{3} \right ) \nf^2 \mp \left (
672 + 80 \left ( 1 \mp \f{1}{3} \right ) \nf \right ) \zetathree \, .
\end{split}
\eeq
Here $\zetax$ is the Riemann zeta function with the value $\zetathree
\approx 1.20206$. Again, we find agreement with the one- and two-loop
results of \cite{Buras:1989xd, Ciuchini:1993vr}, and therefore also
with the findings of \cite{Altarelli:1980fi, Chetyrkin:1997gb,
Gambino:2003zm} that were obtained in different renormalization 
schemes. We also confirm the three-loop results presented recently
\cite{Gorbahn:2004my}. The explicit formulas that allow the conversion
of the latter anomalous dimensions to our scheme can be found in
\App{app:change}.  

\subsection{Threshold Corrections}
\label{subsec:MBCC}

In order to compute the Wilson coefficients for scales $\mul$ much
lower than $\muh$ a proper matching between effective theories
containing $\nf$ and $\nf - 1$ active quark flavors has to be 
performed each time one passes through a flavor threshold. It is
achieved by requiring that the Green's functions in both effective   
theories are the same at the point $\muf = \ord (\mf)$ where the quark
with mass $\mf$ is integrated out. This equality translates into 
\beq \label{eq:thresholdmatching} 
\big \langle \vec{Q}^{\nf - 1} (\muf) \big \rangle^T \vec{C}^{\nf - 1}
(\muf) = \big\langle \vec{Q}^{\nf} (\muf) \big \rangle^T \vec{C}^{\nf}
(\muf) \, ,  
\eeq   
where the labels $\nf$ and $\nf - 1$ indicate to which effective
theory the variable belongs. Including corrections up to NNLO we write
\beq \label{eq:meexpansion} 
\big \langle \vec{Q}^{\nf} (\muf) \big \rangle^T = \big \langle
\vec{Q}^{\nf} \big \rangle^{(0) T} \left ( \hat{1} + \f{\as^{\nf}
(\muf)}{4 \pi} \hat{r}^{\nf (1) \, T} (\muf) + \left ( \f{\as^{\nf}  
(\muf)}{4 \pi} \right )^2 \hat{r}^{\nf (2) \, T} (\muf) \right ) \, ,   
\eeq 
where $\langle \vec{Q}^{\nf} \big \rangle^{(0)}$, $\hat{r}^{\nf (1)}
(\muf)$, and $\hat{r}^{\nf (2)} (\muf)$ codify the tree-level, one-, and
two-loop matrix element of the column vector $\vec{Q}^{\nf}$ containing 
the relevant physical operators. The other quantities entering
\Eq{eq:thresholdmatching} can be expanded in a similar fashion.  

Another subtlety arises when working in mass-independent
renormalization schemes, because the matching conditions connecting
the strong coupling constants of the effective theories with $\nf$ and
$\nf - 1$ active quark flavors are non-trivial in such schemes. In
particular, in the $\MSbar$ scheme one has in the NNLO approximation
\cite{olddecoupling, newdecoupling}\footnote{It should be noted that
the non-logarithmic term in \Eq{eq:asmatching} differs from the result
published in \cite{olddecoupling}. The authors of \cite{olddecoupling}
have revised their original analysis and have found agreement with
\cite{newdecoupling}.}        
\beq \label{eq:asmatching}
\begin{split}
\as^{\nf - 1} (\muf) & = \as^{\nf} (\muf) \left ( 1 - \f{\as^{\nf} 
(\muf)}{4 \pi} \f{2}{3} \ln \f{\muf^2}{\mf^2} \right. \\ 
& \left. + \left ( \f{\as^{\nf} (\muf)}{4 \pi} \right )^2 \left (  
\f{22}{9} - \f{38}{3} \ln \f{\muf^2}{\mf^2} + \f{4}{9} \ln^2
\f{\muf^2}{\mf^2} \right ) \right ) \, ,      
\end{split}
\eeq
where $\mf = \mf (\mf)$ denotes the quark $\MSbar$ mass. Of course,
the appearance of both logarithmic and finite corrections can be
avoided by adjusting the renormalization scheme so that $\as^{\nf - 1}
(\muf) = \as^{\nf} (\muf)$. While this does not affect the physical
amplitudes \cite{Gorbahn:2004my}, one leaves the class of
mass-independent renormalization schemes with the drawback that the
usual RG equations do not hold below the matching point and the
resummation of large logarithms gets obscured. Hence it is much more
convenient to stick to the $\MSbar$ prescription of $\as$ and to apply
\Eq{eq:asmatching} whenever a flavor threshold is crossed. We will
follow this approach below.         

In terms of the discontinuities 
\beq \label{eq:definitiondeltas}
\delta \vec{C}^{(k)} (\muf) = \vec{C}^{f (k)} (\muf) - \vec{C}^{f - 1 
(k)} (\muf) \, , \hspace{1cm} \delta \hat{r}^{(k)} (\muf) = \hat{r}^{f
(k)} (\muf) - \hat{r}^{f - 1 
(k)} (\muf) \, , 
\eeq
the solution of \Eq{eq:thresholdmatching} can be written in a
relatively compact form. Up to the second power in the strong coupling
constant we obtain   
\beq \label{eq:deltaCi}
\begin{split}
\delta \vec{C}^{(0)} (\muf) & = 0 \, , \\[6mm]  
\delta \vec{C}^{(1)} (\muf) & = -\delta \hat{r}^{(1) \, T} (\muf) \, 
\vec{C}^{\nf (0)} (\muf) \, , \\[2mm]    
\delta \vec{C}^{(2)} (\muf) & = -\left ( \delta \hat{r}^{(1) \, T}
(\muf) + \f{2}{3} \ln \f{\muf^2}{\mf^2} \right ) \vec{C}^{\nf (1)} 
(\muf) \\  
& - \left ( \delta \hat{r}^{(2) \, T} (\muf) - \hat{r}^{\nf - 1 (1) \,
T} (\muf) \, \delta \hat{r}^{(1) \, T} (\muf) + \f{2}{3} \ln
\f{\muf^2}{\mf^2} \, \hat{r}^{\nf (1) \, T} (\muf) \right )
\vec{C}^{\nf (0)} (\muf) \, ,     
\end{split}
\eeq
where the second line resembles the NLO result derived in
\cite{Buras:1993dy}. Note that at NNLO the logarithmic $\ord (\as^2)$
correction entering the right-hand side of \Eq{eq:asmatching} starts
to contribute to the matching conditions of the Wilson coefficients at
each flavor threshold.   

{%
\begin{figure}[!t]
\begin{center}
$\begin{array}{c@{\hspace{15mm}}c}
\scalebox{0.65}{
\begin{picture}(138,120) (27,-60)
\SetWidth{0.5}
\SetColor{Black}
\ArrowLine(60,45)(105,0)
\ArrowLine(45,60)(60,45)
\ArrowLine(105,0)(165,-60)
\Vertex(60,45){2.83}
\CBox(110.66,-5.66)(99.34,5.66){Black}{Black}
\Text(105,-25)[]{\Large{\Black{$Q^q_\pm$}}}
\Text(142,-14)[]{\Large{\Black{$d$}}}
\ArrowLine(45,-60)(60,-45)
\ArrowLine(60,-45)(105,0)
\ArrowLine(105,0)(165,60)
\Vertex(60,-45){2.83}
\Text(85,35)[]{\Large{\Black{$s$}}}
\Text(142,13)[]{\Large{\Black{$q$}}}
\Text(85,-38)[]{\Large{\Black{$q$}}}
\Gluon(60,-45)(60,45){6}{8.57}
\Text(42,0)[]{\Large{\Black{$g$}}}
\end{picture}
} & 
\scalebox{0.65}{
\begin{picture}(120,133) (45,-47)
\SetWidth{0.5}
\SetColor{Black}
\ArrowLine(60,58)(105,13)
\ArrowLine(105,13)(150,58)
\ArrowLine(45,73)(60,58)
\ArrowLine(150,58)(165,73)
\ArrowLine(105,13)(165,-47)
\ArrowLine(45,-47)(105,13)
\Vertex(60,58){2.83}
\Vertex(150,58){2.83}
\CBox(110.66,7.34)(99.34,18.66){Black}{Black}
\Text(105,-12)[]{\Large{\Black{$Q^q_\pm$}}}
\Text(69,-7)[b]{\Large{\Black{$q$}}}
\Text(142,-7)[b]{\Large{\Black{$d$}}}
\Text(75,20)[b]{\Large{\Black{$s$}}}
\Text(136,16)[b]{\Large{\Black{$q$}}}
\Gluon(60,58)(150,58){6}{8.57}
\Text(105,78)[]{\Large{\Black{$g$}}}
\end{picture}
}
\end{array}$
\vspace{2mm}
\caption{Examples of diagrams that contribute to the $\ord (\as)$
matrix elements of the current-current operators.}      
\label{fig:ccmatrix}
\end{center}
\end{figure}
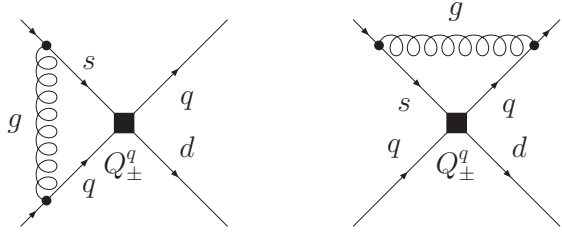
}%

In the case of $C_\pm (\mul)$ the explicit expressions for the
threshold corrections turn out to be much simpler than suggested by
\Eqs{eq:deltaCi}, because the corrections $\delta r^{(1)}_\pm (\muf)$
vanish, as the $\ord (\as)$ matrix elements of the current-current
operators are identical in the effective theories. Sample diagrams can
be seen in \Fig{fig:ccmatrix}. In consequence $\delta C_\pm^{(1)}
(\muf) = 0$. Note that this is in contrast to the case of the QCD and
electroweak penguin operators which receive non-trivial threshold
corrections at NLO \cite{Buras:1993dy, Buras:1991jm}. However,
non-vanishing discontinuities $\delta r^{(2)}_\pm (\muf)$ arise from
the diagrams depicted in \Fig{fig:ccthreshold}. For what concerns the
calculation of the graphs itself, we have adopted two different
methods to regulate IR divergences and found identical results for the
threshold corrections. The first approach mentioned earlier, uses
dimensional regularization for both IR and UV singularities and
calculates on-shell matrix elements with zero external momenta. The
second method uses small quark masses as IR regulators and computes
matrix elements with zero external momenta which are now
off-shell. Useful details on the latter procedure can be found in
\cite{IRquarkmass}. The unphysical coefficients $\hat{r}^{(k)} (\muf)$
differ in both cases and depend on the IR regulators. However, this
dependence cancels in the combination entering $\delta C_\pm^{(2)}
(\muf)$. The correct implementation of the discontinuity in $\as$ of
\Eq{eq:asmatching} and of similar decoupling relations for the gluon
and quark fields are of crucial importance for this cancellation. In
the second method using off-shell matrix elements another subtlety
occurs, as now both sides of \Eq{eq:thresholdmatching} depend on the
gauge parameter, and in order to obtain a gauge-independent and 
IR-safe result one has to take into account that the gauge parameter
is discontinous across flavor thresholds as well. We do not give the
decoupling relations for the gluon, the quark field, and the gauge
parameter here. They can be found for example in \cite{newdecoupling}.

{%
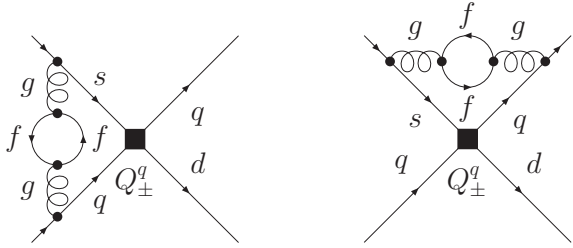
\begin{figure}[!t]
\begin{center}
$\begin{array}{c@{\hspace{15mm}}c}
\scalebox{0.65}{
\begin{picture}(145,120) (20,-60)
\SetWidth{0.5}
\SetColor{Black}
\ArrowLine(60,45)(105,0)
\ArrowLine(45,60)(60,45)
\ArrowLine(105,0)(165,-60)
\Vertex(60,45){2.83}
\CBox(110.66,-5.66)(99.34,5.66){Black}{Black}
\Text(105,-25)[]{\Large{\Black{$Q_\pm^q$}}}
\Text(142,-14)[]{\Large{\Black{$d$}}}
\ArrowLine(45,-60)(60,-45)
\ArrowLine(60,-45)(105,0)
\ArrowLine(105,0)(165,60)
\Gluon(60,15)(60,45){6}{2.57}
\Gluon(60,-45)(60,-15){6}{2.57}
\ArrowArc(60,0)(15,-90,90)
\ArrowArc(60,0)(15,90,270)
\Vertex(60,15){2.83}
\Vertex(60,-15){2.83}
\Vertex(60,-45){2.83}
\Text(35,0)[]{\Large{\Black{$\nf$}}}
\Text(43,30)[]{\Large{\Black{$g$}}}
\Text(43,-30)[]{\Large{\Black{$g$}}}
\Text(85,35)[]{\Large{\Black{$s$}}}
\Text(142,13)[]{\Large{\Black{$q$}}}
\Text(85,-38)[]{\Large{\Black{$q$}}}
\Text(85,0)[]{\Large{\Black{$\nf$}}}
\end{picture}
} & 
\scalebox{0.65}{
\begin{picture}(120,140) (45,-40)
\SetWidth{0.5}
\SetColor{Black}
\Gluon(60,65)(90,65){6}{2.57}
\Gluon(120,65)(150,65){6}{2.57}
\ArrowArc(105,65)(15,-0,180)
\ArrowArc(105,65)(15,-180,0)
\ArrowLine(60,65)(105,20)
\ArrowLine(105,20)(150,65)
\ArrowLine(45,80)(60,65)
\ArrowLine(150,65)(165,80)
\ArrowLine(105,20)(165,-40)
\ArrowLine(45,-40)(105,20)
\Vertex(60,65){2.83}
\Vertex(150,65){2.83}
\CBox(110.66,14.34)(99.34,25.66){Black}{Black}
\Vertex(120,65){2.83}
\Vertex(90,65){2.83}
\Text(105,92)[]{\Large{\Black{$\nf$}}}
\Text(105,38)[]{\Large{\Black{$\nf$}}}
\Text(75,83)[]{\Large{\Black{$g$}}}
\Text(135,83)[]{\Large{\Black{$g$}}}
\Text(105,-5)[]{\Large{\Black{$Q_\pm^q$}}}
\Text(67,0)[b]{\Large{\Black{$q$}}}
\Text(144,0)[b]{\Large{\Black{$d$}}}
\Text(75,27)[b]{\Large{\Black{$s$}}}
\Text(136,23)[b]{\Large{\Black{$q$}}}
\end{picture}
}
\end{array}$
\vspace{2mm}
\caption{Sample diagrams for the $\ord (\as^2)$ matching corrections
to the Wilson coefficients of the current-current operators at each
flavor threshold. At these thresholds the number of active quarks
changes from $\nf$ to $\nf - 1$.}
\label{fig:ccthreshold}
\end{center}
\end{figure}
}%

At the bottom quark threshold scale $\mub$ we find for the non-trivial
matching conditions of the Wilson coefficients of the current-current
operators in the NDR scheme      
\begin{align} \label{eq:deltaCpmmub}
\delta C_\pm^{(2)} (\mub) & = -\eta_b^{ \pm \f{9}{23} \left ( 1 \mp
\f{1}{3} \right )} \Bigg ( \f{2}{3} \ln \f{\mub^2}{\mb^2} \left (
\f{631 \pm 9699}{6348} ( 1 - \eta_b ) + \eta_b C^{(1)}_\pm (\muh)
\right ) \non \\ 
& \mp \left ( 1 \mp \f{1}{3} \right ) \left ( \f{59}{36} + \f{1}{3}
\ln \f{\mub^2}{\mb^2} + \ln^2 \f{\mub^2}{\mb^2} \right )
\hspace{-1mm} \Bigg ) \, ,          
\end{align}
where $\eta_b = \as (\muh)/\as(\mub)$ and $\mb = \mb (\mb)$ denotes
the bottom quark $\MSbar$ mass. We also remark that diagrams like the
one shown on the right of \Figsand{fig:ccmatrix}{fig:ccthreshold},
which correspond to QCD corrections to a current operator, do not
contribute once all counterterms have been included. The coefficients
$\delta C_\pm^{(2)} (\mub)$ depend on the renormalization scheme
chosen for $Q_\pm^q$, in particular on the specific structure of the
evanescent operator defined in the second line of
\Eqs{eq:ccevanescent}. Choices other than these would lead to operator
mixing between $Q_\pm^q$.            

The discontinuities $\delta C_\pm^{(2)} (\muc)$ at the charm quark
threshold scale $\muc$ can be ignored as it is more convenient to
express the final low-energy Wilson coefficient in terms of the $\as$
of the effective theory with four active quark flavors rather than in
terms of the $\as$ of the effective theory with three active quark
flavors, because no RG equation needs to be solved below $\muc$. This
will be explained in more detail at the end of \Sec{sec:Z}.     

\section{Neutral and Charged Currents}
\label{sec:NCCC}

\subsection{Neutral Current: {\boldmath $Z$}-boson Exchange}  
\label{subsec:ZEX}

The low-energy effective Hamiltonian describing the interactions of
neutrinos and quarks mediated by $Z$-boson exchange is given by     
\beq \label{eq:HZ}
\Heff^Z = \f{\pi \aem}{\MW^2 \sws} C_Z (\mul) Q_Z \, , 
\eeq
where
\beq \label{eq:QZ}
Q_Z = \sum_q \left ( \left ( I^3_q \, - 2 e_q \sws \right ) Q_{V}^q -
I^3_q \left ( Q_{A}^q + Q_{CS} \right ) \right ) \, ,         
\eeq 
and the sum over $q$ extends over all active light quark flavors at
the renormalization scale $\mul$, while $I^3_q = (+1/2,-1/2)$ and $e_q
= (+2/3,-1/3)$ denote the third component of the weak-isospin and the
electric charge of the up- and down-type quarks, respectively. The
appropriate normalization of the electromagnetic coupling $\aem$ and
the weak mixing angle $\sws$ will become clear after our discussion in
\Sec{sec:Z}.  

Removing the $Z$-boson as an active degree of freedom from the
effective theory induces a vector as well as an axial-vector coupling
of the left-handed neutrino current to quarks 
\beq \label{eq:QVQA}
Q_V^q = \sum_{\ell = e, \mu, \tau} ( \bar q \gamma_{\mu} q ) ( \bar
{\nu_\ell}_\sL \gamma^{\mu} {\nu_\ell}_\sL ) \, , \hspace{1cm} Q_A^q =
\sum_{\ell = e, \mu, \tau} ( \bar q \gamma_{\mu} \gamma_5 q ) ( \bar
{\nu_\ell}_\sL \gamma^{\mu} {\nu_\ell}_\sL ) \, . 
\eeq 
In the literature on $\Ktopinunu$ the operator $Q_V^q$ is usually
omitted,\footnote{To our knowledge the only exception is the recent
publication \cite{Isidori:2005xm}.} as is does not contribute to the
decay rate through NLO. We keep $Q_V^q$ throughout our NNLO
calculation. While individual diagrams are non-vanishing, we verify
explicitly that both the two-loop matching diagrams and the three-loop
mixing diagrams with $Q_V^q$ sum to zero. We will discuss this issue in
more detail in \Sec{sec:Z} after presenting our final results for the
anomalous dimensions and matrix elements, respectively.

{%
\begin{figure}[!t]
\begin{center}
$\begin{array}{c@{\hspace{10mm}}c}
\scalebox{0.65}{
\begin{picture}(165,146) (75,-77)
\SetWidth{0.5}
\SetColor{Black}
\Photon(75,-4)(135,-4){6}{5}
\ArrowLine(135,-4)(180,-49)
\ArrowLine(180,41)(135,-4)
\Gluon(180,41)(240,41){6}{4.29}
\Gluon(180,-49)(240,-49){6}{4.29}
\Text(154,36)[]{\Large{\Black{$t$}}}
\Text(154,-44)[]{\Large{\Black{$t$}}}
\Text(105,16)[]{\Large{\Black{$Z$}}}
\Text(220,61)[]{\Large{\Black{$g$}}} 
\Text(220,-69)[]{\Large{\Black{$g$}}}
\Vertex(135,-4){2.83}
\Vertex(180,-49){2.83}
\Vertex(180,41){2.83}
\ArrowLine(180,-49)(180,41)
\Text(195,-4)[]{\Large{\Black{$t$}}}
\end{picture}
}  
& 
\scalebox{0.65}{
\begin{picture}(165,146) (75,-77)
\SetWidth{0.5}
\SetColor{Black}
\Photon(75,-4)(135,-4){6}{5}
\ArrowLine(135,-4)(180,-49)
\ArrowLine(180,41)(135,-4)
\Gluon(180,41)(240,41){6}{4.29}
\Gluon(180,-49)(240,-49){6}{4.29}
\Text(154,36)[]{\Large{\Black{$t$}}}
\Text(154,-44)[]{\Large{\Black{$t$}}}
\Text(105,16)[]{\Large{\Black{$Z$}}}
\Text(220,61)[]{\Large{\Black{$g$}}}
\Text(220,-69)[]{\Large{\Black{$g$}}}
\Vertex(135,-4){2.83}
\Vertex(180,-49){2.83}
\Vertex(180,41){2.83}
\ArrowLine(180,-49)(180,-4)
\ArrowLine(180,-4)(180,41)
\Vertex(180,-4){2.83}
\Gluon(180,-4)(240,-4){6}{4.29}
\Text(220,-24)[]{\Large{\Black{$g$}}}
\Text(195,18)[]{\Large{\Black{$t$}}}
\Text(195,-27)[]{\Large{\Black{$t$}}}
\end{picture}
}
\end{array}$
\vspace{2mm}
\caption{The one-loop top quark contribution to the coupling of the
$Z$-boson to two and three gluons. Diagrams obtained by the
interchange of the external gluons are not shown.}           
\label{fig:triangle&box}
\end{center}
\end{figure}
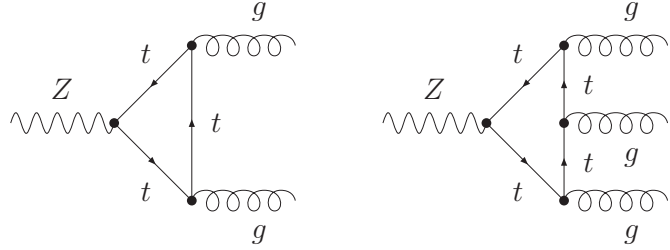
}%

As can be inferred from \Fig{fig:triangle&box}, decoupling the top
quark generates furthermore an effective gauge-variant coupling of the
left-handed neutrino current to two and three gluons which can be
expressed in terms of the following Chern-Simons operator
\beq \label{eq:QCS}
Q_{CS} = \frac{g^2}{8 \pi^2} \, \eps^{\mu_1 \mu_2 \mu_3 \mu_4}
\left(G^a_{\mu_1} \partial_{\mu_2} G^a_{\mu_3} + \frac{1}{3} \gs
f^{abc} G^a_{\mu_1} G^b_{\mu_2} G^c_{\mu_3} \right ) \sum_{\ell = e,
\mu, \tau} ( \bar {\nu_\ell}_\sL \gamma_{\mu_4} {\nu_\ell}_\sL ) \, .          
\eeq
Here $\gs$ denotes the strong coupling constant, $\eps^{\mu_1 \mu_2 
\mu_3 \mu}$ is the fully antisymmetric Levi-Civita tensor defined
with $\eps^{0123} = +1$, $G_{\mu_1}^a$ is the gluon field, and
$f^{abc}$ are the totally antisymmetric structure constants of
$SU(3)$. We remark that the 't Hooft-Veltman (HV) prescription
\cite{HVscheme} and dimensional reduction (DRED) \cite{Siegel:1979wq}
lead to the same result for the triangle diagrams, if the
mathematically consistent formulation of the DRED scheme presented
recently in \cite{Stockinger:2005gx} is employed. A description of the 
HV scheme can be found for example in \cite{Buras:1989xd}.

The inclusion of $Q_{CS}$ in \Eq{eq:QZ} is essential to obtain the
correct anomalous Ward identity for the axial-vector current
\cite{ABJ} and in consequence to guarantee the vanishing of the 
anomalous dimension of $Q_Z$ to all orders in perturbation
theory. We stress that we do not add this contribution in an ad hoc
way, instead $Q_{CS}$ is generated in an unambigous way from the
diagrams in \Fig{fig:triangle&box}. Our effective theory is anomaly
free, because $Q_{CS}$ cancels the anomalous contribution from the
triangle graph with a bottom quark, just as the anomalous effects from
top and bottom quarks cancel in the SM. Let us illustrate how the
cancellation between the contributions from $Q_A^q$ and $Q_{CS}$ to
the anomalous dimension of $Q_Z$ occurs at lowest order. As depicted in
\Fig{fig:anomaly}, the first non-trivial mixing arises at $\ord
(\as^2)$ with $Q_A^q$ mixing into itself through two-loop diagrams and
$Q_{CS}$ mixing into $Q_A^q$ through a one-loop diagram. Choosing the
operator basis as $(Q_Z, Q_{CS})$, we find for the NLO anomalous
dimension matrix\footnote{The calculation has been performed in the
background field gauge for the gluon field \cite{Abbott:1980hw}, which
makes it possible maintain explicit gauge invariance at the level of
off-shell Green's functions \cite{collins}, keeping the gauge
parameter arbitrary.}         
\beq \label{eq:anomalousmixing}
\hat \gamma^{(1)} = \cf  
\begin{pmatrix}
0 & 0 \\
-12 & 12  
\end{pmatrix} \, , 
\eeq
in agreement with \cite{Larin:1993tq}. At NNLO one needs further an
evanescent operator so that $\hat{\gamma}^{(1)}$ is enlarged to a $3
\times 3$ matrix. 

In the chosen operator basis the LO contributions to the initial
values of the Wilson coefficients are 
\beq \label{eq:CZCCS}
(C_Z (\muh), C_{CS} (\muh)) = (1, 0) \, .
\eeq
The particular form of $\hat \gamma^{(1)}$ then ensures that the
Wilson coefficients up to NLO satisfy $(C_Z (\mul), C_{CS} (\mul)) =
(1, 0)$ at any scale $\mul$. In fact, this scale independence is a
striking consequence of the Adler-Bardeen theorem \cite{Adler:1969er},
which states that the Adler-Bell-Jackiw (ABJ) anomaly \cite{ABJ} of
the axial-vector current is not renormalized in perturbation
theory. This theorem is strictly proven to all orders for the abelian
case \cite{collins, Adler:1969er}, while strong arguments suggest that
it holds true for the non-abelian case \cite{Bos:1992nd}. Assuming
that the ABJ anomaly equation survives renormalization, it is easy to
show that $C_Z (\mul)$ is scale independent if and only if $C_{CS}
(\mul)$ does not receive radiative corrections in the chosen operator
basis, where \Eq{eq:CZCCS} holds. In a renormalizable anomaly-free
theory, such as the SM \cite{anomalyfree}, this can always be achieved
by invoking an additional finite renormalization of the axial-vector
current \cite{collins, Larin:1993tq, Trueman:1979en}. For what
concerns $C_Z (\mul)$ this means that one has to perform a finite
renormalization of $Q^q_A$ to obtain the matching condition $C_{CS}
(\muh) = 0$ beyond one loop. The corresponding finite $\ord (\as^2)$
correction to the renormalization constant $Z_{AA}$ will be computed
in \Sec{sec:Z}. Also this finite renormalization is not an ad hoc
addition to our calculation, but originates from the loop diagrams in
\Fig{fig:twoloopZ} containing a top quark. Instead of absorbing these
effects into $Z_{AA}$ one could include them in $C_Z (\muh)$. In this
case one would also find a non-zero anomalous dimension of
$Q_Z$. Since both terms combine to reproduce the effect of $Z_{AA}$ 
the physical result is however unchanged. Furthermore, owing to the
definition of $Q_Z$ in \Eq{eq:HZ}, the Wilson coefficient of $Q_{CS}$
does not receive a matching correction at the bottom quark threshold
scale $\mub$. As the anomalous dimensions of operators in the
effective theory correspond to coefficients of large logarithms in the
full theory, the RG invariance of $C_Z (\mul)$ implies that anomalous
subdiagrams involving the $Z$-boson do not give rise to logarithms
$\ln ( \mb^2/\mt^2 )$ in the associated SM amplitudes to all orders in
perturbation theory. In contrast, large logarithms proportional to
$\mc^2/\MW^2$, which are relevant to our calculation, may arise. Such
terms correspond to higher-dimensional operators, which are a priori
not covered by the Adler-Bardeen theorem. In \Sec{sec:Z} we will,
however, show by an explicit three-loop calculation that the latter
terms are absent in the $\ord (\as^2)$ charm quark contribution to the
$\stodnunu$ transition in the SM.   

{%
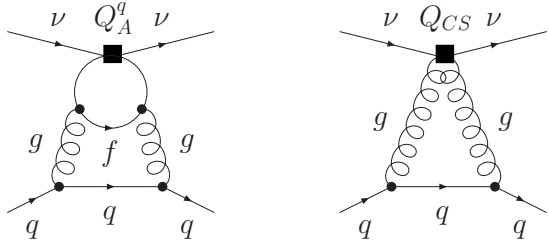
\begin{figure}[!t]
\begin{center}
$\begin{array}{c@{\hspace{15mm}}c}
\scalebox{0.65}{
\begin{picture}(120,139) (60,-29)
\SetWidth{0.5}
\SetColor{Black}
\ArrowLine(60,94)(120,79)
\ArrowLine(120,79)(180,94)
\ArrowLine(90,4)(150,4)
\ArrowLine(150,4)(180,-11)
\ArrowLine(60,-11)(90,4)
\Vertex(90,4){2.83}
\Vertex(150,4){2.83}
\GBox(116,76)(126,86){0.0}
\Text(90,101)[]{\Large{\Black{$\nu$}}}
\Text(150,101)[]{\Large{\Black{$\nu$}}}
\Text(75,-21)[]{\Large{\Black{$q$}}}
\Text(120,-13)[]{\Large{\Black{$q$}}}
\Text(165,-21)[]{\Large{\Black{$q$}}}
\Text(122,102)[]{\Large{\Black{$Q_A^q$}}}
\ArrowArc(120,59)(21.02,87,447)
\Gluon(90,4)(102,49){6}{3.43}
\Gluon(138,49)(150,4){6}{3.43}
\Vertex(102,49){2.83}
\Vertex(138,49){2.83}
\Text(165,29)[]{\Large{\Black{$g$}}}
\Text(77,29)[]{\Large{\Black{$g$}}}
\Text(120,23)[]{\Large{\Black{$\nf$}}}
\end{picture}
} &
\scalebox{0.65}{
\begin{picture}(120,139) (60,-29)
\SetWidth{0.5}
\SetColor{Black}
\ArrowLine(60,94)(120,79)
\ArrowLine(120,79)(180,94)
\Gluon(120,79)(150,4){6}{6.66}
\ArrowLine(90,4)(150,4)
\ArrowLine(150,4)(180,-11)
\ArrowLine(60,-11)(90,4)
\Vertex(90,4){2.83}
\Vertex(150,4){2.83}
\Gluon(90,4)(120,79){6}{6.86}
\GBox(116,76)(126,86){0.0}
\Text(90,101)[]{\Large{\Black{$\nu$}}}
\Text(150,101)[]{\Large{\Black{$\nu$}}}
\Text(75,-21)[]{\Large{\Black{$q$}}}
\Text(120,-13)[]{\Large{\Black{$q$}}}
\Text(165,-21)[]{\Large{\Black{$q$}}}
\Text(157,42)[]{\Large{\Black{$g$}}}
\Text(84,42)[]{\Large{\Black{$g$}}}
\Text(122,102)[]{\Large{\Black{$Q_{CS}$}}}
\GBox(116,76)(126,86){0.0}
\end{picture}
}
\end{array}$
\vspace{2mm}
\caption{Diagrams describing the mixing of $Q_A^q$ and $Q_{CS}$ into
$Q_A^q$ and $E_A^q$ at $\ord (\as^2)$. The fermion loop containing
$\nf$ quark flavors is anomalous, if $\nf$ is odd. The Feynman graph
in which the fermion flow in the closed quark line is opposite to the
one shown in the left diagram is not displayed.}  
\label{fig:anomaly}
\end{center}
\end{figure}
}%

\subsection{Charged Current: {\boldmath $W$}-Exchange}
\label{subsec:WEX}

In contrast to the neutral-current case, the discussion of the
effective charged-current couplings can be kept rather short. The 
interactions between leptons and quarks mediated by $W$-boson exchange
are encoded in the following low-energy effective Hamiltonian 
\beq \label{eq:HW}
\Heff^W = \f{4 \GF}{\sqrt{2}} \, C_{\scriptstyle W} (\mul) \sum_{q =
u, c} \left ( V_{qs} Q^q_3 + V^\ast_{qd} Q^q_4 \right ) \, ,    
\eeq
where 
\beq \label{eq:Q3Q4}
Q^q_3 = \sum_{\ell = e, \mu, \tau} (\bar s_\sL \gamma_{\mu} q_\sL ) (
\bar {\nu_\ell}_\sL \gamma^{\mu} \ell_\sL ) \, , \hspace{1cm} Q^q_4 =
\sum_{\ell = e, \mu, \tau} ( \bar q_\sL \gamma_{\mu} d_\sL ) ( \bar
\ell_\sL \gamma^{\mu} {\nu_\ell}_\sL ) \, .      
\eeq
Since the effective charged-current couplings $Q^q_3$ and $Q^q_4$ do
not mix under renormalization the Wilson coefficient $C_{\scriptstyle
W} (\mul)$ is $\mul$-independent. The normalization of \Eq{eq:HW} is 
chosen such that $C_{\scriptstyle W} (\mul) = 1$.  
 
\section{{\boldmath $Z$}-Penguin Contributions}  
\label{sec:Z}

\subsection{Effective Hamiltonian}
\label{subsec:ZHeff}

After integrating out the top quark and the heavy electroweak gauge
bosons we first encounter an effective Hamiltonian which is valid for
scales $\mul$ in the range $\muc \le \mul \le \muh$ with dynamical
bottom and charm quark fields. The $Z$-penguin contribution involves
$\Heff^{CC}$ and $\Heff^Z$ defined in \Eqsand{eq:HCC}{eq:HZ} as well
as the effective Hamiltonian  
\beq \label{eq:HPv}
\Heff^{\hspace{-0.1mm} P \nu} = \f{\GF}{\sqrt{2}} \lambda_c \f{\pi
\aem}{\MW^2 \sws} C^P_\nu (\mul) Q_\nu \, ,
\eeq
which finally brings the leading dimension-eight operator $Q_\nu$ of
\Eq{eq:Qv} into play.  

The desired matrix element $\langle \pi^+ \nu \bar{\nu} | {\cal T} |
K^+ \rangle$ involves the transition operator ${\cal T} = {\cal T}^P +
{\cal T}^B$, where ${\cal T}^P$ and ${\cal T}^B$ denote the
$Z$-penguin and electroweak box contribution, respectively. The
$Z$-penguin contribution to the transition operator takes the
following form 
\beq \label{eq:TP}
\begin{split}
-{\cal T}^P & =  \Heff^{\hspace{-0.1mm} P \nu} - i \int \!
d^4 x \hspace{1mm} T \big ( \Heff^{CC} (x) \Heff^Z (0) \big ) 
\\[1mm]    
& = \f{\GF}{\sqrt{2}} \lambda_c \f{\pi \aem}{\MW^2 \sws} \big (
C^P_\nu (\mul) Q_\nu + 4 \hspace{0.4mm} C_+ (\mul) Q^P_+ + 4
\hspace{0.4mm} C_- (\mul) Q^P_- \big ) \, .
\end{split}
\eeq
Notice that in passing from the first to the second line we have used
$C_Z (\mul) = 1$. The last two terms in \Eqs{eq:TP} are the bilocal 
composite operators $Q_\pm^P$ that involve the effective current- and
neutral-current couplings $Q_\pm$ and $Q_Z$. The former operator has
already been introduced in \Eq{eq:QPpm}. In contrast to
\cite{Buchalla:1993wq} we have defined it in terms of chiral and not
``$V - A$'' fermion fields. This results in the factors of $4$
multiplying $C_\pm (\mul)$ in the above equation.
          
As the normalization of $\aem$ is determined by the short distance
interactions at $\muh$, the Wilson coefficients are appropriately
expressed in terms of $\GF$ using the relation $g^2 (\muh)/4\pi = \aem
(\muh)/\sws (\muh) = \sqrt{2}/\pi \, \GF \MW^2$, where all running
couplings are defined in the $\MSbar$ scheme. The typical case is that
of electroweak box diagrams, to be discussed in the next section,
which are clearly proportional to $g^4 (\muh)/\MW^2 \propto \GF^2
\MW^2$. After decoupling, all short distance information is encoded in
the Wilson coefficients and in $\GF$, which does not evolve in the
effective theory. Hence the electromagnetic coupling $\aem$ and the
weak mixing angle $\sws$ entering \Eqsand{eq:TP}{eq:TB} are naturally
evaluated at the weak scale \cite{Bobeth:2003at}.      
     
{%
\begin{figure}[!t]
\begin{center}
$\begin{array}{c@{\hspace{15mm}}c}
\scalebox{0.65}{
\begin{picture}(150,134) (30,-31)
\SetWidth{0.5}
\SetColor{Black}
\Photon(75,74)(135,74){6}{5}
\ArrowLine(135,74)(180,89)
\ArrowLine(105,14)(135,74)
\Photon(105,14)(105,-31){6}{4}
\Vertex(75,74){2.83}
\Vertex(135,74){2.83}
\Vertex(105,14){2.83}
\ArrowLine(75,74)(90,44)
\ArrowLine(90,44)(105,14)
\Vertex(90,44){2.83}
\ArrowLine(30,89)(52,82)
\GlueArc(75,67)(27.46,146.89,303.11){6}{6.86}
\Vertex(52,82){2.83}
\ArrowLine(52,82)(75,74)
\Text(125,-8)[]{\Large{\Black{$Z$}}}
\Text(131,42)[]{\Large{\Black{$c$}}}
\Text(88,22)[]{\Large{\Black{$c$}}}
\Text(105,94)[]{\Large{\Black{$W$}}}
\Text(45,34)[]{\Large{\Black{$g$}}}
\Text(71,59)[]{\Large{\Black{$c$}}}
\Text(157,96)[]{\Large{\Black{$d$}}}
\Text(44,97)[]{\Large{\Black{$s$}}}
\Text(66,90)[]{\Large{\Black{$s$}}}
\end{picture}
} & 
\scalebox{0.65}{
\begin{picture}(161,173) (25,8)
\SetWidth{0.5}
\SetColor{Black}
\Photon(75,113)(135,113){6}{5}
\ArrowLine(105,53)(135,113)
\Photon(105,53)(105,8){6}{4}
\Vertex(75,113){2.83}
\Vertex(135,113){2.83}
\Vertex(105,53){2.83}
\ArrowLine(30,128)(52,121)
\Vertex(52,121){2.83}
\ArrowLine(52,121)(75,113)
\Text(125,31)[]{\Large{\Black{$Z$}}}
\Text(131,81)[]{\Large{\Black{$c$}}}
\Text(105,133)[]{\Large{\Black{$W$}}}
\ArrowLine(75,113)(105,53)
\ArrowLine(135,113)(158,121)
\ArrowLine(158,121)(180,128)
\Text(79,81)[]{\Large{\Black{$c$}}}
\GlueArc(105,93.11)(59.89,27.76,152.24){6}{11.14}
\Vertex(158,121){2.83}
\Text(40,112)[]{\Large{\Black{$s$}}}
\Text(62,106)[]{\Large{\Black{$s$}}}
\Text(148,104)[]{\Large{\Black{$d$}}}
\Text(171,111)[]{\Large{\Black{$d$}}}
\Text(105,173)[]{\Large{\Black{$g$}}}
\end{picture}
}
\end{array}$
\vspace{2mm}
\caption{Examples of $Z$-penguin diagrams that contribute to the
initial value of the Wilson coefficient of the leading dimension-eight
operator $Q_\nu$ at $\ord (\as^2)$.} 
\label{fig:penguinmatching}
\end{center}
\end{figure}
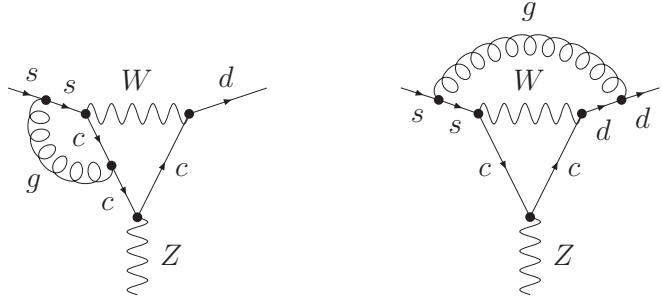
}%

\subsection{Initial Conditions}
\label{subsec:ZIC}

The matching corrections for $C^P_\nu (\mul)$ are again found by  
requiring equality of perturbative amplitudes generated by the full
and the effective theory. Examples of two-loop $Z$-penguin SM diagrams
can be seen in \Fig{fig:penguinmatching}. Regulating spurious IR
divergences dimensionally we obtain for the non-zero matching
conditions in the NDR scheme    
\beq \label{eq:CPpminitial}
\begin{split}
C^{P (1)}_\nu (\muh) & = 8 \left ( 2 + \ln \f{\muh^2}{\MW^2} \right )
\, , \\  
C^{P (2)}_\nu (\muh) & = 4 \hspace{0.4mm} \cf \left ( 33 + 4 \pi^2 +
34 \ln \f{\muh^2}{\MW^2} + 12 \ln^2 \f{\muh^2}{\MW^2} \right ) \, ,  
\end{split}
\eeq
where the first line recalls the NLO result \cite{Buchalla:1993wq},
while the second one represents the new NNLO expression. 

\subsection{Anomalous Dimensions: Non-Anomalous Contributions}
\label{subsec:ZADMNON}

{%
\begin{figure}[!t]
\begin{center}
$\begin{array}{c@{\hspace{15mm}}c}
\scalebox{0.65}{
\begin{picture}(120,137) (45,-30)
\SetWidth{0.5}
\SetColor{Black}
\ArrowLine(45,92)(105,77)
\ArrowLine(105,77)(165,92)
\ArrowLine(105,2)(165,-13)
\ArrowArc(97.5,39.5)(38.24,-78.69,78.69)
\Text(65,99)[]{\Large{\Black{$\nu$}}}
\Text(145,99)[]{\Large{\Black{$\nu$}}}
\Text(150,38)[]{\Large{\Black{$c$}}}
\Text(75,-22)[]{\Large{\Black{$s$}}}
\Text(135,-20)[]{\Large{\Black{$d$}}}
\GBox(100,-3)(110,7){0.0}
\GBox(100,72)(110,82){0.0}
\Text(106,-19)[]{\Large{\Black{$Q^c_\pm$}}}
\Text(106,97)[]{\Large{\Black{$Q^c_A, E^c_A$}}}
\ArrowLine(45,-13)(105,2)
\ArrowArc(112.5,39.5)(38.24,101.31,258.69)
\Text(60,38)[]{\Large{\Black{$c$}}}
\end{picture}
} & 
\scalebox{0.65}{
\begin{picture}(120,138) (45,-30)
\SetWidth{0.5}
\SetColor{Black}
\ArrowLine(45,93)(105,78)
\ArrowLine(165,93)(105,78)
\ArrowLine(105,3)(45,-12)
\ArrowLine(105,3)(165,-12)
\ArrowArcn(97.5,40.5)(38.24,78.69,-78.69)
\Text(75,100)[]{\Large{\Black{$s$}}}
\Text(135,99)[]{\Large{\Black{$\nu$}}}
\Text(150,39)[]{\Large{\Black{$\ell$}}}
\Text(75,-21)[]{\Large{\Black{$d$}}}
\Text(135,-22)[]{\Large{\Black{$\nu$}}}
\GBox(100,-2)(110,8){0.0}
\GBox(100,73)(110,83){0.0}
\Text(106,-18)[]{\Large{\Black{$Q^c_4$}}}
\Text(106,98)[]{\Large{\Black{$Q^c_3$}}}
\ArrowArc(112.5,40.5)(38.24,101.31,258.69)
\Text(60,39)[]{\Large{\Black{$c$}}}
\end{picture}
}
\end{array}$
\vspace{2mm}
\caption{Diagrams involving the double operator insertions $(Q^q_\pm,
Q^q_A)$, $(Q^q_\pm, E^q_A)$, and $(Q^q_3, Q^q_4)$.} 
\label{fig:oneloopmixing}
\end{center}
\end{figure}
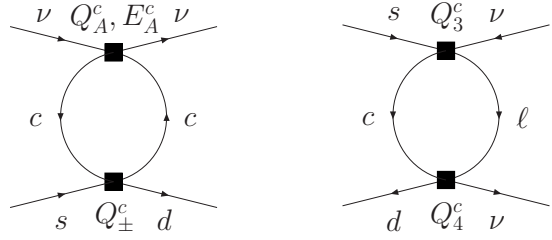
}%

The mixing of lower- into higher-dimensional operators through double
insertions leads in general to inhomogeneous RG equations
\cite{Herrlich:1994kh}. In the case of the Wilson coefficient $C^P_\nu
(\mul)$ introduced in \Eq{eq:TP} one has explicitly       
\beq \label{eq:RGCPv}
\mul \f{d}{d \mul} C^P_\nu (\mul) = \gamma_\nu (\mul) C^P_\nu (\mul) +
4 \sum_{i = \pm} \gamma^P_{i, \, \nu} (\mul) C_i (\mul) \, ,   
\eeq  
where $\gamma_\nu (\mul)$ encodes the self-mixing of $Q_\nu$, while
the anomalous dimension tensor $\gamma^P_{\pm, \, \nu} (\mul)$
describes the mixing of the bilocal composite operators $Q^P_\pm$ into
$Q_\nu$. The factor of $4$ in the above relation is of course a direct
consequence of the factors of $4$ in \Eq{eq:TP}.

Since the conserved current $\bar{s}_\sL \gamma_\mu d_\sL$ in $Q_\nu$
is not renormalized, the RG evolution of $Q_\nu$ stems solely from the
prefactor $\mc^2/\gs^2$ in \Eq{eq:Qv}. In terms of the expansion
coefficients of the anomalous dimension of the charm quark  $\MSbar$
mass and of the QCD $\beta$-function, the corresponding anomalous
dimension reads           
\beq \label{eq:gammav}
\gamma^{(k)}_\nu = 2 \left ( \gammamck - \betak \right ) \, . 
\eeq
In the particular case of QCD one has up to the NNLO level\footnote{We
have calculated the anomalous dimension of the quark mass and the
strong coupling constant in the $\MSbar$ scheme up to the three-loop
level, finding perfect agreement with the literature
\cite{Larin:1993tq, threeloopQCD}.}   
\beq \label{eq:gammambeta}
\begin{split}
\gammamczero = 8 \, , \hspace{5mm} \gammamcone = \f{404}{3} -
\f{40}{9} \nf \, , \hspace{5mm} \gammamctwo = 2498 -
\left(\f{4432}{27} +  \f{320}{3} \zetathree \right) \nf - \f{280}{81}
\nf^2 \, , \\[2mm]      
\betazero = 11 - \f{2}{3} \nf \, , \hspace{5mm} \betaone = 102 -
\f{38}{3} \nf \, , \hspace{5mm} \betatwo = \f{2857}{2} - \f{5033}{18}
\nf + \f{325}{54} \nf^2 \, . \hspace{12.5mm}
\end{split}
\eeq

{%
\begin{figure}[!t]
\begin{center}
$\begin{array}{c@{\hspace{15mm}}c}
\scalebox{0.65}{
\begin{picture}(143,137) (22,-41)
\SetWidth{0.5}
\SetColor{Black}
\ArrowLine(45,92)(105,77)
\ArrowLine(105,77)(165,92)
\ArrowLine(45,-13)(105,2)
\ArrowLine(105,2)(165,-13)
\ArrowArc(97.5,39.5)(38.24,-78.69,78.69)
\Text(75,99)[]{\Large{\Black{$\nu$}}}
\Text(135,99)[]{\Large{\Black{$\nu$}}}
\GlueArc(87.79,9.56)(33.42,114.38,213.73){6}{5.14}
\Vertex(60,-9){2.83}
\Vertex(74,40){2.83}
\Text(150,38)[]{\Large{\Black{$c$}}}
\Text(75,-22)[]{\Large{\Black{$s$}}}
\Text(135,-20)[]{\Large{\Black{$d$}}}
\GBox(100,-3)(110,7){0.0}
\GBox(100,72)(110,82){0.0}
\Text(106,-19)[]{\Large{\Black{$Q_\pm^{c}$}}}
\Text(106,97)[]{\Large{\Black{$Q_A^c$}}}
\Text(37,19)[]{\Large{\Black{$g$}}}
\ArrowArc(114.17,37.83)(40.23,103.17,176.91)
\ArrowArc(113.93,40.93)(39.94,-178.67,-102.92)
\Text(70,67)[]{\Large{\Black{$c$}}}
\Text(70,12)[]{\Large{\Black{$c$}}}
\end{picture}
} & 
\scalebox{0.65}{
\begin{picture}(135,148) (30,-30)
\SetWidth{0.5}
\SetColor{Black}
\ArrowLine(45,103)(105,88)
\ArrowLine(105,88)(165,103)
\Text(65,112)[]{\Large{\Black{$\nu$}}}
\Text(145,112)[]{\Large{\Black{$\nu$}}}
\GBox(100,83)(110,93){0.0}
\Text(106,108)[]{\Large{\Black{$Q_A^d, Q_V^d$}}}
\ArrowLine(105,88)(165,-2)
\ArrowLine(45,-2)(75,43)
\ArrowLine(75,43)(105,88)
\ArrowArc(90,28)(21.02,177,537)
\GlueArc(115.61,37.86)(40.1,-129.68,-28.06){6}{6.86}
\Vertex(90,7){2.83}
\Vertex(151,19){2.83}
\GBox(67,34)(77,44){0.0}
\Text(123,-22)[]{\Large{\Black{$g$}}}
\Text(124,26)[]{\Large{\Black{$c$}}}
\Text(54,52)[]{\Large{\Black{$Q_\pm^{c}$}}}
\Text(45,23)[]{\Large{\Black{$s$}}}
\Text(73,70)[]{\Large{\Black{$d$}}}
\Text(150,48)[]{\Large{\Black{$d$}}}
\end{picture}
}
\end{array}$
\vspace{2mm}
\caption{Sample diagrams for the mixing of the double insertion
$(Q^q_\pm, Q_Z)$ into $Q_\nu$ at $\ord (\as^2)$.}
\label{fig:penguinnlo}
\end{center}
\end{figure}
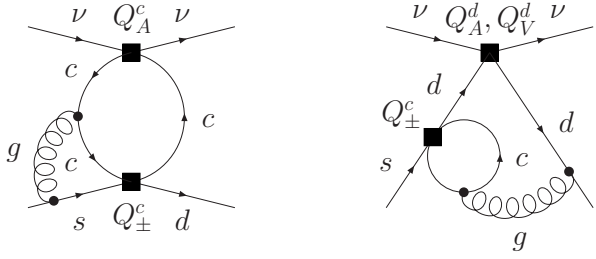
}%

The contributions to the anomalous dimension tensor $\gamma^P_{\pm,
\nu} (\mul)$ stemming from non-anomalous diagrams can be decomposed in
the following way  
\beq \label{eq:gammaPpmvdecomposition}
\gamma^{P (k)}_{\pm, \nu} = -\frac{1}{2} \gamma^{A (k)}_{\pm, \nu} -
\left ( \f{1}{2} - \f{2}{3} \sws \hspace{-0.5mm} \right ) \gamma^{V
(k)}_{\pm, \nu} \, ,    
\eeq
where the superscript $A$ and $V$ marks the corrections arising from
diagrams with a double operator insertion $(Q^q_\pm, Q^q_A)$ and
$(Q^q_\pm, Q^q_V)$. In the NDR scheme supplemented by the definition
of evanescent operators given in \Eqs{eq:ccevanescent} we obtain after
setting $\ca = 3$ and $\cf = 4/3$ the following coefficients     
\begin{align} \label{eq:gammaAVpmv}
\gamma^{A (0)}_{\pm, \nu} & = -4 \left ( 1 \pm 3 \right ) \, , & 
\gamma^{V (0)}_{\pm, \nu} & = 0 \, , \non \\[3mm]
\gamma^{A (1)}_{\pm, \nu} & = 16 \left ( 2 \mp 11 \right ) \, , &   
\gamma^{V (1)}_{\pm, \nu} & = 0 \, , \\[2mm]
\gamma^{A (2)}_{\pm, \nu} & = -\f{2}{225} \left ( 45124 \pm 484917
\right ) + 32 \left ( 13 \pm 15 \right ) \zetathree \pm 144 \nf \, , &
\gamma^{V (2)}_{\pm, \nu} & = 0 \, .
\non 
\end{align}
The results in the second line agree with the findings for
$\gamma_{\pm 3}^{(1)}$ of the prior NLO calculation
\cite{Buchalla:1993wq} if one takes into account $i)$ a factor of 
$-1/2$ arising from the decomposition of $\gamma_{\pm, \nu}^{P (k)}$
in \Eq{eq:gammaPpmvdecomposition} and $ii)$ a factor of $4$ that stems
from the fact that our operators $Q_\pm^P$ and $Q_\nu$ are defined in
terms of chiral and not as traditionally done ``$V - A$'' fermion
fields. The third line shows our new NNLO results. We stress that also
at NNLO the part of the double operator insertion $(Q^q_\pm, Q_Z)$
proportional to $Q^c_\pm Q^c_A - Q^u_\pm Q^u_A$ accounts for the 
complete mixing. In order to understand this feature it is important
to realize that one can distinguish two kinds of contributions: $i)$
diagrams in which $Q_Z$ couples to an up-type quark as on the left of
\Figs{fig:oneloopmixing}, \fig{fig:penguinnlo}, and
\fig{fig:penguinnnlo}, and $ii)$ diagrams in which $Q_Z$ couples to a
down-type quark as on the right of
\Figsand{fig:penguinnlo}{fig:penguinnnlo}. This classification holds 
true to all orders in QCD. Diagrams of type $i)$ containing an
insertion of the vector part of $Q_Z$ do not contribute to the
anomalous dimensions. Further diagrams of type $ii)$ vanish in LO and
are UV-finite at NLO, but do have UV poles at NNLO. However, their
overall contribution is cancelled by diagrams like the ones shown in
\Fig{fig:offwave}, which induce a flavor off-diagonal wave function
renormalization. Of course, this additional wave function
renormalization has to be included in the renormalization of
$Q_Z$. Finally let us mention that the contributions from diagrams of
type $ii)$ containing an insertion of the axial- and vector part of
$Q_Z$ have opposite signs, which is a consequence of $Q_\pm^q$
containing only left-handed down and strange quark fields.

{%
\begin{figure}[!t]
\begin{center}
$\begin{array}{c@{\hspace{15mm}}c}
\scalebox{0.65}{
\begin{picture}(143,137) (22,-23)
\SetWidth{0.5}
\SetColor{Black}
\GlueArc(105,97.44)(25.44,-160.62,-19.38){6}{5.14}
\ArrowLine(45,122)(105,107)
\ArrowLine(105,107)(165,122)
\ArrowLine(45,17)(105,32)
\ArrowLine(105,32)(165,17)
\ArrowArc(97.5,69.5)(38.24,-78.69,78.69)
\Text(75,129)[]{\Large{\Black{$\nu$}}}
\Text(135,129)[]{\Large{\Black{$\nu$}}}
\GlueArc(87.79,39.56)(33.42,114.38,213.73){6}{5.14}
\Vertex(60,21){2.83}
\Vertex(74,70){2.83}
\Text(150,68)[]{\Large{\Black{$c$}}}
\Text(75,8)[]{\Large{\Black{$s$}}}
\Text(135,10)[]{\Large{\Black{$d$}}}
\GBox(100,27)(110,37){0.0}
\GBox(100,102)(110,112){0.0}
\Text(106,11)[]{\Large{\Black{$Q_\pm^{c}$}}}
\Text(106,127)[]{\Large{\Black{$Q_A^c$}}}
\Text(37,49)[]{\Large{\Black{$g$}}}
\CArc(114.17,67.83)(40.23,103.17,176.91)
\ArrowArc(113.93,70.93)(39.94,-178.67,-102.92)
\Text(70,42)[]{\Large{\Black{$c$}}}
\Vertex(81,89){2.83}
\Vertex(129,89){2.83}
\Text(105,54)[]{\Large{\Black{$g$}}}
\end{picture}
} & 
\scalebox{0.65}{
\begin{picture}(135,146) (30,-30)
\SetWidth{0.5}
\SetColor{Black}
\ArrowLine(45,116)(105,101)
\ArrowLine(105,101)(165,116)
\Text(65,125)[]{\Large{\Black{$\nu$}}}
\Text(145,125)[]{\Large{\Black{$\nu$}}}
\GBox(100,96)(110,106){0.0}
\Text(106,121)[]{\Large{\Black{$Q_A^d, Q_V^d$}}}
\ArrowLine(105,101)(165,11)
\ArrowLine(45,11)(75,56)
\ArrowLine(75,56)(105,101)
\ArrowArc(90,41)(21.02,177,537)
\GlueArc(115.61,50.86)(40.1,-129.68,-28.06){6}{6.86}
\Vertex(90,20){2.83}
\Vertex(151,32){2.83}
\GBox(67,47)(77,57){0.0}
\Text(124,39)[]{\Large{\Black{$c$}}}
\Text(54,65)[]{\Large{\Black{$Q_\pm^{c}$}}}
\Text(45,36)[]{\Large{\Black{$s$}}}
\Text(73,83)[]{\Large{\Black{$d$}}}
\Text(150,61)[]{\Large{\Black{$d$}}}
\GlueArc(91.34,36.71)(42,-159.5,-46.98){6}{8.57}
\Vertex(120,6){2.83}
\Vertex(52,22){2.83}
\Text(73,-22)[]{\Large{\Black{$g$}}}
\Text(137,-4)[]{\Large{\Black{$g$}}}
\end{picture}
}
\end{array}$
\vspace{2mm}
\caption{Sample diagrams for the $\ord (\as^3)$ mixing of $(Q^q_\pm,
Q_Z)$ into $Q_\nu$.}          
\label{fig:penguinnnlo}
\end{center}
\end{figure}
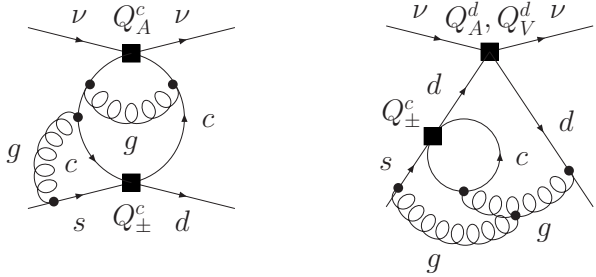
}%

\subsection{Anomalous Dimensions: Anomalous Contributions}
\label{subsec:ZADMANO}

In order to calculate the $\ord (\as^3)$ mixing of the double
insertions $(Q^q_\pm, Q_Z)$ into $Q_\nu$, we will need the
renormalization constants of $Q^q_A$ and $Q_{CS}$ defined
in \Eqsand{eq:QVQA}{eq:QCS} up to $\ord (\as^2)$, because these
operators appear as subloop counterterms in the effective
theory. Diagrams involving such subgraphs can be seen in
\Fig{fig:penguinanomaly}. The renormalized operators $Q^q_A$ and
$Q_{CS}$ defined in \Eqsand{eq:QVQA}{eq:QCS} may be expressed in terms
of the bare ones as    
\beq \label{eq:QAQCS}
\begin{split}
Q^q_A & = Z_{AA} Q^q_{A, \bare} + Z_{AE} E^q_{A, \bare} \, , \\[1mm]  
Q_{CS} & = \mul^{-2 \eps} \, Q_{CS, \bare} + Z_{CA} Q^q_{A, \bare} +
Z_{CE} E^q_{A, \bare} \, ,  
\end{split}
\eeq
where the unexpected factor $\mu^{-2 \eps}$ stems from the relation
between the bare and the renormalized strong coupling constant. Note
first that $Q^q_A$ is protected by its gauge invariance from
non-diagonal renormalization involving the gauge-variant operator
$Q_{CS}$, and second that $Q_{CS}$ has no diagonal renormalization due
to the factor $\gs^2$ in its definition.  

{%
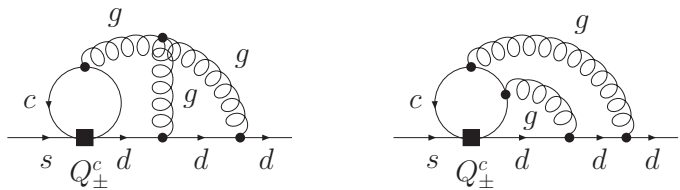
\begin{figure}[!t]
\begin{center}
$\begin{array}{c@{\hspace{5mm}}c}
\scalebox{0.65}{
\begin{picture}(195,109) (40,-17)
\SetWidth{0.5}
\SetColor{Black}
\Gluon(150,62)(150,4){6}{5.43}
\ArrowLine(195,4)(225,4)
\Vertex(195,4){2.83}
\ArrowLine(60,4)(105,4)
\CBox(109.66,-0.66)(100.34,8.66){Black}{Black}
\GlueArc(137.95,1.37)(56.26,127.11,3.96){-6}{12.43}
\ArrowLine(105,4)(150,4)
\ArrowLine(150,4)(195,4)
\Vertex(150,4){2.83}
\Vertex(105,45){2.83}
\Vertex(150,62){2.83}
\ArrowArc(105,24)(21.02,3,363)
\Text(83,-10)[c]{\Large{\Black{$s$}}}
\Text(128,-8)[c]{\Large{\Black{$d$}}}
\Text(73,25)[c]{\Large{\Black{$c$}}}
\Text(173,-8)[c]{\Large{\Black{$d$}}}
\Text(211,-8)[c]{\Large{\Black{$d$}}}
\Text(197,50)[c]{\Large{\Black{$g$}}}
\Text(167,27)[c]{\Large{\Black{$g$}}}
\Text(123,74)[c]{\Large{\Black{$g$}}}
\Text(108,-18)[c]{\Large{\Black{$Q^c_\pm$}}}
\end{picture}
} & 
\scalebox{0.65}{
\begin{picture}(195,109) (40,-17)
\SetWidth{0.5}
\SetColor{Black}
\ArrowLine(195,4)(225,4)
\Vertex(195,4){2.83}
\ArrowLine(60,4)(105,4)
\CBox(109.66,-0.66)(100.34,8.66){Black}{Black}
\GlueArc(137.95,1.37)(56.26,127.11,3.96){-6}{12.43}
\GlueArc(132.95,3.87)(28.26,106.11,3.96){-6}{4.43}
\ArrowLine(105,4)(165,4)
\ArrowLine(165,4)(195,4)
\Vertex(162,4){2.83}
\Vertex(105,45){2.83}
\Vertex(125,29){2.83}
\ArrowArc(105,24)(21.02,3,363)
\Text(83,-10)[c]{\Large{\Black{$s$}}}
\Text(135,-8)[c]{\Large{\Black{$d$}}}
\Text(73,25)[c]{\Large{\Black{$c$}}}
\Text(180,-8)[c]{\Large{\Black{$d$}}}
\Text(211,-8)[c]{\Large{\Black{$d$}}}
\Text(140,14)[c]{\Large{\Black{$g$}}}
\Text(170,70)[c]{\Large{\Black{$g$}}}
\Text(108,-18)[c]{\Large{\Black{$Q^c_\pm$}}}
\end{picture}
}
\end{array}$
\vspace{3mm}
\caption{Sample diagrams for the flavor off-diagonal wave function
renormalization at $\ord (\as^2)$.}                
\label{fig:offwave}
\end{center}
\end{figure}
}%

Within dimensional regularization the renormalization of $Q^q_A$ is
not exhausted by a multiplicative factor, but involves mixing with
the following evanescent operator as well  
\beq \label{eq:EA}
E^q_A = i \, \eps^{\mu_1 \mu_2 \mu_3 \mu_4} \left ( \bar q
\gamma_{\mu_1} \gamma_{\mu_2} \gamma_{\mu_3} q \right ) \sum_{\ell =
e, \mu, \tau} \left (\bar {\nu_\ell}_\sL \gamma_{\mu_4}
{\nu_{\ell}}_\sL \right ) + 6 Q^q_A \, .  
\eeq
In our case the quark fields in $Q^q_A$ and $E^q_A$ correspond to an
open fermion line. We will calculate the parts of the anomalous
dimensions that involve anomalous subloops with insertions of $Q^q_A$
using three different prescriptions for $\gamma_5$ for this open
line, namely NDR, HV, and DRED. Together with the two possibilities to
treat $\gamma_5$ in the closed fermion loop, which are HV and DRED,
this amounts to six renormalization prescriptions in total. It 
is instructive to see how different scheme-dependent pieces combine
into a scheme-independent result for $\PcX$. Note that diagrams with
an insertion $E^q_A$ must be included not only in the NDR and HV
schemes \cite{Buras:1989xd, Dugan:1990df, Herrlich:1994kh, collins},
but also in DRED, which is unexpected at first sight. The crucial
point here is that a mathematically consistent definition of DRED
involves infinite-dimensional spaces just as NDR and HV: the DRED
scheme entails $i)$ a formally $4$-dimensional, but really
infinite-dimensional, space for the gauge fields and Dirac matrices,
and $ii)$ a formally $n$-dimensional space for the momenta, which is a
subspace of the former one \cite{Stockinger:2005gx}. In consequence,
$E^q_A$ is not identical to zero in DRED, as it belongs to the
formally $n-4$-dimensional complement of the $n$-dimensional space.  

{%
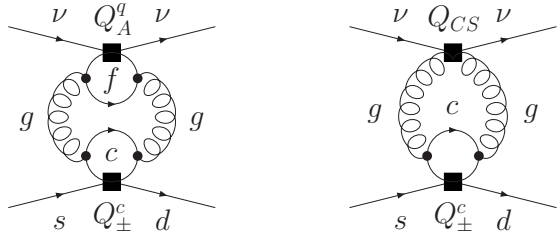
\begin{figure}[!t]
\begin{center}
$\begin{array}{c@{\hspace{15mm}}c}
\scalebox{0.65}{
\begin{picture}(129,137) (41,-30)
\SetWidth{0.5}
\SetColor{Black}
\ArrowLine(45,92)(105,77)
\ArrowLine(105,77)(165,92)
\ArrowLine(45,-13)(105,2)
\ArrowLine(105,2)(165,-13)
\Text(75,99)[]{\Large{\Black{$\nu$}}}
\Text(135,99)[]{\Large{\Black{$\nu$}}}
\Text(75,-22)[]{\Large{\Black{$s$}}}
\Text(135,-20)[]{\Large{\Black{$d$}}}
\GBox(100,-3)(110,7){0.0}
\GBox(100,72)(110,82){0.0}
\Text(106,-19)[]{\Large{\Black{$Q_\pm^{c}$}}}
\Text(106,97)[]{\Large{\Black{$Q^q_A$}}}
\ArrowArc(105,62)(15,90,450)
\ArrowArcn(105,17)(15,630,270)
\Vertex(90,62){2.83}
\Vertex(120,62){2.83}
\Vertex(120,17){2.83}
\Vertex(90,17){2.83}
\GlueArc(97.5,39.5)(23.72,108.43,251.57){6}{5.14}
\GlueArc(112.5,39.5)(23.72,-71.57,71.57){6}{5.14}
\Text(56,37)[]{\Large{\Black{$g$}}}
\Text(155,37)[]{\Large{\Black{$g$}}}
\Text(105,18)[]{\Large{\Black{$c$}}}
\Text(105,61)[]{\Large{\Black{$\nf$}}}
\end{picture}
} & 
\scalebox{0.65}{
\begin{picture}(121,137) (45,-30) 
\SetWidth{0.5}
\SetColor{Black}
\ArrowLine(45,92)(105,77)
\ArrowLine(105,77)(165,92)
\ArrowLine(45,-13)(105,2)
\ArrowLine(105,2)(165,-13)
\Text(75,99)[]{\Large{\Black{$\nu$}}}
\Text(135,99)[]{\Large{\Black{$\nu$}}}
\Text(75,-22)[]{\Large{\Black{$s$}}}
\Text(135,-20)[]{\Large{\Black{$d$}}}
\GBox(100,-3)(110,7){0.0}
\GBox(100,72)(110,82){0.0}
\Text(106,-19)[]{\Large{\Black{$Q_\pm^{c}$}}}
\Text(106,97)[]{\Large{\Black{$Q_{CS}$}}}
\ArrowArcn(105,17)(15,630,270)
\Vertex(120,17){2.83}
\Vertex(90,17){2.83}
\Text(105,45)[]{\Large{\Black{$c$}}}
\GlueArc(112.19,41.06)(32.74,102.69,227.32){6}{6.86}
\GlueArc(97.81,41.06)(32.74,-47.32,77.31){6}{6.86}
\Text(60,42)[]{\Large{\Black{$g$}}}
\Text(151,42)[]{\Large{\Black{$g$}}}
\end{picture}
}
\end{array}$
\vspace{2mm}
\caption{Examples of diagrams involving the double operator
insertion $(Q^q_\pm, Q^q_A)$ and $(Q^q_\pm, Q_{CS})$. In the
$(Q^q_\pm, Q^q_A)$ case the fermion loop containing $\nf$ quark
flavors is anomalous, if $\nf$ is odd.}                   
\label{fig:penguinanomaly}
\end{center}
\end{figure}
}%

The renormalization constants $Z_{AA}$, $Z_{AE}$, $Z_{CA}$, and
$Z_{CE}$ entering \Eqs{eq:QAQCS} are found by calculating the UV
divergent parts of Feynman diagrams in the effective theory. Sample 
graphs encoding the first non-trivial mixing of the set $(Q^q_A,
Q_{CS})$ into $Q^q_A$ and $E^q_A$ are displayed in
\Fig{fig:anomaly}. In the $\MSbar$ scheme we obtain    
\beq \label{eq:ZAAZCE}
\begin{split}
\begin{aligned}
Z_{AA}^{(2, 1)} & = 3 \hspace{0.4mm} \cf \, , & \hspace{1cm}
Z_{AE}^{(2, 1)} & = -\f{1}{2} \hspace{0.4mm} \cf \, , \\[1mm]
Z_{CA}^{(2, 1)} & = -6 \hspace{0.4mm} \cf \, , & \hspace{1cm}
Z_{CE}^{(2, 1)} & = \cf \, ,   
\end{aligned}
\end{split}
\eeq
where the symbol $Z_{ij}^{(k,l)}$ denotes the coefficient of the
$1/\eps^l$ pole of the $\ord (\as^k)$ term of the associated
renormalization constant. By taking into account the factor $\mu^{-2 
\eps}$ present in the second line of \Eqs{eq:QAQCS} and switching to
the basis $(Q_Z, Q_{CS})$ one recovers the ADM given in
\Eq{eq:anomalousmixing}. All the one-, two-, and  three-loop results  
presented in \Eqs{eq:ZAAZCE}, \eq{eq:finiteaxial}, \eq{eq:Z11pmAv} and
\eq{eq:ZNDRHVDRED} are again calculated using two different methods. In
the first approach, IR singularities are regulated by introducing a
common mass parameter into all the propagator denominators including
the gluon ones \cite{Misiak:1994zw, Chetyrkin:1997fm}, while in the
second one only the mass of the open quark line is kept non-zero
\cite{IRquarkmass}. The two methods give the same results for the
$\MSbar$ renormalization constants.     

The aforementioned finite renormalization of $Q^q_A$ is most easily
found by insisting that the $\ord (\as^2)$ correction to its initial
condition is identical to zero. This matching requires the calculation
of the two graphs shown in \Fig{fig:twoloopZ}. The UV divergences from
these diagrams are canceled by a counterterm proportional to
$Z^{(2,1)}_{AA}$. The leftover finite contribution can be either
absorbed into the initial value of the Wilson coefficient of $Q^q_A$
or into a finite renormalization constant $Z^{(2, 0)}_{AA}$. The
latter possibility is more convenient, as it avoids a spurious RG
running of the effective neutral-current coupling  $Q_Z$, which would
otherwise occur beyond NLO. We find   
\beq \label{eq:finiteaxial}
Z_{AA}^{\Delta (2,0)} = \f{3}{2} \hspace{0.4mm} \cf \, ,   
\eeq
for both HV and DRED. In the former scheme we reproduce the result of
\cite{Larin:1993tq}. With the superscript $\Delta$ we indicate that we
have only considered the contributions related to anomalous
graphs. The contributions from all other diagrams, which have only
open fermion lines, can be discussed separately. If the HV scheme is
used for the latter diagrams, an additional finite renormalization
constant is needed \cite{collins, Larin:1993tq, Trueman:1979en}.    

{%
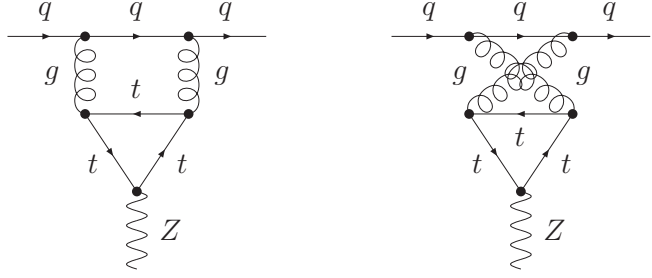
\begin{figure}[!t]
\begin{center}
$\begin{array}{c@{\hspace{15mm}}c}
\scalebox{0.65}{
\begin{picture}(150,158) (30,-22)
\SetWidth{0.5}
\SetColor{Black}
\ArrowLine(75,113)(135,113)
\Gluon(75,68)(75,113){6}{3.43}
\ArrowLine(135,68)(75,68)
\ArrowLine(75,68)(105,23)
\ArrowLine(105,23)(135,68)
\Photon(105,23)(105,-22){6}{4}
\Vertex(75,68){2.83}
\Vertex(135,68){2.83}
\Vertex(135,113){2.83}
\Vertex(75,113){2.83}
\Vertex(105,23){2.83}
\Gluon(135,113)(135,68){6}{3.43}
\Text(105,83)[]{\Large{\Black{$t$}}}
\Text(132,39)[]{\Large{\Black{$t$}}}
\Text(80,39)[]{\Large{\Black{$t$}}}
\Text(125,1)[]{\Large{\Black{$Z$}}}
\Text(155,91)[]{\Large{\Black{$g$}}}
\Text(56,91)[]{\Large{\Black{$g$}}}
\Text(52,128)[]{\Large{\Black{$q$}}}
\Text(105,128)[]{\Large{\Black{$q$}}}
\Text(157,128)[]{\Large{\Black{$q$}}}
\ArrowLine(30,113)(75,113)
\ArrowLine(135,113)(180,113)
\end{picture}
} & 
\scalebox{0.65}{
\begin{picture}(150,158) (30,-22)
\SetWidth{0.5}
\SetColor{Black}
\ArrowLine(75,113)(135,113)
\ArrowLine(135,68)(75,68)
\ArrowLine(75,68)(105,23)
\ArrowLine(105,23)(135,68)
\Photon(105,23)(105,-22){6}{4}
\Vertex(75,68){2.83}
\Vertex(135,68){2.83}
\Vertex(135,113){2.83}
\Vertex(75,113){2.83}
\Vertex(105,23){2.83}
\Text(105,55)[]{\Large{\Black{$t$}}}
\Text(132,39)[]{\Large{\Black{$t$}}}
\Text(80,39)[]{\Large{\Black{$t$}}}
\Text(125,1)[]{\Large{\Black{$Z$}}}
\Text(142,91)[]{\Large{\Black{$g$}}}
\Text(70,91)[]{\Large{\Black{$g$}}}
\Text(52,128)[]{\Large{\Black{$q$}}}
\Text(105,128)[]{\Large{\Black{$q$}}}
\Text(157,128)[]{\Large{\Black{$q$}}}
\Gluon(75,68)(135,113){6}{6.86}
\Gluon(75,113)(135,68){6}{6.86}
\ArrowLine(135,113)(180,113)
\ArrowLine(30,113)(75,113)
\end{picture}
}
\end{array}$
\vspace{2mm}
\caption{The SM diagrams one has to compute in order to find the $\ord
(\as^2)$ correction to the initial value of the Wilson coefficient of
$Q^q_A$.}                   
\label{fig:twoloopZ}
\end{center}
\end{figure}
}%

While we could rely on the Adler-Bardeen theorem to ensure that the
contributions of $Q^q_A$ and $Q_{CS}$ cancel in the renormalization of
$Q_Z$, the situation is more complicated in the case of the transition
operator encountered in \Eq{eq:TP}, because it involves double
insertions of the type $(Q^q_\pm, Q^q_A)$ and $(Q^q_\pm,
Q_{CS})$. Typical examples of such graphs are shown in
\Fig{fig:penguinanomaly}. Apparently, their renormalization requires
counterterms proportional to $Q_\nu$. The associated renormalization
constants $Z^P_{\pm j, \nu} (\mul)$ can be extracted at any given
order in $\as$ by requiring  
\beq \label{eq:renormalization} 
\langle Q^P_{\pm j} (\mul) \rangle + Z^P_{\pm j, \, \nu} (\mul)
\hspace{0.4mm} \langle Q_{\nu, \bare} (\mul) \rangle \, ,     
\eeq
to be UV finite. Here
\beq \label{eq:Qpmj}
Q^P_{\pm j} = -i \int \! d^4 x \hspace{1mm} T \big (  Q^c_\pm (x) Q_j
(0) - Q^u_\pm (x) Q_j (0) \big ) \, ,  
\eeq
and $j = A, C, E$, while $\langle \, \ldots \, \rangle$ denotes matrix 
elements which include the proper QCD renormalization of the coupling,
the masses and the fields. Note that since $Q^q_\pm$ and $Q_j$ are
renormalized operators, all subloop divergences are properly canceled
in \Eq{eq:renormalization}. Hence the renormalization constants
$Z^P_{\pm j, \nu} (\mul)$ are sufficient to achieve a finite result.

The general form of the anomalous dimension tensor for double
insertions has been derived in \cite{Herrlich:1994kh}. In the
following discussion we will only need the explicit expression for the
part of $\gamma^P_{\pm, \, \nu} (\mul)$ related to anomalous diagrams
given by  
\beq \label{eq:gammapmvDelta}
\gamma^\Delta_{\pm, \, \nu} (\mul) = - \hspace{-2mm} \sum_{k = A,
C} \hspace{-1mm} \Big( \mul \f{d}{d \mul} Z^{\Delta}_{\pm k, \, \nu}
(\mul) + \Big ( \sum_{i = \pm} \gamma_i (\mul) Z^\Delta_{i k, \, \nu}
(\mul) + \hspace{-3mm} \sum_{j = A, C, E} \hspace{-2mm}
\gamma_{kj} (\mul) Z^\Delta_{\pm j, \, \nu} (\mul) \, \Big ) \Big )
Z^{-1}_\nu (\mul) \, .     
\eeq
where $\gamma_i (\mu)$ and $\gamma_{kj} (\mu)$ are the elements of the
ADM in the $(Q^q_+, Q^q_-)$ and $(Q^q_A, Q_{CS}, E^q_A)$ sector, while
$Z_\nu (\mu)$ denotes the renormalization constant of $Q_\nu$. 

The renormalization constants $Z^I_{i j, \, \nu} (\mul)$ have the
following perturbative expansion 
\beq \label{eq:Zexpansion}
Z^I_{i j, \, \nu} (\mul) = \sum_{k = 1}^\infty \left ( \f{\as
(\mul)}{4 \pi}\right )^k Z^{I (k)}_{i j, \, \nu} \, , \hspace{1cm}
Z^{I (k)}_{i j, \, \nu} = \sum_{l = 0}^k  \f{1}{\eps^l} Z^{I (k,
l)}_{i j, \, \nu} \, ,    
\eeq
for $I = P, \Delta$. Following the standard $\MSbar$ prescription,
$Z^I_{i j, \, \nu} (\mul)$ is given by pure $1/\eps^l$ poles, except
when $i = \pm$ and  $j = E$. In the latter case, the renormalization
constant is finite to make sure that the matrix elements of double
insertions involving evanescent operators vanish in $n = 4$ dimensions
\cite{Buras:1989xd, Dugan:1990df, Herrlich:1994kh}. 

The finite parts of \Eq{eq:gammapmvDelta} in the limit $\eps$ going to
zero gives the anomalous dimension tensor. Performing an expansion in
powers of the strong coupling one recognizes that the first
non-trivial correction to $\gamma^{\Delta}_{\pm, \, \nu} (\mul)$
arises at the third order. We obtain         
\beq \label{eq:gammapmvDeltaNNLO}
\gamma^{\Delta (2)}_{\pm, \, \nu} = 6 Z^{\Delta (3, 1)}_{\pm A, \,
\nu} + 4 Z^{P (3, 1)}_{\pm C, \, \nu} - 4 Z^{(2, 1)}_{AE} Z^{P (1,
0)}_{\pm E, \, \nu} - 2 Z^{(2, 1)}_{CE} Z^{P (1, 0)}_{\pm E, \, \nu} -
4 Z^{\Delta (2, 0)}_{AA} Z^{P (1, 1)}_{\pm A, \, \nu} \, ,     
\eeq
which clearly verifies the impact of the finite renormalization of the
evanescent operator $E^q_A$ and the axial-vector coupling
$Q^q_A$. Note that in the above equation the superscript $\Delta$ has
been replaced by $P$ whenever possible. The renormalization constants
$Z^{\Delta (3, 1)}_{\pm A, \, \nu}$ and $Z^{P (3, 1)}_{\pm C, \, \nu}$
are found by calculating the three- and two-loop diagrams shown in
\Fig{fig:penguinanomaly}, whereas the determination of $Z^{P (1,
1)}_{\pm A, \, \nu}$ and $Z^{P (1, 0)}_{\pm E, \, \nu}$ requires only
a one-loop computation. The relevant Feynman graphs are displayed on
the left-hand side of \Fig{fig:oneloopmixing}.        

On the other hand the pole parts of \Eq{eq:gammapmvDelta} must
vanish. From this condition one obtains relations between single,
double and triple $1/\eps$ poles of the renormalization constants. In
our case the non-trivial ones read   
\beq \label{eq:checks}
\begin{split}
6 Z^{\Delta (3, 2)}_{\pm A, \, \nu} - 4 Z^{(2, 1)}_{AA} Z^{P (1,
1)}_{\pm A, \, \nu} & = 0 \, , \\[1mm]  
4 Z^{P (3, 2)}_{\pm C, \, \nu} - 2 Z^{(2, 1)}_{CA} Z^{P (1, 1)}_{\pm
A, \, \nu} & = 0 \, .   
\end{split}
\eeq
These equations constitute a powerful check of our three-loop 
calculation. For instance, an erroneous omission of the factor
$\mu^{-2 \eps}$ in the second line of \Eqs{eq:QAQCS} changes
\Eq{eq:gammapmvDeltaNNLO} as well as \Eqs{eq:checks}, and indeed leads
to a failure of the check.   

We now give the values of the renormalization constants entering
\Eq{eq:gammapmvDeltaNNLO} for the three possible renormalization
prescriptions for the open fermion line. The quantities of
\Eqs{eq:ZAAZCE} and 
\beq \label{eq:Z11pmAv}
Z^{P (1, 1)}_{\pm A, \, \nu} = -2 \left ( 1 \pm \ca \right ) \, , 
\eeq  
do not depend on the treatment of $\gamma_5$ in the open fermion
line. This is not the case for the remaining ones. In the $\MSbar$
scheme we find    
\bea \label{eq:ZNDRHVDRED}
Z^{\Delta (3, 1)}_{\pm A, \nu} = \begin{cases} -4 \cf \left ( 3 \pm
\ca \right ) \\[1mm] -8 \cf \end{cases} \hspace{-5mm} \, ,
\hspace{2.5mm} Z^{P (3, 1)}_{\pm C, \nu} = \begin{cases} 3 \cf 
\left ( 5 \pm \ca \right ) \\[1mm] 3 \cf \left ( 3 \mp \ca \right )
\end{cases} \hspace{-5mm} \, , \hspace{2.5mm} Z^{P (1, 0)}_{\pm E, 
\nu} = \begin{cases} 36 \left ( 1 \pm \ca \right ) \\[1mm] 0
\end{cases} \hspace{-5mm} \, ,
\hspace{2.5mm}  
\eea
where the expressions in the first line correspond to the NDR
scheme, while the second line shows the HV and DRED
results. Amazingly, HV and DRED defined as in 
\cite{Stockinger:2005gx} give exactly the same results for all
renormalization constants through NNLO.

Inserting \Eqs{eq:ZAAZCE}, \eq{eq:finiteaxial}, \eq{eq:Z11pmAv}, and
\eq{eq:ZNDRHVDRED} into \Eq{eq:gammapmvDeltaNNLO}, we see that
$\gamma^{\Delta (2)}_{\pm, \nu}$ vanishes in all six renormalization
schemes. This non-trivial result implies that anomalous subloops
involving the $Z$-boson do not give rise to a NNLO logarithm $\ln
(\mb^2/\mt^2)$ proportional to $\mc^2/\MW^2$ in the decay amplitude of
$\Ktopinunu$. We have checked the absence of these $\ord (\as^2)$ terms
explicitly by calculating the three-loop SM diagrams containing an
anomalous bottom quark loop and verifying that in the limit $\mb$ going
to zero no IR divergence appear in the corresponding amplitude. Beyond
NNLO the non-logarithmic pieces of three-loop diagrams containing
anomalous subgraphs will be relevant and it is highly non-trivial
whether the cancellation between the effects from top and bottom quark
triangles carries over to this and higher orders.

\subsection{RG Evolution}
\label{subsec:ZRGE}

Since in our renormalization scheme specified by the evanescent 
operators in \Eqs{eq:ccevanescent} the Wilson coefficients $C_\pm
(\mul)$ evolve independently from each other, \Eq{eq:RGCPv} splits
into two inhomogeneous differential equations. Using
\Eq{eq:renormalizationgroupequation} the RG evolution of the Wilson 
coefficients entering the $Z$-penguin contribution may then be recast
into the following homogeneous differential equation
\cite{Buchalla:1993wq, Herrlich:1996vf}     
\beq \label{eq:simpleRGCPv} 
\mul \f{d}{d \mul} \vec{C}_P (\mul) = \hat \gamma^T_P (\mul) \vec{C}_P
(\mul) \, ,      
\eeq
where 
\beq \label{eq:CPgP} 
\vec{C}_P (\mul) = \begin{pmatrix} 4 \hspace{0.4mm} C_+ (\mul) \\ 4
\hspace{0.4mm} C_- (\mul) \\ C^P_\nu (\mul) \end{pmatrix} \, ,
\hspace{1cm} \hat \gamma_P (\mul) = \begin{pmatrix} \gamma_+ (\mul) &
0 & \gamma^P_{+, \nu} (\mul) \\ 0 & \gamma_- (\mul) & \gamma^P_{-,
\nu} (\mul) \\ 0 & 0 & \gamma_\nu (\mul) \end{pmatrix} \, , 
\eeq
which can be solved by the standard techniques \cite{Ciuchini:1993vr,
Buras:1993dy, Buras:1991jm} introduced for single operator
insertions. Since $\hat{\gamma}_P (\mul)$ and $\as (\mul)$ depend on the
number of active quark flavors $\nf$, we have to solve
\Eq{eq:simpleRGCPv} separately for $\mub \le \mul \le \muh$ and $\muc
\le \mul \le \mub$. At the bottom quark threshold scale $\mub$
additional matching corrections, which will be discussed later in this
section, have to be taken into account. The Wilson coefficients
$\vec{C}_P (\mul)$ are given by    
\beq \label{eq:RGEsolution}
\vec{C}_P (\mul) = \hat{U}_P (\mul, \muh) \vec{C}_P (\muh) \, . 
\eeq
Keeping the first three terms in the expansions of $\hat{\gamma}_P
(\mul)$ and of the QCD $\beta$-function, one finds for the evolution
matrix $\hat{U}_P (\mul, \muh)$ in the NNLO approximation
\cite{Gorbahn:2004my, NNLOmatrixkernel}      
\beq \label{eq:asevolutionmatrixexpansion}
\hat{U}_P (\mul, \muh) = \hat{K}_P (\mul) \hspace{0.25mm}
\hat{U}^{(0)}_P (\mul, \muh) \hat{K}^{-1}_P (\muh) \, ,    
\eeq
where
\beq \label{eq:kmatrices}
\begin{split}
\hat{K}_P (\mul) & = \hat{1} + \f{\as (\mul)}{4 \pi} \hat{J}^{(1)}_P +
\left ( \f{\as (\mul)}{4 \pi} \right )^2 \hat{J}^{(2)}_P \, , \\[1mm]    
\hat{K}^{-1}_P (\muh) & = \hat{1} - \f{\as (\muh)}{4 \pi} 
\hat{J}^{(1)}_P - \left ( \f{\as (\muh)}{4 \pi} \right )^2 \left (
\hat{J}^{(2)}_P - \big ( \hat{J}^{(1)}_P \big )^2 \right ) \, ,      
\end{split}
\eeq
and 
\beq \label{eq:loevolutionmatrix}
\hat{U}^{(0)}_P (\mul, \muh) = \hat{V}_P \, {\rm diag} \left ( \f{\as 
(\muh)}{\as (\mul)} \right )^{a^i_P} \hat{V}_P^{-1} \, , 
\eeq  
denotes the LO evolution matrix, which is expressed through the
eigenvalues $a^i_P$ of $\hat{\gamma}^{(0) \hspace{0.2mm} T}_P$ and the
corresponding diagonalizing matrix $\hat{V}_P$:   
\beq \label{eq:magicnumbers}
\left ( \hat{V}^{-1}_P \hat{\gamma}^{(0) \hspace{0.2mm} T}_P \hat{V}_P
\right )_{i j} = 2 \betazero a^i_P \delta_{i j} \, .      
\eeq
In order to give the explicit expressions for the matrices
$\hat{J}^{(1)}_P$ and $\hat{J}^{(2)}_P$ we define   
\beq \label{eq:jandgmatrices}
\hat{J}^{(k)}_P = \hat{V}_P \hat{S}^{(k)}_P \hat{V}^{-1}_P \, ,
\hspace{1cm} \hat{G}^{(k)}_P = \hat{V}^{-1}_P  \hat{\gamma}^{(k)
\hspace{0.2mm} T}_P \hat{V}_P \, ,   
\eeq
for $k = 1, 2$. The entries of the matrix kernels $\hat{S}^{(1)}_P$
and $\hat{S}^{(2)}_P$ are given by 
\beq \label{eq:smatrices}
\begin{split}
\big ( \hat S^{(1)}_P \big)_{i j} & = \f{\betaone}{\betazero} a^i_P
\delta_{i j} - \f{\big ( \hat G^{(1)}_P \big )_{i j}}{2 \betazero 
\left ( 1 + a^i_P - a^j_P \right )} \, , \\[2mm]        
\big ( \hat S^{(2)}_P \big)_{i j} & = \f{\betatwo}{2 \betazero} a^i_P 
\delta_{i j} + \sum_k \f{1 + a^i_P - a^k_P}{2 + a^i_P - a^j_P} \left (
\big ( \hat S^{(1)}_P \big )_{i k} \big ( \hat S^{(1)}_P \big )_{k j}
- \f{\betaone}{\betazero} \big ( \hat S^{(1)}_P \big )_{i j} \delta_{j
k} \right ) \\ 
& - \f{\big ( \hat G^{(2)}_P \big )_{i j}}{2 \betazero \left ( 2 +
a^i_P - a^j_P \right )} \, ,   
\end{split} 
\eeq
where the first line recalls the familiar NLO result
\cite{Buras:1991jm}, while the second and third represent the
corresponding NNLO expression derived in \cite{Gorbahn:2004my,
NNLOmatrixkernel}.  
 
We will now collect the various expressions that enter the RG analysis
of the $Z$-penguin contribution. The LO evolution from $\muh$ down to
$\mub$ is described by 
\beq \label{eq:U0Pfive}
\hat{U}_P^{\nfive \, (0)} = \begin{pmatrix} \eta_b^{\f{6}{23}} & 0 &
0 \\[2mm] 0 & \eta_b^{-\f{12}{23}} & 0 \\[2mm] \f{12}{5} \left ( 
\eta_b^{\f{6}{23}} - \eta_b^{\f{1}{23}} \right ) & \f{6}{13} \left ( 
\eta_b^{-\f{12}{23}} - \eta_b^{\f{1}{23}} \right ) &
\eta_b^{\f{1}{23}} \end{pmatrix} \, .
\eeq
Adding an extra index for the number of flavors, the corresponding
matrices $\hat{J}_P^{(k)}$ read   
\beq \label{eq:JP1five}
\hat{J}_P^{\nfive \, (1)} = \begin{pmatrix} \f{5165}{3174} & 0 & 0
\\[2mm] 0 & -\f{2267}{1587} & 0 \\[2mm] -\f{15857}{1587} &
\f{15305}{3174} & -\f{14924}{1587} \end{pmatrix} \, ,  
\eeq
and 
\beq \label{eq:JP2five}
\hat{J}_P^{\nfive \, (2)} = \begin{pmatrix}
-7.35665 & 0 & 0 \\[2mm] 0 & -54.9107 & 0 \\[2mm] 
17.7699 & -1.7514 & 18.3025 \end{pmatrix} \, ,      
\eeq
where in the latter matrix we have employed the numerical value of
$\zetathree$.  

The LO evolution from $\mub$ down to $\muc$ is characterized by
\beq \label{eq:U0Pfour}
\hat{U}_P^{\nfour \, (0)} = \begin{pmatrix} \eta_{cb}^{\f{6}{25}} &
0 & 0 \\[2mm] 0 & \eta_{cb}^{-\f{12}{25}} & 0 \\[2mm] \f{12}{7} \left
( \eta_{cb}^{\f{6}{25}} - \eta_{cb}^{-\f{1}{25}} \right ) & \f{6}{11}
\left ( \eta_{cb}^{-\f{12}{25}} - \eta_{cb}^{-\f{1}{25}} \right ) &
\eta_{cb}^{-\f{1}{25}} \end{pmatrix} \, ,  
\eeq
where $\eta_{cb} = \as (\mub)/\as (\muc)$. The corresponding matrices
$\hat{J}_P^{(k)}$ take the following form    
\beq \label{eq:JP1four}
\hat{J}_P^{\nfour \, (1)} = \begin{pmatrix} \f{6719}{3750} & 0 & 0
\\[2mm] 0 & -\f{3569}{1875} & 0 \\[2mm] -\f{15931}{1875} &
\f{5427}{1250} & -\f{15212}{1875} \end{pmatrix} \, ,  
\eeq
and
\beq \label{eq:JP2four}
\hat{J}_P^{\nfour \, (2)} = \begin{pmatrix}
-10.2451 & 0 & 0 \\[2mm] 0 & -50.3422 & 0 \\[2mm] 8.0325 & -0.3657 &
4.91177 \end{pmatrix} \, .      
\eeq
In the last relation terms proportional to $\zetathree$ have not been
spelled out explicitly again. The unbracketed superscripts of the
above matrices indicates whether the object belongs to the effective
theory with five or four active quark flavors. 

\subsection{Threshold Corrections}
\label{subsec:ZMB}

Since $\vec{C}_P (\mul)$ contains the Wilson coefficients $C_\pm
(\mul)$ it receives a non-trivial $\ord (\as^2)$ matching correction 
when passing from the effective theory with five active quark flavors
to the one with only four. The explicit expression for the
discontinuities $\delta C^{(2)}_\pm (\mub)$ can be found in 
\Eq{eq:deltaCpmmub}. In the case of $C^P_\nu (\mul)$ one has to
distinguish two possible sources of discontinuities, corresponding to
the two terms in the first line of \Eqs{eq:TP}: $i)$ radiative
corrections to $Q_\nu$ alone and $ii)$ diagrams with double operator
insertions $(Q^q_\pm, Q_Z)$. In the first case only a single one-
and a single two-loop diagram similar to the ones shown on the right
of \Figsand{fig:ccmatrix}{fig:ccthreshold} can be drawn. These
contributions are canceled by counterterms and the matrix elements are
zero. Sample diagrams of the second type are displayed on the left of
\Fig{fig:oneloopmixing} and in \Fig{fig:penguinnlo}. Since none of
them contains a virtual bottom quark the discontinuities of the
corresponding matrix elements vanish identically. By matching the
effective theories at the bottom quark threshold scale $\mub$ we
obtain from \Eqs{eq:deltaCi} in the NDR scheme   
\beq \label{eq:Cpvthreshold}
\begin{split}
\delta C^{P (2)}_\nu (\mub) & = -\f{2}{3} \ln \f{\mub^2}{\mb^2} \Bigg
( \hspace{-1mm} \left ( \f{284704}{2645} + \f{694522}{20631} \eta_b
\right ) \eta^{\f{1}{23}}_b - \left ( \f{1033492}{7935} +
\f{8264}{529} \eta_b \right ) \eta^{\f{6}{23}}_b \\[1mm] 
& + \left ( \f{3058}{1587} + \f{18136}{6877} \eta_b \right )
\eta^{-\f{12}{23}}_b + \eta_b \left (   \eta^{\f{1}{23}}_b C^{P 
(1)}_\nu (\muh) \right. \\[1mm]  
& + \left. \f{48}{5} \left ( \eta^{\f{6}{23}}_b - \eta^{\f{1}{23}}_b
\right ) C^{(1)}_+ (\muh) + \f{24}{13} \left ( \eta^{-\f{12}{23}}_b -
\eta^{\f{1}{23}}_b \right ) C^{(1)}_- (\muh) \right ) \hspace{-1mm}
\Bigg ) \, ,       
\end{split}
\eeq
where $\mb = \mb (\mb)$ denotes the bottom quark $\MSbar$ mass.

\subsection{Matrix Elements}
\label{subsec:ZME}

For scales $\mul$ below $\muc$ the transition operator in the case of
the $Z$-penguin contribution is simply given by $-{\cal T}^P =
\Heff^P$ with    
\beq \label{eq:HeffP}
\Heff^P = \f{4 \GF}{\sqrt{2}} \f{\aem}{2 \pi \sws} \, \lambda_c \, 
\sum_{\ell = e, \mu, \tau}  C_P (\mul) ( \bar{s}_\sL \gamma_\mu
d_\sL ) ( \bar {\nu_\ell}_\sL \gamma^{\mu} {\nu_\ell}_\sL ) \, ,       
\eeq
which is part of the low-energy effective Hamiltonian that we have
encountered already in \Eq{eq:Heff}. The bilocal contribution to
${\cal T}^P$ in \Eq{eq:TP} has disappeared, because the charm quark
field is integrated out and the effect from the charm quark loop is
absorbed into $C_P (\muc)$ through the matching calculation at
$\muc$. There is still a bilocal contribution from the up quark loop,
but its contribution is suppressed by a factor of $\Lambda_{\rm
QCD}^2/\mc^2$ with respect to the one stemming from the charm
quark. The former is not included in our formalism. This 
power-suppressed contribution contains genuine long-distance effects
and has been computed in \cite{Isidori:2005xm}. It will be included in
our numerical analysis presented in \Sec{sec:numerics}.   

The local operator entering \Eq{eq:HeffP} has zero anomalous
dimension. Therefore $C_P (\mul)$ is $\mul$-independent for scales 
$\mul$ below $\muc$. Since we do not need to solve a RG equation for
$\mul \le \muc$, there is no need to express the result in terms of
the $\as$ of the effective theory containing three active quark
flavors and we can avoid to include the non-trivial matching
corrections of \Eq{eq:asmatching} at the charm quark threshold scale
$\muc$. In terms of the $\as$ of the effective theory with four active
quark flavors the product $C_P (\muc) \left \langle Q_\nu (\muc)
\right \rangle$ takes the following form 
\beq \label{eq:CP}
C_P (\muc) \left \langle Q_\nu \right \rangle^{(0)} = \f{x_c
(\muc)}{32} \f{4 \pi}{\as (\muc)} \left ( C_\nu^P (\muc) \left \langle
Q_\nu \right \rangle^{(0)} + 4 \sum_{i = \pm} C_i^P (\muc) \left
\langle Q_i^P (\muc) \right \rangle \right ) \, ,  
\eeq
where we have made use of the fact that the renormalized matrix
element of $Q_\nu$ does not receive radiative corrections to  all
orders in $\as$. Notice furthermore the factor of $4$ which is a 
result of our definition of $Q^P_\pm$ and $Q_\nu$ given in
\Eqsand{eq:QPpm}{eq:Qv}.  

In order to complete the evaluation of $C_P (\muc)$ the renormalized
matrix elements of the bilocal composite operators $Q^P_\pm$ are
needed. Including corrections up to NNLO, we write them in terms of
the tree-level matrix element $\langle Q_\nu \rangle^{(0)}$ in the
following way  
\beq \label{eq:QpmP}
\left \langle Q_\pm^P(\muc) \right \rangle = \left ( \f{\as (\muc)}{4
\pi} r^{P (1)}_\pm (\muc) + \left ( \f{\as (\muc)}{4 \pi} \right )^2
r^{P (2)}_\pm (\muc) \right ) \langle Q_\nu \rangle^{(0)} \, ,
\eeq
where $r^{P (1)}_\pm (\muc)$ and $r^{P (2)}_\pm (\muc)$ codify the one-
and two-loop corrections, respectively. Like in the case
$\gamma^P_{\pm, \nu} (\mul)$ it is useful to decompose the
coefficients further into        
\beq \label{eq:rPpmvdecomposition}
r^{P (k)}_\pm (\muc) = -\frac{1}{2} r^{A (k)}_\pm (\muc) - \left (
\f{1}{2} - \f{2}{3} \sws \hspace{-0.5mm} \right ) r^{V (k)}_\pm (\muc)
\, ,    
\eeq
where the superscript $A$ and $V$ marks the corrections arising from
diagrams with a double operator insertion $(Q^q_\pm, Q^q_A)$ and
$(Q^q_\pm, Q^q_V)$.

{%
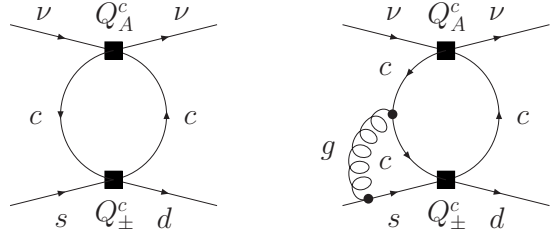
\begin{figure}[!t]
\begin{center}
$\begin{array}{c@{\hspace{15mm}}c}
\scalebox{0.65}{
\begin{picture}(120,137) (45,-30)
\SetWidth{0.5}
\SetColor{Black}
\ArrowLine(45,92)(105,77)
\ArrowLine(105,77)(165,92)
\ArrowLine(105,2)(165,-13)
\ArrowArc(97.5,39.5)(38.24,-78.69,78.69)
\Text(65,99)[]{\Large{\Black{$\nu$}}}
\Text(145,99)[]{\Large{\Black{$\nu$}}}
\Text(150,38)[]{\Large{\Black{$c$}}}
\Text(75,-22)[]{\Large{\Black{$s$}}}
\Text(135,-20)[]{\Large{\Black{$d$}}}
\GBox(100,-3)(110,7){0.0}
\GBox(100,72)(110,82){0.0}
\Text(106,-19)[]{\Large{\Black{$Q^c_\pm$}}}
\Text(106,97)[]{\Large{\Black{$Q_A^c$}}}
\ArrowLine(45,-13)(105,2)
\ArrowArc(112.5,39.5)(38.24,101.31,258.69)
\Text(60,38)[]{\Large{\Black{$c$}}}
\end{picture}
} & 
\scalebox{0.65}{
\begin{picture}(120,137) (45,-30)
\SetWidth{0.5}
\SetColor{Black}
\ArrowLine(45,92)(105,77)
\ArrowLine(105,77)(165,92)
\ArrowLine(45,-13)(105,2)
\ArrowLine(105,2)(165,-13)
\ArrowArc(97.5,39.5)(38.24,-78.69,78.69)
\Text(75,99)[]{\Large{\Black{$\nu$}}}
\Text(135,99)[]{\Large{\Black{$\nu$}}}
\GlueArc(87.79,9.56)(33.42,114.38,213.73){6}{5.14}
\Vertex(60,-9){2.83}
\Vertex(74,40){2.83}
\Text(150,38)[]{\Large{\Black{$c$}}}
\Text(75,-22)[]{\Large{\Black{$s$}}}
\Text(135,-20)[]{\Large{\Black{$d$}}}
\GBox(100,-3)(110,7){0.0}
\GBox(100,72)(110,82){0.0}
\Text(106,-19)[]{\Large{\Black{$Q_\pm^{c}$}}}
\Text(106,97)[]{\Large{\Black{$Q_A^c$}}}
\Text(37,19)[]{\Large{\Black{$g$}}}
\ArrowArc(114.17,37.83)(40.23,103.17,176.91)
\ArrowArc(113.93,40.93)(39.94,-178.67,-102.92)
\Text(70,67)[]{\Large{\Black{$c$}}}
\Text(70,12)[]{\Large{\Black{$c$}}}
\end{picture}
}
\end{array}$
\vspace{2mm}
\caption{Typical diagrams contributing to the matrix elements of
$Q^P_\pm$ at $\ord (\as)$ (left) and $\ord (\as^2)$ (right).}   
\label{fig:Zmatrix}
\end{center}
\end{figure}
}%

Regulating spurious IR divergences dimensionally we obtain after
setting $\ca = 3$ and $\cf = 4/3$ in the NDR scheme
\begin{align} \label{eq:rpmAV}
r^{A (1)}_\pm (\muc) & = -2 \left ( 1 \pm 3 \right ) \left ( 1 - \ln
\f{\muc^2}{\mc^2} \right ) \, , & r^{V (1)}_\pm (\muc) & = 0 \, , \non
\non \\[-2mm] \\[-2mm]
r^{A (2)}_\pm (\muc) & = -2 \left (2 \pm 9 \right ) - 8 \left ( 4 \pm 
1 \right ) \ln \f{\muc^2}{\mc^2} \pm 24 \ln^2 \f{\muc^2}{\mc^2} \, , &
r^{V (2)}_\pm (\muc) & = 0 \, , \non  
\end{align}
where $\mc = \mc (\muc)$ denotes the charm quark $\MSbar$ mass. The
first line of the above equations agrees with the known NLO results
\cite{Buchalla:1993wq} if one takes the normalizations of the
operators and of $r_{\pm, \nu}^{A (k)}$ in \Eq{eq:rPpmvdecomposition}
into account. See the discussion after \Eq{eq:gammaAVpmv}. The second
line represents the new NNLO expressions. We emphasize that also at
the NNLO level only the part of the double operator insertion
$(Q^q_\pm, Q_Z)$ proportional to $Q^c_\pm Q^c_A - Q^u_\pm Q^u_A$ gives
a non-zero contribution to the matrix element of $Q^P_\pm$. To understand
this feature we again distinguish Feynman graphs with coupling of
$Q_Z$ to up-type quarks as in \Fig{fig:Zmatrix} from those with
down-type coupling of $Q_Z$ as on the left of
\Fig{fig:wavematrix}. The former diagrams do not contribute to the
matrix elements at all. In addition the latter diagrams  do not arise
at NLO while at NNLO they give a finite contribution. These terms are
again cancelled by corrections involving a flavor off-diagonal wave
function renormalization as shown on the right of
\Fig{fig:wavematrix}. Finally we remark that diagrams with coupling to
down-type quarks and an insertion of $Q^q_A$ differ only by a sign
from those with an insertion of $Q^q_V$, because $Q_\pm^q$ contains
only left-handed down and strange quark fields. We recall that in the
full theory electromagnetic gauge invariance requires that terms
proportional to $\sws$ in the $\stodnunu$ amplitude add to zero in the
limit of vanishing external momenta. The fact that the vector part of
$Q_Z$ does not contribute to $\PcX$ is thus nothing else but the
realization of this Ward identity in the effective theory.

{%
\begin{figure}[!t]
\begin{center}
$\begin{array}{c@{\hspace{15mm}}c}
\scalebox{0.65}{
\begin{picture}(135,148) (10,-18)
\SetWidth{0.5}
\SetColor{Black}
\ArrowLine(45,103)(105,88)
\ArrowLine(105,88)(165,103)
\Text(65,112)[]{\Large{\Black{$\nu$}}}
\Text(145,112)[]{\Large{\Black{$\nu$}}}
\GBox(100,83)(110,93){0.0}
\Text(106,108)[]{\Large{\Black{$Q_A^d, Q_V^d$}}}
\ArrowLine(105,88)(165,-2)
\ArrowLine(45,-2)(75,43)
\ArrowLine(75,43)(105,88)
\ArrowArc(90,28)(21.02,177,537)
\GlueArc(115.61,37.86)(40.1,-129.68,-28.06){6}{6.86}
\Vertex(90,7){2.83}
\Vertex(151,19){2.83}
\GBox(67,34)(77,44){0.0}
\Text(123,-22)[]{\Large{\Black{$g$}}}
\Text(124,26)[]{\Large{\Black{$c$}}}
\Text(54,52)[]{\Large{\Black{$Q_\pm^{c}$}}}
\Text(45,23)[]{\Large{\Black{$s$}}}
\Text(73,70)[]{\Large{\Black{$d$}}}
\Text(150,48)[]{\Large{\Black{$d$}}}
\end{picture}
}
&
\scalebox{0.65}{
\begin{picture}(195,109) (60,-40)
\SetWidth{0.5}
\SetColor{Black}
\ArrowLine(195,4)(225,4)
\Vertex(195,4){2.83}
\ArrowLine(60,4)(105,4)
\CBox(109.66,-0.66)(100.34,8.66){Black}{Black}
\GlueArc(137.95,1.37)(56.26,127.11,3.96){-6}{12.43}
\ArrowLine(105,4)(195,4)
\Vertex(105,45){2.83}
\ArrowArc(105,24)(21.02,3,363)
\Text(83,-10)[c]{\Large{\Black{$s$}}}
\Text(150,-8)[c]{\Large{\Black{$d$}}}
\Text(73,25)[c]{\Large{\Black{$c$}}}
\Text(211,-8)[c]{\Large{\Black{$d$}}}
\Text(170,70)[c]{\Large{\Black{$g$}}}
\Text(108,-18)[c]{\Large{\Black{$Q^c_\pm$}}}
\end{picture}
}
\end{array}$
\vspace{2mm}
\caption{Typical diagrams contributing to the $\ord (\as^2)$
matrix element of $Q_\pm^P$ (left) and the $\ord (\as)$ flavor
off-diagonal wave function renormalization (right).} 
\label{fig:wavematrix}
\end{center}
\end{figure}
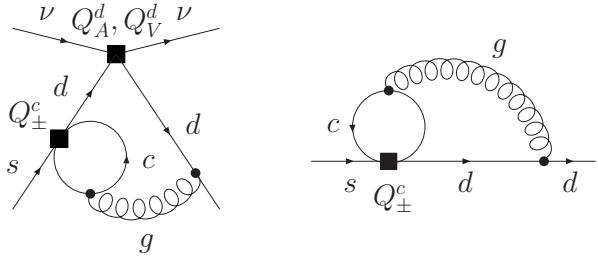
}%

\subsection{Final Result}
\label{subsec:Zfinal}

Having calculated all the necessary ingredients of the RG analysis at
the NNLO level, we are now in a position to present the final result
for the $Z$-penguin contribution to $X^\ell (x_c)$ that enters the
definition of $\PcX$ in \Eqsand{eq:defPcX}{eq:Xlxc}. 

In the NNLO approximation the Wilson coefficient $\vec{C}_P (\muc)$ of
\Eq{eq:CPgP} has the following perturbative expansion 
\beq \label{eq:CPPE}
\vec{C}_P (\muc) = \vec{C}_P^{(0)} (\muc) + \f{\as (\muc)}{4 \pi}
\vec{C}_P^{(1)} (\muc) + \left ( \f{\as (\muc)}{4 \pi} \right )^2
\vec{C}_P^{(2)} (\muc) \, , 
\eeq
where 
\begin{align} \label{eq:CPvfinal}
\vec{C}_P^{(0)} (\muc) & = \hat{U}_P^{\nfour \, (0)} \hat{U}_P^{\nfive
\, (0)} \vec{C}_P^{(0)} \Qmuh \, , \non \\[2mm]   
\vec{C}_P^{(1)} (\muc) & = \hat{J}_P^{\nfour \, (1)} \hat{U}_P^{
\nfour \, (0)} \hat{U}_P^{\nfive \, (0)} \vec{C}_P^{(0)} \Qmuh \non \\
& + \eta_{cb} \hspace{0.4mm} \hat{U}_P^{\nfour \, (0)} \left (
\hat{J}_P^{\nfive \, (1)} - \hat{J}_P^{\nfour \, (1)} \right)
\hat{U}_P^{\nfive \, (0)} \vec{C}_P^{(0)} \Qmuh \non \\ 
& + \eta_b \hspace{0.4mm} \eta_{cb} \hspace{0.4mm} \hat{U}_P^{\nfour
\, (0)} \hat{U}_P^{\nfive \, (0)} \left ( \vec{C}_P^{(1)} \Qmuh -
\hat{J}_P^{\nfive \, (1)} \vec{C}_P^{(0)} \Qmuh \right ) \, , \non
\\[2mm]   
\vec{C}_P^{(2)} (\muc) & = \hat{J}_P^{\nfour \, (2)} \hat{U}_P^{\nfour
\, (0)} \hat{U}_P^{\nfive \, (0)} \vec{C}_P^{(0)} \Qmuh \\ 
& + \eta_{cb} \hspace{0.4mm} \hat{J}_P^{\nfour \, (1)}
\hat{U}_P^{\nfour \, (0)} \left ( \hat{J}_P^{\nfive \, (1)} -
\hat{J}_P^{\nfour \, (1)} \right) \hat{U}_P^{\nfive \, (0)}
\vec{C}_P^{(0)} \Qmuh \non \\  
& + \eta_b \hspace{0.4mm} \eta_{cb} \hspace{0.4mm} \hat{J}_P^{\nfour 
\, (1)} \hat{U}_P^{\nfour \, (0)} \hat{U}_P^{\nfive \, (0)} \left (
\vec{C}_P^{(1)} \Qmuh -  \hat{J}_P^{\nfive \, (1)} \vec{C}_P^{(0)}
\Qmuh \right ) \non \\  
& + \eta_{cb}^2 \hspace{0.4mm} \hat{U}_P^{\nfour \, (0)} \left (
\hat{J}_P^{\nfive \, (2)} - \hat{J}_P^{\nfour \, (2)} -
\hat{J}_P^{\nfour \, (1)} \left ( \hat{J}_P^{\nfive \, (1)} -
\hat{J}_P^{\nfour \, (1)} \right ) - \delta \vec{C}_P^{(2)} \Qmub
\right ) \hat{U}_P^{\nfive \, (0)} \vec{C}_P^{(0)} \Qmuh \non  \\ 
& + \eta_b \hspace{0.4mm} \eta_{cb}^2 \hspace{0.4mm} \hat{U}_P^{\nfour
\, (0)} \left ( \hat{J}_P^{\nfive \, (1)} - \hat{J}_P^{\nfour \, (1)}
\right ) \hat{U}_P^{\nfive \, (0)} \left (  \vec{C}_P^{(1)}  \Qmuh -
\hat{J}_P^{\nfive \, (1)} \vec{C}_P^{(0)} \Qmuh \right ) \non \\ 
& + \eta_b^2 \hspace{0.4mm} \eta_{cb}^2 \hspace{0.4mm}
\hat{U}_P^{\nfour \, (0)} \hat{U}_P^{\nfive \, (0)} \left (
\vec{C}_P^{(2)} \Qmuh - \hat{J}_P^{\nfive \, (1)} \vec{C}_P^{(1)}
\Qmuh - \left ( \hat{J}_P^{\nfive \, (2)} - \big (  \hat{J}_P^{\nfive
\, (1)} \big )^2 \right ) \vec{C}_P^{(0)} \Qmuh \right ) \, . \non 
\end{align}
The explicit expressions for $\vec{C}_P^{(k)} (\muh)$, $\hat{U}^{\nf
\, (0)}_P$, $\hat{J}^{\nf \, (k)}_P$, and $\delta \vec{C}_P^{(2)}
(\mub)$ have been given in \Eqsto{eq:U0Pfive}{eq:Cpvthreshold}. We
also recall that $\eta_b = \as (\muh)/\as (\mub)$ and $\eta_{cb} =
\as (\mub)/\as (\muc)$.

It is useful to express the running charm quark $\MSbar$ mass $\mc
(\mul)$ entering the definition of the dimension-eight operator
$Q_\nu$ of \Eq{eq:Qv} in terms of the input parameter $\mc (\mc)$. At 
the scale $\muc$ the required NNLO relation reads
\beq \label{eq:xc}
x_c (\muc) = \kappa_c \left ( 1 + \f{\as (\muc)}{4 \pi}
\xi_c^{(1)} + \left ( \f{\as (\muc)}{4 \pi} \right )^2 \xi_c^{(2)}
\right ) x_c (\mc) \, .     
\eeq 
Here $\kappa_c = \eta_c^{24/25}$ with $\eta_c = \as (\muc)/\as(\mc)$
and  
\beq \label{eq:xccoeff}
\begin{split}
\xi_c^{(1)} & = \f{15212}{1875} \left ( 1 - \eta_c^{-1} \right ) \, ,
\\[2mm] 
\xi_c^{(2)} & = \f{966966391}{10546875} - \f{231404944}{3515625}
\eta_c^{-1} - \f{272751559}{10546875} \eta_c^{-2}  - \f{128}{5} \left 
( 1 - \eta_c^{-2} \right ) \zetathree \, .
\end{split}
\eeq

Including corrections up to third order in perturbation theory one
finds from \Eq{eq:CP}: 
\beq \label{eq:CPPT}
C_P (\muc) = \kappa_c \, \f{x_c (\mc)}{32} \left ( \f{4 \pi}{\as
(\muc)} C_P^{(0)} (\muc) + C_P^{(1)} (\muc) + \f{\as (\muc)}{4 \pi}
C_P^{(2)} (\muc) \right ) \, ,     
\eeq
where 
\begin{align} \label{eq:CPcoeff}
C_P^{(0)} (\muc) & = C_\nu^{P (0)} (\muc) \, , \non \\[5mm]
C_P^{(1)} (\muc) & = C_\nu^{P (1)} (\muc) + 4 \sum_{i = \pm} C_i^{P (0)}
(\muc) \, \rho_i^{P (1)} (\muc) + \xi_c^{(1)} \, C_\nu^{P (0)}
(\muc) \, , \non \\[1mm]     
C_P^{(2)} (\muc) & = C_\nu^{P (2)} (\muc) + 4 \sum_{i = \pm} \left (
C_i^{P (1)} (\muc) \, \rho_i^{P (1)} (\muc)  + C_i^{P (0)} (\muc)
\, \rho_i^{P (2)} (\muc) \right ) \non \\   
& + \xi_c^{(1)} \bigg ( C_\nu^{P (1)} (\muc)+ 4 \sum_{i = \pm} C_i^{P
(0)} (\muc) \, \rho_i^{P (1)} (\muc) \bigg ) + \xi_c^{(2)} \,
C_\nu^{P (0)} (\muc) \, .     
\end{align}
The factors of $4$ are again a result of the definition of $Q^P_\pm$
and $Q_\nu$ in \Eqsand{eq:QPpm}{eq:Qv}.    

The coefficients $\rho_\pm^{P (k)} (\muc)$ are obtained from $r_\pm^{P
(k)} (\muc)$ by expanding the charm quark $\MSbar$ mass $\mc (\muc)$
entering \Eqs{eq:rpmAV} in $\as$ around $\mc (\mc)$. Explicitly one
finds    
\bea \label{eq:rhoPs} 
\begin{split}
\rho_\pm^{P (1)} (\muc) & = r_\pm^{P (1)} (\muc, \mc) + ( 1 \pm 3 )
\ln \kappa_c \, , \\[2mm]         
\rho_\pm^{P (2)} (\muc) & = r_\pm^{P (2)} (\muc, \mc) - 4
\hspace{0.4mm} ( 4 \pm 1 ) \ln \kappa_c \mp 12 \ln^2 \kappa_c \pm 24
\ln \kappa_c \ln \f{\muc^2}{\mc^2} + ( 1 \pm 3 ) \, \xi_c^{(1)} \, .       
\end{split}
\eea
Here the additional argument in $r_\pm^{P (k)} (\muc, \mc)$ indicates
that the expansion coefficients of \Eqs{eq:rPpmvdecomposition} have to
be evaluated at $\mc (\mc)$ and not at $\mc (\muc)$. Note that the
second term in the first line of the above equations is absent in the
analytic NLO formulas of $\PcX$ presented in \cite{Buchalla:1993wq,
Buchalla:1998ba}. The relevance of this $\muc$-dependent term will be
discussed in \Sec{sec:numerics}.   

\section{Electroweak Box Contributions}   
\label{sec:B}

\subsection{Effective Hamiltonian}
\label{subsec:HB}

Apart from the presence of a non-trivial matching correction at the
bottom quark threshold scale $\mub$ the NNLO correction in the
electroweak box sector does not involve new conceptual features
compared to the LO and NLO. This simplifies the following discussion
notably. 
  
For scales $\mul$ in the range $\muc \le \mul \le \muh$ the
electroweak box contribution involves $\Heff^W$ defined in \Eq{eq:HW}
as well as the effective Hamiltonian given by  
\beq \label{eq:HBv}
\Heff^{\hspace{-0.1mm} B \nu} = \f{\GF}{\sqrt{2}} \lambda_c \f{2
\hspace{0.4mm} \pi \aem}{\MW^2 \sws} C^B_\nu (\mul) Q_\nu \, .
\eeq

In terms of these building blocks the part of the transition operator
${\cal T}$ stemming from electroweak boxes can be written as    
\beq \label{eq:TB}
\begin{split}
-{\cal T}^B & = \Heff^{B \hspace{0.25mm} \nu} - i \int \! d^4 x
\hspace{1mm} T \big ( \Heff^W (x) \Heff^W (0) \big ) \\[1mm]    
& = \f{\GF}{\sqrt{2}} \lambda_c \f{2 \hspace{0.2mm} \pi \aem}{\MW^2
\sws} \big ( C^B_\nu (\mul) Q_\nu + 4 \hspace{0.4mm} Q^B \big ) \, .  
\end{split}
\eeq
Notice that in passing from the first to the second line we used the
fact that the Wilson coefficients of the effective charged-current
couplings $Q_3^q$ and $Q_4^q$ equal one at all scales. In particular
they do not receive matching corrections at any scale. The last term
in \Eqs{eq:TB} is the bilocal composite operators $Q^B$ which has been
introduced in \Eq{eq:QB} already. The factor of $4$ originates once
again from the use of chiral fermion fields in $Q^B$ and $Q_\nu$.

\subsection{Initial Conditions}
\label{subsec:ICB}

{%
\begin{figure}[!t]
\begin{center}
$\begin{array}{c@{\hspace{15mm}}c}
\scalebox{0.65}{
\begin{picture}(198,152) (12,-26)
\SetWidth{0.5}
\SetColor{Black}
\ArrowLine(75,2)(30,-13)
\ArrowLine(210,107)(165,92)
\ArrowLine(165,2)(210,-13)
\Vertex(75,2){2.83}
\Vertex(75,92){2.83}
\Vertex(165,92){2.83}
\Vertex(165,2){2.83}
\Text(120,112)[]{\Large{\Black{$W$}}}
\Text(186,113)[]{\Large{\Black{$\nu$}}}
\Text(180,47)[]{\Large{\Black{$\ell$}}}
\Text(120,-18)[]{\Large{\Black{$W$}}}
\Text(55,-26)[b]{\Large{\Black{$d$}}}
\Text(185,-23)[b]{\Large{\Black{$\nu$}}}
\Photon(75,92)(165,92){6}{8}
\Photon(75,2)(165,2){6}{8}
\ArrowLine(165,92)(165,2)
\ArrowLine(75,92)(75,47)
\ArrowLine(75,47)(75,2)
\Vertex(75,47){2.83}
\ArrowLine(30,107)(52,100)
\ArrowLine(52,100)(75,92)
\GlueArc(71.16,76.83)(30.07,129.59,277.33){6}{6}
\Vertex(52,100){2.83}
\Text(63,25)[]{\Large{\Black{$c$}}}
\Text(63,70)[]{\Large{\Black{$c$}}}
\Text(27,56)[]{\Large{\Black{$g$}}}
\Text(65,112)[]{\Large{\Black{$s$}}}
\Text(42,118)[]{\Large{\Black{$s$}}}
\end{picture}
} & 
\scalebox{0.65}{
\begin{picture}(193,157) (17,-26)
\SetWidth{0.5}
\SetColor{Black}
\ArrowLine(210,112)(165,97)
\ArrowLine(165,7)(210,-8)
\Vertex(75,7){2.83}
\Vertex(75,97){2.83}
\Vertex(165,97){2.83}
\Vertex(165,7){2.83}
\Text(120,117)[]{\Large{\Black{$W$}}}
\Text(186,118)[]{\Large{\Black{$\nu$}}}
\Text(180,52)[]{\Large{\Black{$\ell$}}}
\Text(120,-13)[]{\Large{\Black{$W$}}}
\Text(185,-18)[b]{\Large{\Black{$\nu$}}}
\Photon(75,97)(165,97){6}{8}
\Photon(75,7)(165,7){6}{8}
\ArrowLine(165,97)(165,7)
\ArrowLine(30,112)(52,105)
\ArrowLine(52,105)(75,97)
\Vertex(52,105){2.83}
\Text(65,117)[]{\Large{\Black{$s$}}}
\Text(42,123)[]{\Large{\Black{$s$}}}
\ArrowLine(75,97)(75,7)
\ArrowLine(75,7)(52,0)
\ArrowLine(52,0)(30,-8)
\Gluon(52,0)(52,105){6}{9.43}
\Vertex(52,0){2.83}
\Text(88,52)[]{\Large{\Black{$c$}}}
\Text(32,52)[]{\Large{\Black{$g$}}}
\Text(65,-12)[]{\Large{\Black{$d$}}}
\Text(42,-18)[]{\Large{\Black{$d$}}}
\end{picture}
}
\end{array}$
\vspace{2mm}
\caption{Examples of electroweak box diagrams that contribute to the
$\ord (\as^2)$ correction to the initial value of the Wilson
coefficient of the leading dimension-eight operator $Q_\nu$.} 
\label{fig:boxmatching}
\end{center}
\end{figure}
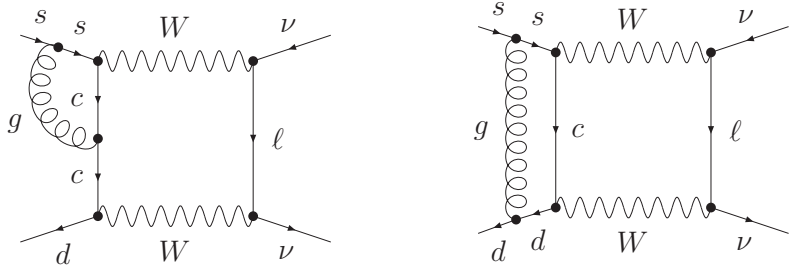
}%

The initial conditions of $C^B_\nu (\mul)$ are as usual found by
matching perturbative amplitudes in the full and the effective
theory. Examples of two-loop electroweak box diagrams can be seen in
\Fig{fig:boxmatching}. Regulating spurious IR divergences once
dimensionally and once with small quark masses we found identical
results for the initial conditions. In the NDR scheme supplemented by
the definition of the evanescent operator given in
\Eq{eq:boxevanescent}, the non-zero matching conditions read   
\beq \label{eq:CBinitial}
\begin{split}
C^{B (1)}_\nu (\muh) & = -4 \left ( 9  + 4 \ln \f{\muh^2}{\MW^2}
\right ) \, , \\ 
C^{B (2)}_\nu (\muh) & = -8 \hspace{0.4mm} \cf \left ( 20 + 2 \pi^2 +
25 \ln \f{\muh^2}{\MW^2} + 6 \ln^2 \f{\muh^2}{\MW^2} \right ) \, , 
\end{split}
\eeq
where the first line agrees with the literature \cite{Buchalla:1993wq,
Buchalla:1998ba}, while the second one is the new NNLO
expression.\footnote{We remark that the logarithmic term in the
second line of \Eqs{eq:CBinitial} differs from the expression one
would expect from the results published in \cite{Buchalla:1993wq}. The
disagreement is due to a subtlety in regulating spurious IR divergences
\cite{Buchalla:1998ba, Misiak:1999yg}. This mistake has been corrected
in \cite{Buchalla:1998ba}.}  

\subsection{Anomalous Dimensions}
\label{subsec:ADMB}

In the case of the Wilson coefficient $C^B_\nu (\mul)$ the RG equation
takes the following form 
\beq \label{eq:RGCBv}
\mul \f{d}{d \mul} C^B_\nu (\mul) = \gamma_\nu (\mul) C^B_\nu (\mul) +
4 \hspace{0.4mm} \gamma^B_\nu (\mul) \, ,   
\eeq
with $\gamma_\nu (\mul)$ given in \Eq{eq:gammav}. The anomalous
dimension tensor $\gamma^B_\nu (\mul)$ encodes the mixing of the
bilocal composite structures $Q^B$ into $Q_\nu$. Sample diagrams are
shown in \Fig{fig:boxadm}. The UV pole parts of these Feynman graphs
are evaluated using the method that has been described
earlier. The factor of $4$ in the above equation is a direct result of
the factor of $4$ in \Eq{eq:TB}. 

In the NDR scheme supplemented by the definition of the evanescent
operator given in \Eq{eq:boxevanescent} the expansion 
coefficients of $\gamma^B_\nu (\mul)$ read
\beq \label{eq:gammaBv}
\begin{split}
\gamma^{B (0)}_{\nu} & = -8 \, \\[4mm]
\gamma^{B (1)}_{\nu} & = 8 \hspace{0.4mm} \cf \, , \\[1mm]
\gamma^{B (2)}_{\nu} & = 2 \hspace{0.4mm} \cf \left ( \f{69}{\ca} -
\f{458}{3} \ca - \left ( \f{48}{\ca} - 96 \hspace{0.4mm} \ca \right )
\zetathree + \f{38}{3} \nf \right ) \, ,  
\end{split}
\eeq
where the second line differs from the findings for
$\gamma_{12}^{(1)}$ of the original NLO calculation
\cite{Buchalla:1993wq} even after taking into account a factor of $4$
stemming from the different normalization of $Q^B$ and $Q_\nu$ used
here and therein. It however agrees with the results of
\cite{Buchalla:1998ba} where the error made in \cite{Buchalla:1993wq} 
has been corrected. The third line represents our new NNLO result.    
  
{%
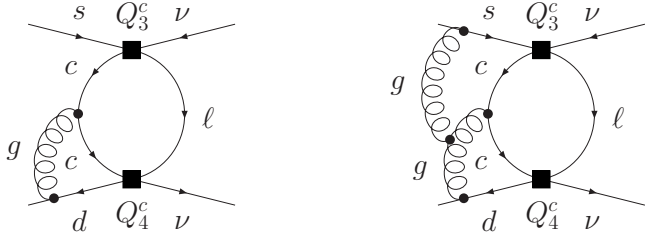
\begin{figure}[!t]
\begin{center}
$\begin{array}{c@{\hspace{15mm}}c}
\scalebox{0.65}{
\begin{picture}(165,168) (0,0)
\SetWidth{0.5}
\SetColor{Black}
\ArrowArc(113.93,71.93)(39.94,-178.67,-102.92)
\ArrowArc(114.17,68.83)(40.23,103.17,176.91)
\GlueArc(87.79,40.56)(33.42,114.38,213.73){6}{5.14}
\Vertex(74,71){2.83}
\ArrowLine(45,123)(105,108)
\ArrowLine(165,123)(105,108)
\ArrowLine(105,33)(45,18)
\ArrowLine(105,33)(165,18)
\ArrowArcn(97.5,70.5)(38.24,78.69,-78.69)
\Text(75,130)[]{\Large{\Black{$s$}}}
\Text(135,129)[]{\Large{\Black{$\nu$}}}
\Vertex(60,22){2.83}
\Text(150,69)[]{\Large{\Black{$\ell$}}}
\Text(75,9)[]{\Large{\Black{$d$}}}
\Text(135,8)[]{\Large{\Black{$\nu$}}}
\GBox(100,28)(110,38){0.0}
\GBox(100,103)(110,113){0.0}
\Text(106,12)[]{\Large{\Black{$Q_4^{c}$}}}
\Text(106,128)[]{\Large{\Black{$Q_3^{c}$}}}
\Text(37,50)[]{\Large{\Black{$g$}}}
\Text(70,98)[]{\Large{\Black{$c$}}}
\Text(70,43)[]{\Large{\Black{$c$}}}
\end{picture}
} & 
\scalebox{0.65}{
\begin{picture}(165,168) (0,0)
\SetWidth{0.5}
\SetColor{Black}
\ArrowArc(113.93,71.93)(39.94,-178.67,-102.92)
\ArrowArc(114.17,68.83)(40.23,103.17,176.91)
\GlueArc(87.79,40.56)(33.42,114.38,213.73){6}{5.14}
\Vertex(74,71){2.83}
\ArrowLine(45,123)(105,108)
\ArrowLine(165,123)(105,108)
\ArrowLine(105,33)(45,18)
\ArrowLine(105,33)(165,18)
\ArrowArcn(97.5,70.5)(38.24,78.69,-78.69)
\Text(75,130)[]{\Large{\Black{$s$}}}
\Text(135,129)[]{\Large{\Black{$\nu$}}}
\Vertex(60,22){2.83}
\Text(150,69)[]{\Large{\Black{$\ell$}}}
\Text(75,9)[]{\Large{\Black{$d$}}}
\Text(135,8)[]{\Large{\Black{$\nu$}}}
\GBox(100,28)(110,38){0.0}
\GBox(100,103)(110,113){0.0}
\Text(106,12)[]{\Large{\Black{$Q_4^{c}$}}}
\Text(106,128)[]{\Large{\Black{$Q_3^{c}$}}}
\Text(70,98)[]{\Large{\Black{$c$}}}
\GlueArc(83.69,83.98)(42.28,124.08,221.44){6}{6.86}
\Vertex(52,56){2.83}
\Vertex(60,119){2.83}
\Text(35,38)[]{\Large{\Black{$g$}}}
\Text(22,88)[]{\Large{\Black{$g$}}}
\Text(70,43)[]{\Large{\Black{$c$}}}
\end{picture}
}
\end{array}$
\vspace{2mm}
\caption{Typical examples of diagrams that describe the mixing of the 
double insertion $(Q^q_3, Q^q_4)$ into the leading dimension-eight
operators $Q_\nu$ at $\ord (\as^2)$ (left) and $\ord (\as^3)$
(right).}    
\label{fig:boxadm}
\end{center}
\end{figure}
}%

\subsection{RG Evolution}
\label{subsec:RGEB}

Obviously the RG evolution of the electroweak box contribution may be
recast into the following homogeneous differential equation 
\beq \label{eq:simpleRGCBv} 
\mul \f{d}{d \mul} \vec{C}_B (\mul) = \hat \gamma^T_B (\mul) \vec{C}_B
(\mul) \, ,      
\eeq
where
\beq \label{eq:CBgB} 
\vec{C}_B (\mul) = \begin{pmatrix} 4 \\ C^B_\nu (\mul) \end{pmatrix}
\, , \hspace{1cm} \hat \gamma_B (\mul) = 
\begin{pmatrix} 0 & \gamma^B_\nu (\mul) \\  
0 & \gamma_\nu (\mul) 
\end{pmatrix} \, . 
\eeq

As both $\hat{\gamma}_B (\mul)$ and $\as (\mul)$ depend on the number
of active quark flavors $\nf$, we have to solve \Eq{eq:simpleRGCBv}
separately for $\mub \le \mul \le \muh$ and $\muc \le \mul \le
\mub$. At the bottom quark threshold scale $\mub$ additional matching
corrections, which will be discussed in the next subsection, have to
be taken into account.   

In the following we will detail the different expressions that enter
the RG analysis of the electroweak box contribution. Our notation
derives from \Eqsto{eq:asevolutionmatrixexpansion}{eq:smatrices}
thereby. The LO evolution from $\muh$ down to $\mub$ is related to 
\beq \label{eq:U0Bfive}
\hat{U}_B^{\nfive \, (0)} = \begin{pmatrix} 1 & 0 \\[2mm] 12 \left (
1 - \eta_b^{\f{1}{23}} \right ) & \eta_b^{\f{1}{23}} \end{pmatrix} \,
. 
\eeq
The corresponding matrices $\hat{J}_B^{(k)}$ are given by 
\beq \label{eq:JB1five}
\hat{J}_B^{\nfive \, (1)} = \begin{pmatrix} 0 & 0 \\[2mm]
\f{2402}{1587} & -\f{14924}{1587} \end{pmatrix} \, ,  
\eeq
and
\beq \label{eq:JB2five}
\hat{J}_B^{\nfive \, (2)} = \begin{pmatrix} 0 & 0 \\[2mm] 
\f{1296371522}{39457581} - \f{34624}{1081} \zetathree &
-\f{177621017}{7555707} + \f{800}{23} \zetathree \end{pmatrix}
\, .      
\eeq

The LO evolution from $\mub$ down to $\muc$ is induced by 
\beq \label{eq:U0Bfour}
\hat{U}_B^{\nfour \, (0)} = \begin{pmatrix} 1 & 0 \\[2mm] -12 \left (
1 - \eta_{cb}^{-\f{1}{25}} \right ) & \eta_{cb}^{-\f{1}{25}}
\end{pmatrix} \, .    
\eeq
The corresponding matrices $\hat{J}_B^{(k)}$ read  
\beq \label{eq:JB1four}
\hat{J}_B^{\nfour \, (1)} = \begin{pmatrix} 0 & 0 \\[2mm]
\f{581}{1875} & -\f{15212}{1875} \end{pmatrix} \, ,  
\eeq
and
\beq \label{eq:JB2four}
\hat{J}_B^{\nfour \, (2)} = \begin{pmatrix} 0 & 0 \\[2mm] 
\f{684990354}{19140625} - \f{6976}{245} \zetathree &
-\f{272751559}{10546875} + \f{128}{5} \zetathree \end{pmatrix}
\, .      
\eeq

\subsection{Threshold Corrections}
\label{subsec:TCB}

In analogy to the case of $C^P_\nu (\mul)$ all discontinuities of the
matrix elements that could potentially contribute to the threshold
correction of $C^B_\nu (\mul)$ vanish identically. By matching the
effective theories at the bottom quark threshold scale $\mub$ we
obtain from \Eqs{eq:deltaCi} in the NDR scheme supplemented by the
definition of the evanescent operator given in \Eq{eq:boxevanescent}
the following non-trivial correction     
\beq \label{eq:Cbvthreshold}
\begin{split}
\delta C^{B (2)}_\nu (\mub) & = -\f{2}{3} \ln \f{\mub^2}{\mb^2} \left
( \left ( \f{238784}{529} - \f{9608}{1587} \eta_b \right )
\eta^{\f{1}{23}}_b - \f{1336}{3} + \eta^{\f{24}{23}}_b C^{B (1)}_\nu
(\muh) \right ) \, ,        
\end{split}
\eeq
where $\mb = \mb (\mb)$ is the bottom quark $\MSbar$ mass.

\subsection{Matrix Elements}
\label{subsec:MEB}

For scales $\mul$ below $\muc$ the transition operator in the case of
the electroweak box contributions takes the form $-{\cal T}^B =
\Heff^B$ with      
\beq \label{eq:HeffB}
\Heff^B = \f{4 \GF}{\sqrt{2}} \f{\aem}{2 \pi \sws} \, \lambda_c \, 
\sum_{\ell = e, \mu, \tau}  C_B^\ell (\mul) ( \bar{s}_\sL \gamma_\mu 
d_\sL ) ( \bar {\nu_\ell}_\sL \gamma^{\mu} {\nu_\ell}_\sL ) \, ,       
\eeq
which is part of the low-energy effective Hamiltonian of
\Eq{eq:Heff}. Again the bilocal contribution to ${\cal T}^B$ in
\Eq{eq:TB} has disappeared, because the charm quark field is
integrated out and the effect from its loop is absorbed into $C_B^\ell
(\muc)$. The leftover contribution from the up quark loop to ${\cal
T}^B$ is like in ${\cal T}^P$ power-suppressed. The numerical size of
these corrections has been calculated in \cite{Isidori:2005xm}
and is included in our numerical analysis of \Sec{sec:numerics}.

{%
\begin{figure}[!t]
\begin{center}
$\begin{array}{c@{\hspace{15mm}}c}
\scalebox{0.65}{
\begin{picture}(120,138) (45,-30)
\SetWidth{0.5}
\SetColor{Black}
\ArrowLine(45,93)(105,78)
\ArrowLine(165,93)(105,78)
\ArrowLine(105,3)(45,-12)
\ArrowLine(105,3)(165,-12)
\ArrowArcn(97.5,40.5)(38.24,78.69,-78.69)
\Text(75,100)[]{\Large{\Black{$s$}}}
\Text(135,99)[]{\Large{\Black{$\nu$}}}
\Text(150,39)[]{\Large{\Black{$\ell$}}}
\Text(75,-21)[]{\Large{\Black{$d$}}}
\Text(135,-22)[]{\Large{\Black{$\nu$}}}
\Text(106,-18)[]{\Large{\Black{$Q_4^{c}$}}}
\Text(106,98)[]{\Large{\Black{$Q_3^{c}$}}}
\GBox(100,-2)(110,8){0.0}
\GBox(100,73)(110,83){0.0}
\ArrowArc(112.5,40.5)(38.24,101.31,258.69)
\Text(60,39)[]{\Large{\Black{$c$}}}
\end{picture}
} & 
\scalebox{0.65}{
\begin{picture}(165,168) (0,0)
\SetWidth{0.5}
\SetColor{Black}
\ArrowArc(113.93,71.93)(39.94,-178.67,-102.92)
\ArrowArc(114.17,68.83)(40.23,103.17,176.91)
\GlueArc(87.79,40.56)(33.42,114.38,213.73){6}{5.14}
\Vertex(74,71){2.83}
\ArrowLine(45,123)(105,108)
\ArrowLine(165,123)(105,108)
\ArrowLine(105,33)(45,18)
\ArrowLine(105,33)(165,18)
\ArrowArcn(97.5,70.5)(38.24,78.69,-78.69)
\Text(75,130)[]{\Large{\Black{$s$}}}
\Text(135,129)[]{\Large{\Black{$\nu$}}}
\Vertex(60,22){2.83}
\Text(150,69)[]{\Large{\Black{$\ell$}}}
\Text(75,9)[]{\Large{\Black{$d$}}}
\Text(135,8)[]{\Large{\Black{$\nu$}}}
\GBox(100,28)(110,38){0.0}
\GBox(100,103)(110,113){0.0}
\Text(106,12)[]{\Large{\Black{$Q_4^{c}$}}}
\Text(106,128)[]{\Large{\Black{$Q_3^{c}$}}}
\Text(37,50)[]{\Large{\Black{$g$}}}
\Text(70,98)[]{\Large{\Black{$c$}}}
\Text(70,43)[]{\Large{\Black{$c$}}}
\end{picture}
}
\end{array}$
\vspace{2mm}
\caption{Typical examples of diagrams that contribute to the matrix
elements of $Q^B$ at $\ord (\as)$ (left diagram) and $\ord (\as^2)$
(right diagram).}           
\label{fig:Bmatrix}
\end{center}
\end{figure}
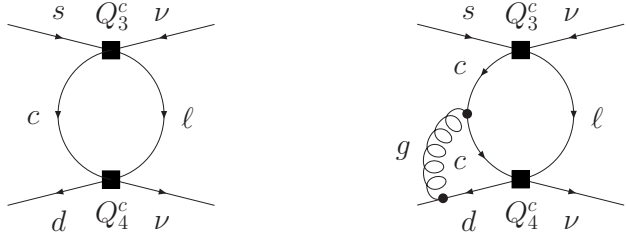
}%

It is again favorable to express the final low-energy Wilson
coefficient in terms of the $\as$ of the effective theory with four
active quark flavors rather than in terms of the $\as$ of the
effective theory with three active quark flavors. Proceeding in this
way the product $C_B^\ell (\muc) \left \langle Q_\nu (\muc) \right
\rangle$ can be written in the form  
\beq \label{eq:CB}
C_B^\ell (\muc) \left \langle Q_\nu \right \rangle^{(0)} = \f{x_c
(\muc)}{16} \f{4 \pi}{\as (\muc)} \left ( C_\nu^B (\muc) \left \langle
Q_\nu \right \rangle^{(0)} + 4 \left \langle Q^B_\ell (\muc) \right
\rangle \right ) \, ,  
\eeq
where $Q^B_\ell$ denotes the part of $Q^B$ that contains a lepton of
flavor $\ell$. The factor of $4$ is again a consequence of the
definition of $Q^B$ and $Q_\nu$ in \Eqsand{eq:QB}{eq:Qv}. 

The renormalized matrix elements $\left \langle Q^B_\ell (\muc) \right
\rangle$ are found by computing the finite parts of one- and two-loop
diagrams. Examples can be seen in \Fig{fig:Bmatrix}. Regulating
spurious IR divergences once dimensionally and once with small quark
masses we found identical results for the matrix elements. In the NDR
scheme supplemented by the definition of the evanescent operator given
in \Eq{eq:boxevanescent} the expansion coefficients for the tau lepton
contribution to the matrix element take the following form
\beq \label{eq:rBtau}
\begin{split}
r^{B (1)}_\tau (\muc) & = 5 + \f{4 \hspace{0.4mm} x_\tau}{1 - x_\tau}
\ln x_\tau + 4 \ln \f{\muc^2}{\mc^2} \, , \\[2mm] 
r^{B (2)}_\tau (\muc) & = -2 \hspace{0.4mm} \cf \left ( \f{9 + 7
\hspace{0.4mm} x_\tau}{1 - x_\tau} + \f{x_\tau \left ( 3 + 13
\hspace{0.4mm} x_\tau \right )}{\left ( 1 - x_\tau \right)^2} \,  \ln x_\tau - \f{12
\hspace{0.4mm} x_\tau}{ 1 - x_\tau} \hspace{0.4mm} \li \left ( 1 -
x_\tau \right ) \right. \\[1mm] 
& - \left. \left ( \f{1 - 13 \hspace{0.4mm} x_\tau}{1 - x_\tau} -
\f{12 \hspace{0.4mm} x_\tau^2}{\left ( 1 - x_\tau \right )^2} \ln
x_\tau \right ) \ln \f{\muc^2}{\mc^2} - 6 \ln^2 \f{\muc^2}{\mc^2}
\right ) \, .        
\end{split}
\eeq
Here $x_\tau = \mtau^2/\mc^2$ with $\mc = \mc (\muc)$ denotes the
ratio of the tau lepton and the charm quark $\MSbar$ mass squared.    
The first line of the above equations agrees with the known NLO result
\cite{Buchalla:1993wq}, after including the aforementioned factor of
$4$, while the second one represents the new NNLO expression.  

In the case of the electron and the muon the lepton mass can be
neglected compared to the charm quark mass. In the limit $x_\tau$
going to zero \Eqs{eq:rBtau} simplify to  
\beq \label{eq:rBe}
\begin{split}
r^{B (1)}_{e, \mu} (\muc) & = 5 + 4 \ln \f{\muc^2}{\mc^2} \, , \\[2mm]
r^{B (2)}_{e, \mu} (\muc) & = -2 \hspace{0.4mm} \cf \left ( 9  - \ln
\f{\muc^2}{\mc^2} - 6 \ln^2 \f{\muc^2}{\mc^2} \right ) \, .    
\end{split}
\eeq

\subsection{Final Result}
\label{subsec:finalB}

The analytic expression for the electroweak box contribution to
$X^\ell (x_c)$ that enters the definition of $\PcX$ in \Eq{eq:defPcX}
can be obtained in a straightforward manner following the detailed
exposition presented at the end of \Sec{sec:Z}. 

In particular the perturbative expansion of the Wilson coefficient
$\vec{C}_B (\muc)$ in \Eq{eq:CBgB} is given by
\Eqsand{eq:CPPE}{eq:CPvfinal} after replacing all subscripts $P$ by 
$B$. The explicit expressions for $\vec{C}_B^{(k)} (\muh)$, 
$\hat{U}^{\nf \, (0)}_B$, $\hat{J}^{\nf \, (k)}_B$, and $\delta
\vec{C}_B^{(2)} (\mub)$ can be found in
\Eqsto{eq:U0Bfive}{eq:Cbvthreshold}.   

Keeping terms up to third order in the strong coupling expansion
\Eq{eq:CB} can be written as  
\beq \label{eq:CBPT}
C_B^\ell (\muc) = \kappa_c \, \f{x_c (\mc)}{16} \left ( \f{4 \pi}{\as
(\muc)} C_B^{\ell \hspace{0.2mm} (0)} (\muc) + C_B^{\ell
\hspace{0.2mm} (1)} (\muc) + \f{\as (\muc)}{4 \pi} C_B^{\ell
\hspace{0.2mm} (2)} (\muc) \right ) \, ,     
\eeq
where $\kappa_c$ is defined after \Eq{eq:xc} and
\bea \label{eq:CBcoeff}
\begin{split}
C_B^{\ell \hspace{0.2mm} (0)} (\muc) & = C_\nu^{B (0)} (\muc) \, ,
\\[2mm]
C_B^{\ell \hspace{0.2mm} (1)} (\muc) & = C_\nu^{B (1)} (\muc) +
4 \hspace{0.4mm} \rho_\ell^{B (1)} (\muc) + \xi_c^{(1)} \, C_\nu^{B
(0)} (\muc) \, , \\[2mm]    
C_B^{\ell \hspace{0.2mm} (2)} (\muc) & = C_\nu^{B (2)} (\muc)  +
4 \hspace{0.4mm} \rho_\ell^{B (2)} (\muc) + \xi_c^{(1)} \, C_\nu^{B
(1)} (\muc) +  4 \hspace{0.4mm} \xi_c^{(1)}  \rho_\ell^{B (1)} (\mu_c)  + \xi_c^{(2)} \, C_\nu^{B (0)} (\muc) \, .  
\end{split}
\eea
Here the factors of $4$ arise again from the use of chiral fermion
fields in $Q^B$ and $Q_\nu$ of \Eqsand{eq:QB}{eq:Qv}. 

The coefficients $\rho_\ell^{B (k)} (\muc)$ are obtained from
$r_\ell^{B (k)} (\muc)$ by expanding the charm quark $\MSbar$ mass 
$\mc (\muc)$ entering \Eqsand{eq:rBtau}{eq:rBe} in $\as$ around $\mc
(\mc)$. Explicitly we find in the case of the tau lepton    
\beq \label{eq:rhoBtaus} 
\begin{split}
\rho_\tau^{B (1)} (\muc) & = r_\tau^{B (1)} (\muc, \mc) +
\f{4}{x_\tau - \kappa_c} \left ( \kappa_c \ln \kappa_c - \f{x_\tau
\left ( 1 - \kappa_c \right )}{1 - x_\tau} \ln x_\tau \right ) \, ,
\\[2mm]         
\rho_\tau^{B (2)} (\muc) & = r_\tau^{B (2)} (\muc, \mc) +
\f{32}{\left ( x_\tau - \kappa_c \right )} \left ( \f{4
\hspace{0.4mm} x_\tau \left ( 1 - \kappa_c \right )}{3 \left ( 1 -
x_\tau \right )} \right. \\[1mm]    
&  \left. -\f{x_\tau \left ( x_\tau \left ( 13 - 29 \hspace{0.4mm}
x_\tau \right ) + \kappa_c \left ( 3 + 29 \hspace{0.4mm} x_\tau^2
\right ) - \kappa_c^2 \left ( 3 + 13 \hspace{0.4mm} x_\tau \right )
\right )}{12 \left ( x_\tau - \kappa_c \right ) \left ( 1 - x_\tau
\right )^2} \ln x_\tau \right. \\[1mm]
& \left. + \f{\kappa_c \left ( 17 x_\tau - \kappa_c \right )}{12 \left
( x_\tau - \kappa_c \right )} \ln \kappa_c + \f{x_\tau^2}{x_\tau -
\kappa_c} \ln x_\tau \ln \kappa_c - \f{x_\tau^2 + 2 \hspace{0.4mm}
x_\tau \kappa_c - \kappa_c^2}{2 \left ( x_\tau - \kappa_c \right )}
\ln^2 \kappa_c \right. \\[1mm]
& \left.  -\f{x_\tau \left ( x_\tau - \kappa_c \right )}{ 1 -
x_\tau} \li \left ( 1 - x_\tau \right ) - x_\tau
\hspace{0.4mm} \li \left ( 1 - \f{x_\tau}{\kappa_c} \right )\right )
\\[1mm] 
& + \f{32}{\left ( x_\tau - \kappa_c \right ) \left ( 1 - x_\tau
\right )} \left ( x_\tau \left ( 1 - \kappa_c \right ) 
- \f{x_\tau^2 \left ( 1 - \kappa_c \right ) \left ( 1 - 2
\hspace{0.4mm} x_\tau + \kappa_c \right )}{\left ( x_\tau - \kappa_c
\right ) \left ( 1 - x_\tau \right )} \ln x_\tau \right. \\[1mm] 
&   + \f{\kappa_c
\left ( 1 - x_\tau \right ) \left ( 2 \hspace{0.4mm} x_\tau - \kappa_c
\right )}{x_\tau - \kappa_c} \ln \kappa_c   \bigg ) \ln
\f{\muc^2}{\mc^2} \\[1mm]
& + \f{4 \hspace{0.4mm} \kappa_c}{x_\tau - \kappa_c} \left ( 1 -
\f{x_\tau}{x_\tau - \kappa_c} \ln x_\tau + \f{x_\tau}{x_\tau -
\kappa_c} \ln \kappa_c \right ) \xi_c^{(1)} \, .     
\end{split}
\eeq

In the limit $x_\tau$ going to zero which is the relevant one in the
case of the electron and the muon one arrives at
\beq \label{eq:rhoBes} 
\begin{split}
\rho_{e, \mu}^{B (1)} (\muc) & = r_{e, \mu}^{B (1)} (\muc, \mc) - 4
\ln \kappa_c \, , \\[2mm]       
\rho_{e, \mu}^{B (2)} (\muc) & = r_{e, \mu}^{B (2)} (\muc, \mc) -
\f{8}{3} \ln \kappa_c + 16 \ln^2 \kappa_c - 32 \ln \kappa_c \ln
\f{\muc^2}{\mc^2} - 4 \hspace{0.4mm} \xi_c^{(1)} \, . 
\end{split}
\eeq

The additional argument in $r_\ell^{B (k)} (\muc, \mc)$ signals that
the expansion coefficients of \Eqsand{eq:rBtau}{eq:rBe} have to be
evaluated at $\mc (\mc)$ and not at $\mc (\muc)$. The same applies to
the variables $x_\tau$ and $\mc$ appearing explicitly in
\Eqsand{eq:rhoBtaus}{eq:rhoBes}. We remark that the second terms in 
the first lines of the latter equations are not present in the NLO
formulas of $\PcX$ given in \cite{Buchalla:1993wq,
Buchalla:1998ba}. The importance of this $\muc$-dependent terms will 
be discussed in \Sec{sec:numerics}. 

\section{Final Result for {\boldmath $\PcX$} at NNLO}
\label{sec:final}

The function $X^\ell (x_c)$ that enters the definition of $\PcX$ in
\Eq{eq:defPcX} is given in terms of the contribution of the
$Z$-penguin and the electroweak boxes by  
\beq \label{eq:Xlxc}
X^\ell (x_c) = C_P (\muc) + C_B^\ell (\muc) \, ,
\eeq 
where the analytic NNLO expression for $C_P (\muc)$ and $C_B^\ell
(\muc)$ can be found in \Eqsand{eq:CPPT}{eq:CBPT}, respectively. The
latter equations together with \Eqsand{eq:defPcX}{eq:Xlxc} then enable
one to find the analytic formula for $\PcX$ through $\ord (\as)$.

The explicit analytic expression for $\PcX$ including the complete NNLO
corrections is so complicated and long that we derive an approximate
formula. Setting $\lambda = 0.2248$, $\mt (\mt) = 163.0 \! \GeV$
and $\muh = 80.0 \! \GeV$ we derive an approximate formula for $\PcX$
that summarizes the dominant parametric and theoretical uncertainties
due to $\mc (\mc)$, $\as (\MZ)$, $\muc$, and $\mub$. It reads
\bea \label{eq:masterformula}
\PcX = 0.3832 \left ( \f{\mc (\mc)}{1.30 \! \GeV} \right )^{1.3750}
\left ( \f{\as (\MZ)}{0.1187} \right )^{1.9480} \left ( 1 +
\sum_{i,j,k,l} \kappa_{ijlm} L_{\mc}^i L_{\as}^j L_{\muc}^k L_{\mub}^l
\right ) ,\hspace{2mm}       
\eea
where 
\beq \label{eq:defLs}
\begin{split}
\begin{aligned} 
L_{\mc} & = \ln \left ( \f{\mc (\mc)}{1.30 \! \GeV} \right ) \, , & 
\hspace{1cm} L_{\as} & = \ln \left ( \f{\as (\MZ)}{0.1187} \right ) \,
, \\[2mm]
L_{\muc} & = \ln \left ( \f{\muc}{1.5 \! \GeV} \right ) \, , & 
\hspace{1cm} L_{\mub} & = \ln \left ( \f{\mub}{5.0 \! \GeV} \right )
\, , 
\end{aligned}
\end{split}
\eeq
and the sum includes the expansion coefficients $\kappa_{ijkl}$ given
in \Tab{tab:kappas}. The above formula approximates the exact NNLO
result with an accuracy of $\pm 0.6 \%$ in the ranges $1.15 \! \GeV
\le \mc (\mc) \le 1.45 \! \GeV$, $0.1150 \le \as (\MZ) \le 0.1230$,
$1.0 \! \GeV \le \muc \le 3.0 \! \GeV$, and $2.5 \! \GeV \le \mub \le
10.0 \! \GeV$. The uncertainties due to $\mt (\mt)$, $\muh$, and the
different methods of computing $\as (\muc)$ from $\as (\MZ)$, which
are not quantified above, are all below $\pm 0.2 \%$. Their actual
size at NLO and NNLO will be discussed in the next section.     

{%
\renewcommand{\arraystretch}{1.25}
\begin{table}[!t]
\begin{center}
\begin{tabular}{|l|l|l|}
\hline
$\kappa_{1000} = 0.7432$ & $\kappa_{0100} = -3.3790$ & $\kappa_{0010} = 
-0.0001$ \\ 
\hline
$\kappa_{0001} = 0.0028$ & $\kappa_{1100} = -0.4350$ & $\kappa_{1010} =
0.1669$  \\
\hline 
$\kappa_{0020} = -0.0903$ & $\kappa_{0020} = -0.0065$ & $\kappa_{0030}
= 0.0330$ \\
\hline 
\end{tabular} 
\vspace{2mm}
\caption{The coefficients $\kappa_{ijkl}$ arising in the approximate
formula for $\PcX$ at NNLO.}                  
\label{tab:kappas}
\end{center}
\end{table}
}%

\section{Numerical Analysis}
\label{sec:numerics}

\subsection{Theoretical Uncertainties of {\boldmath $\PcX$} at NLO} 
\label{subsec:TEPCXNLO}

Before presenting the numerical analysis of the NNLO correction to
$\PcX$, it is instructive to display the theoretical uncertainties
present at the NLO level. These originate in the leftover unphysical
dependence on $\muc$ but are also due to the dependence on $\mub$ and
$\muh$, and to higher order terms that arise in the evaluation of $\as
(\muc)$ from the experimental input $\as (\MZ)$. The latter
uncertainties have not been scrutinized in previous NLO analyses of
the charm quark contribution \cite{Buchalla:1993wq, Buchalla:1998ba}.

\begin{figure}[!p]
\begin{center}
\scalebox{1.1}{
\begingroup%
  \makeatletter%
  \newcommand{\GNUPLOTspecial}{%
    \@sanitize\catcode`\%=14\relax\special}%
  \setlength{\unitlength}{0.1bp}%
\begin{picture}(3600,2160)(0,0)%
\special{psfile=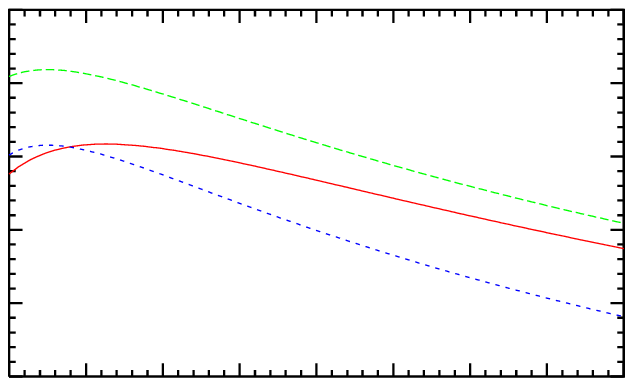 llx=0 lly=0 urx=360 ury=216 rwi=3600}
\put(1975,0){\makebox(0,0){\scalebox{1.0}{$\muc$ [GeV]}}}%
\put(100,1180){%
\special{ps: gsave currentpoint currentpoint translate
270 rotate neg exch neg exch translate}%
\makebox(0,0)[b]{\shortstack{\scalebox{1.0}{$\PcX$}}}%
\special{ps: currentpoint grestore moveto}%
}%
\put(3450,200){\makebox(0,0){ 3}}%
\put(3081,200){\makebox(0,0){ 2.75}}%
\put(2713,200){\makebox(0,0){ 2.5}}%
\put(2344,200){\makebox(0,0){ 2.25}}%
\put(1975,200){\makebox(0,0){ 2}}%
\put(1606,200){\makebox(0,0){ 1.75}}%
\put(1238,200){\makebox(0,0){ 1.5}}%
\put(869,200){\makebox(0,0){ 1.25}}%
\put(500,200){\makebox(0,0){ 1}}%
\put(450,2060){\makebox(0,0)[r]{ 0.42}}%
\put(450,1708){\makebox(0,0)[r]{ 0.4}}%
\put(450,1356){\makebox(0,0)[r]{ 0.38}}%
\put(450,1004){\makebox(0,0)[r]{ 0.36}}%
\put(450,652){\makebox(0,0)[r]{ 0.34}}%
\put(450,300){\makebox(0,0)[r]{ 0.32}}%
\end{picture}%
\endgroup
 }\\[5mm]
\scalebox{1.1}{
\begingroup%
  \makeatletter%
  \newcommand{\GNUPLOTspecial}{%
    \@sanitize\catcode`\%=14\relax\special}%
  \setlength{\unitlength}{0.1bp}%
\begin{picture}(3600,2160)(0,0)%
\special{psfile=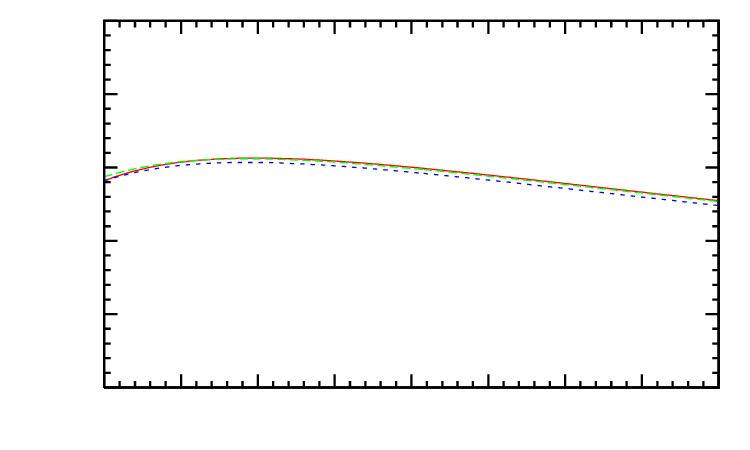 llx=0 lly=0 urx=360 ury=216 rwi=3600}
\put(1975,0){\makebox(0,0){\scalebox{1.0}{$\muc$ [GeV]}}}%
\put(100,1180){%
\special{ps: gsave currentpoint currentpoint translate
270 rotate neg exch neg exch translate}%
\makebox(0,0)[b]{\shortstack{\scalebox{1.0}{$\PcX$}}}%
\special{ps: currentpoint grestore moveto}%
}%
\put(3450,200){\makebox(0,0){ 3}}%
\put(3081,200){\makebox(0,0){ 2.75}}%
\put(2713,200){\makebox(0,0){ 2.5}}%
\put(2344,200){\makebox(0,0){ 2.25}}%
\put(1975,200){\makebox(0,0){ 2}}%
\put(1606,200){\makebox(0,0){ 1.75}}%
\put(1238,200){\makebox(0,0){ 1.5}}%
\put(869,200){\makebox(0,0){ 1.25}}%
\put(500,200){\makebox(0,0){ 1}}%
\put(450,2060){\makebox(0,0)[r]{ 0.42}}%
\put(450,1708){\makebox(0,0)[r]{ 0.4}}%
\put(450,1356){\makebox(0,0)[r]{ 0.38}}%
\put(450,1004){\makebox(0,0)[r]{ 0.36}}%
\put(450,652){\makebox(0,0)[r]{ 0.34}}%
\put(450,300){\makebox(0,0)[r]{ 0.32}}%
\end{picture}%
\endgroup
 }
\end{center}
\vspace{0mm}
\caption{$\PcX$ as a function of $\muc$ at NLO (upper plot) and
NNLO (lower plot). The three different lines correspond to three
different methods of computing $\as (\muc)$ from $\as (\MZ)$ (see
text).}        
\label{fig:NLONNLOplot}
\end{figure}

The dependence of $\PcX$ on $\muc$ can be seen in
\Fig{fig:NLONNLOplot}. The solid red line in the upper plot shows the
NLO result obtained by evaluating $\as (\muc)$ from $\as (\MZ)$
solving the RG equation of $\as$ numerically, while the long-
green and short-dashed blue lines are obtained by first determining
the scale parameter $\Lambda_{\MSbar}$ from $\as (\MZ)$, either using
the explicit solution of the RG equation of $\as$ or by solving the RG
equation of $\as$ iteratively for $\Lambda_{\MSbar}$, and subsequently
calculating $\as (\muc)$ from $\Lambda_{\MSbar}$. The corresponding
two-loop values for $\as (\muc)$ have been obtained with the program
{\tt RunDec} \cite{Chetyrkin:2000yt}. Obviously, the difference
between the three curves is due to higher order terms and has to be
regarded as part of the theoretical error. With its size of $\pm
0.012$ it is comparable to the variation of the NLO result due to
$\muc$, amounting to $\pm 0.020$. 

In \cite{Buchalla:1993wq, Buchalla:1998ba} values for the latter
uncertainty have been quoted that are more than twice as large. The
observed difference is related to the definition of the charm quark
mass. While in the latter publications the value $\mc (\mc)$ has been
employed in the logarithms $\ln ( \muc^2/\mc^2 )$ of the one-loop
matrix elements, we consistently apply $\mc (\muc)$ throughout our NLO
analysis.  

Using the $\MSbar$ scheme and integrating out the charm quark field at
the scale $\muc$, the mass $\mc (\muc)$ appears in all intermediate
steps of the computation. In particular, the bare one-loop matrix
elements computed at NLO are proportional to $\mc^{2 - 2 \eps} (\muc)$.
After multiplying this with the $1/\eps$ poles and finite parts of the
loop integrals, adding the counterterm diagrams and expanding in $\eps$
one finds the results given in the first lines of \Eqs{eq:rpmAV},
\eq{eq:rBtau}, and \eq{eq:rBe}. Switching by hand to $\mc (\mc)$ in the
argument of the logarithms $\ln ( \muc^2/\mc^2 )$ as done in
\cite{Buchalla:1993wq}, amounts thus to treating the factors of $\mc^2
(\muc)$ and $\mc^{-2 \eps} (\muc)$ differently, although they stem from
the very same analytical term in $n = 4 - 2 \eps$ dimensions. It is
important to realize that such a replacement $i)$ introduces a
correction which has no diagrammatic counterpart at two loops, and $ii)$
implies a mass counterterm at NNLO which is not $\MSbar$, as it contains
an explicit $\ln ( \muc^2/\mc^2 )$ term. Using $\mc (\mc)$ in the $\ln
(\muc^2/\mc^2)$ terms of the one-loop matrix elements is for these
reasons disputable, although it leads to results that differ from the
one obtained with $\mc (\muc)$ by terms that are formally of
NNLO. These $\ord (\as)$ terms lead to an artificially large
$\muc$-dependence at NLO and are hence not a good estimate of the size
of the uncalculated higher order terms: substituting all factors $\mc
(\muc)$ entering the one-loop matrix elements in a consistent way by
$\mc (\mc)$ results in an uncertainty from $\muc$ which is close to
the one quoted above. In practice this replacement is achieved by
employing $\mc (\mc)$ in the logarithms $\ln ( \muc^2/\mc^2 )$ of
\Eqs{eq:rpmAV}, \eq{eq:rBtau}, and \eq{eq:rBe}, and by rescaling the
NLO terms stemming from the second terms in \Eqsand{eq:CP}{eq:CB} by a
factor $\kappa_c^{-1}$. Of course our conclusion is also based on the
actual NNLO calculation, which indeed finds a correction that is much
smaller than the theoretical uncertainty at NLO reported in
\cite{Buchalla:1993wq, Buchalla:1998ba}.

Finally, while in \cite{Buchalla:1993wq, Buchalla:1998ba} only $\muc$
was varied, the theoretical error given in \Eq{eq:PcXNLO} includes
also the dependence on $\mub$ and $\muh$ of $\pm 0.001$ each. The
specified scale uncertainties correspond to the ranges $1.0 \! \GeV
\le \muc \le 3.0 \! \GeV$, $2.5 \! \GeV \le \mub \le 10.0 \! \GeV$, and
$40.0 \! \GeV \le \muh \le 160.0 \! \GeV$, and the quoted theoretical
error has been obtained by varying them independently.

\begin{table}[!t] 
\begin{center}
\begin{tabular}{|l|c|c|}
\hline 
&&\\[-3.0mm]
Parameter & Value $\pm$ Error & Reference \\[0.5mm]
\hline \hline
&&\\[-3.0mm]
$\mc (\mc)$ \hspace{0.2mm} [GeV] & $1.30 \pm 0.05$ & \cite{charm}, our
average \\[1mm]     
$\as (\MZ)$ & $0.1187 \pm 0.0020$ & \cite{PDG} \\[1mm]
$\im \lambda_t$ \hspace{0.2mm} $[10^{-4}]$ & $1.407^{+0.096}_{-0.098}$
& \cite{Charles:2004jd} \\[1mm] 
$\re \lambda_t$ \hspace{0.25mm} $[10^{-4}]$ & $-3.13^{+0.20}_{-0.17}$ 
& \cite{Charles:2004jd} \\[1mm] 
$\re \lambda_c$ & $-0.22006^{+0.00093}_{-0.00091}$ &
\cite{Charles:2004jd} \\[2mm]   
\hline 
\end{tabular}
\caption{Input parameters used in the numerical analysis of $\PcX$,
$\BR (\Ktopinunu)$, $| V_{td} |$, $\sin 2 \beta$, and $\gamma$.} 
\label{tab:input}
\end{center}
\end{table}

\subsection{Branching Ratio for {\boldmath $\Ktopinunu$} at NLO} 
\label{subsec:BRKPNLO}

Using the input parameters listed in \Tab{tab:input}, we find from
\Eqs{eq:PcXNLO}, \eq{eq:BR}, \eq{eq:kappap}, and \eq{eq:Xnum} at the
NLO level     
\beq \label{eq:BRNLO}
\BR (\Ktopinunu) = \left ( 7.96 \pm 0.76_{\PcX} \pm 0.84_{\rm other}
\right ) \times 10^{-11} \, , 
\eeq
where the second error collects the uncertainties due to $\kappa_+$,
$\dPcu$, $X (x_t)$, and the CKM elements. Numerically, the enhancement
of $\BR (\Ktopinunu)$ coming from $\dPcu$ \cite{Isidori:2005xm} has
been compensated by the suppression due to the decrease of $\mtpole$
\cite{Group:2005cc}.    

\subsection{Branching Ratio for {\boldmath $\KLtopinunu$} at NLO}
\label{subsec:BRKLNLO}

Employing the value and errors of $\im \lambda_t$ as given in
\Tab{tab:input}, we obtain from \Eqs{eq:Xnum}, \eq{eq:BRL}, and
\eq{eq:kappaL} in the NLO approximation   
\beq \label{eq:BRKLNLO}
\BR (\KLtopinunu) = \left ( 2.85 \pm 0.05_{\mut} \pm 0.39_{\rm
other} \right ) \times 10^{-11} \, ,    
\eeq
where the second error collects the uncertainties due to $\kappa_L$,
$\mt (\mt)$, and the CKM elements. 

\subsection{Theoretical Uncertainties of {\boldmath $\PcX$} at NNLO}
\label{subsec:TEPcXNNLO}
     
Having described the details of our calculation in the previous
sections, we now present our results for $\PcX$. From \Eqs{eq:CPPT},
\eq{eq:CBPT}, and \eq{eq:Xlxc} we find at NNLO   
\beq \label{eq:PcXNNLO}
\PcX  = 0.375 \pm 0.009_{\rm theory} \pm 0.031_{\mc} \pm 0.009_{\as}
\, .
\eeq 
Obviously the error on the charm quark contribution to $\Ktopinunu$ is
now fully dominated by the uncertainty in $\mc (\mc)$. Comparing these
numbers with \Eq{eq:PcXNLO} we observe that our NNLO calculation
reduces the theoretical uncertainty by a factor of $4$.

As can be nicely seen in \Tab{tab:pcasmuc} and in the lower plot of
\Fig{fig:NLONNLOplot}, $\PcX$ depends very weakly on $\muc$ at NNLO,
varying by only $\pm 0.006$. Furthermore, the three different
treatments of $\as$ affect the NNLO result by as little as $\pm
0.001$. The three-loop values of $\as (\muc)$ used in the numerical
analysis have been obtained with the program {\tt RunDec}
\cite{Chetyrkin:2000yt}. The theoretical error quoted in
\Eq{eq:PcXNNLO} includes also the dependence on $\mub$ and $\muh$ of
$\pm 0.001$ each. The presented scale uncertainties correspond to the
ranges given earlier, and the specified theoretical error has again
been obtained by varying them independently.

\subsection{Branching Ratio for {\boldmath $\Ktopinunu$} at NNLO}
\label{subsec:BRKPNNLO}

Using \Eqs{eq:BR}, \eq{eq:kappap}, \eq{eq:Xnum}, and \eq{eq:PcXNNLO}
the result in \Eq{eq:BRNLO} is modified to the NNLO value
\beq \label{eq:BRNNLO}
\BR (\Ktopinunu) = \left ( 8.01 \pm 0.49_{\PcX} \pm 0.84_{\rm other}
\right ) \times 10^{-11} \, .   
\eeq
                
As the value of $\mc (\mc)$ is, besides the CKM parameters, the main
leftover parametric uncertainty in the evaluation of $\BR (\Ktopinunu)$,
we show in the upper plot of \Fig{fig:BRmcmtplot}, $\BR (\Ktopinunu)$ as
a function of $\mc (\mc)$. At present the errors from the CKM
parameters veils the benefit of the NNLO calculation of $\PcX$
presented in this paper. For completeness in the lower plot of the
same figure we also show the dependence on $\mtpole$, which is
significantly smaller, because $\mtpole$ is much better known than
$\mc (\mc)$.  

\subsection{Statistical Analyses of the Branching Ratios of {\boldmath
$\Ktopinunus$}}  
\label{subsec:ckmfittervsutfit}

The partial uncertainties given in \Eqsand{eq:BRKLNLO}{eq:BRNNLO} are
not statistically distributed. A very important issue in determining
the central SM values and errors of $\BR (\KLtopinunu)$ and $\BR
(\Ktopinunu)$ is thus the treatment of the experimental and especially
the theoretical uncertainties entering these observables. The
increasing accuracy in the global analysis of the standard UT and the
achieved reduction of the theoretical uncertainty of $\PcX$ clearly
calls for a closer look at the matter in question.  

To this end it is of interest to see what results are obtained by the
two most developed statistical methods, namely the
\rfit approach used by the \ckmfitter and the Bayesian approach
employed by the \utfit and 
to identify those experimental and
theoretical uncertainties for which a reduction of errors would
contribute the most to the quality of the determination of the
$\Ktopinunus$ branching ratios. 
In this context we would like to caution the reader
 that a direct comparison of the results obtained by the two groups in 
Tab.~\ref{tab:ckmfittervsutfit}
 is quite challenging and the comments given below are hopefully of help
 for
 the reader to make her or his unbiased judgment of the situation. Our
 final result is given subsequently in \Sec{subsec:finalpredictions}.

{%
\renewcommand{\arraystretch}{1.25}
\begin{table*}[!t]
\small{
\begin{center}
\begin{tabular}{|l|c|c|c|}
\hline
Observable &  Central $\pm \; 68 \%$ CL & $95 \%$ CL & $99 \%$ CL\\
\hline \hline 
$\BR (\KLtopinunu)~~[10^{-11}]$ & $2.77^{+0.58}_{-0.49} $ & $[1.80,3.82]$ &
$[1.62,4.23]$ \\ \hline
$\PcX$ & $0.367^{+0.050}_{-0.049}$ & $[0.306,0.427]$ & $[0.293,0.439]$
\\ \hline 
$\BR (\Ktopinunu)~~[10^{-11}]$ & $7.95^{+1.81}_{-1.67}$ & $[5.61,10.50]$ &
$[5.28,11.18]$ \\ \hline \hline
$\BR (\KLtopinunu)~~[10^{-11}]$ & $2.56 \pm 0.30$ & $[2.02,3.16]$  &
$[1.88,3.38]$ \\ \hline
$\PcX$ & $0.375 \pm 0.024$ & $[0.333,0.418]$ & $[0.324,0.429]$ \\ \hline
$\BR (\Ktopinunu)~~[10^{-11}]$ & $7.68 \pm 0.58$ & $[6.61,8.87]$ &
$[6.32,9.26]$ \\ \hline
\end{tabular}
\vspace{2mm}
\caption[]{SM predictions for $\BR (\KLtopinunu)$, $\PcX$, and $\BR
(\Ktopinunu)$ at the $68 \%$, $95 \%$, and $99 \%$ CL obtained by the
\ckmfitter (upper half) and the \utfit (lower half) incorporating all
constraints on the CKM elements following from a global analysis of
the standard UT.}   
\label{tab:ckmfittervsutfit}
\end{center}
}
\end{table*}
}%

\begin{figure}[!p]
\begin{center}
\scalebox{0.5}{\includegraphics{klpi0nn_CKMFitter.eps}} 
\scalebox{0.5}{\includegraphics{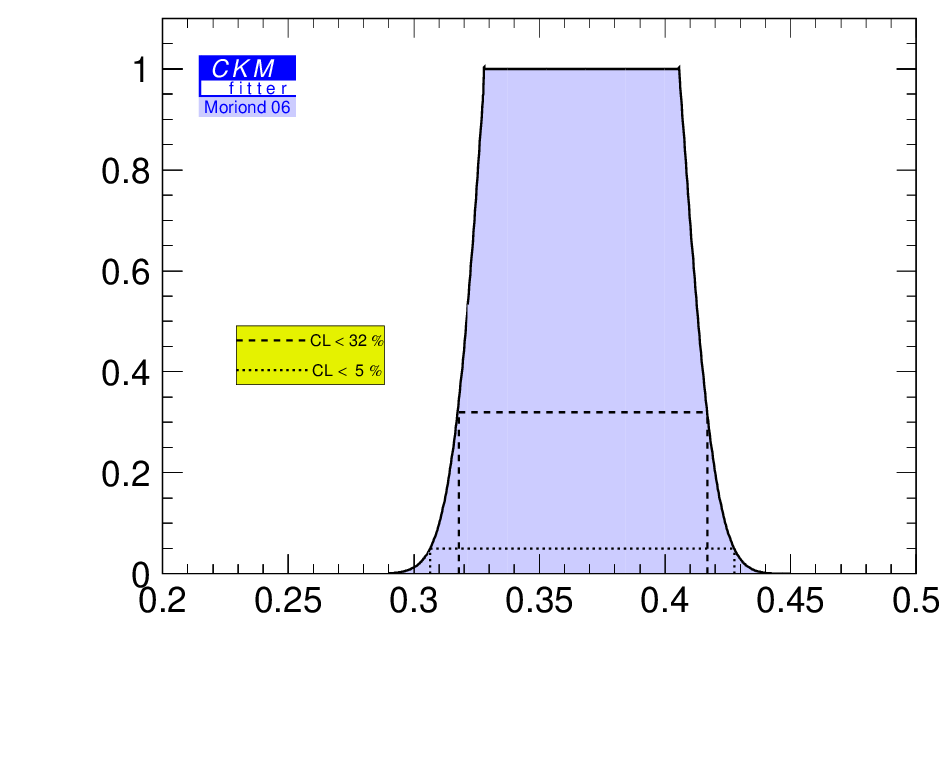}}

\vspace{5mm}

\scalebox{0.5}{\includegraphics{kpinn_CKMFitter.eps}}
\end{center}
\begin{picture}(0,0)%
\setlength{\unitlength}{1pt}%
\put(14,325){{\raisebox{0mm}[0mm][0mm]{%
           \makebox[0mm]{\scalebox{0.9}{\rotatebox{90}{1 - CL}}}}}}%
\put(130,253){{\raisebox{0mm}[0mm][0mm]{%
           \makebox[0mm]{\scalebox{0.9}{$\BR (\KLtopinunu)~~[10^{-11}]$}}}}}%
\put(243,325){{\raisebox{0mm}[0mm][0mm]{%
           \makebox[0mm]{\scalebox{0.9}{\rotatebox{90}{1 - CL}}}}}}%
\put(361,253){{\raisebox{0mm}[0mm][0mm]{%
           \makebox[0mm]{\scalebox{0.9}{$\PcX$}}}}}%
\put(131,125){{\raisebox{0mm}[0mm][0mm]{%
           \makebox[0mm]{\scalebox{0.9}{\rotatebox{90}{1 - CL}}}}}}%
\put(247,52){{\raisebox{0mm}[0mm][0mm]{%
           \makebox[0mm]{\scalebox{0.9}{$\BR (\Ktopinunu)~~[10^{-11}]$}}}}}%
\end{picture}
\caption{Likelihood of $\BR (\KLtopinunu)$, $\PcX$, and $\BR
(\Ktopinunu)$ following from a global CKM analysis performed by the
\ckmfitter\!\!. The borders of the $68 \%$ ($95 \%$) probability
regions are indicated by the long (short) dashed lines.}            
\label{fig:CKMFitterpdfs}
\end{figure}

\begin{figure}[!p]
\begin{center}
\scalebox{0.4}{\includegraphics{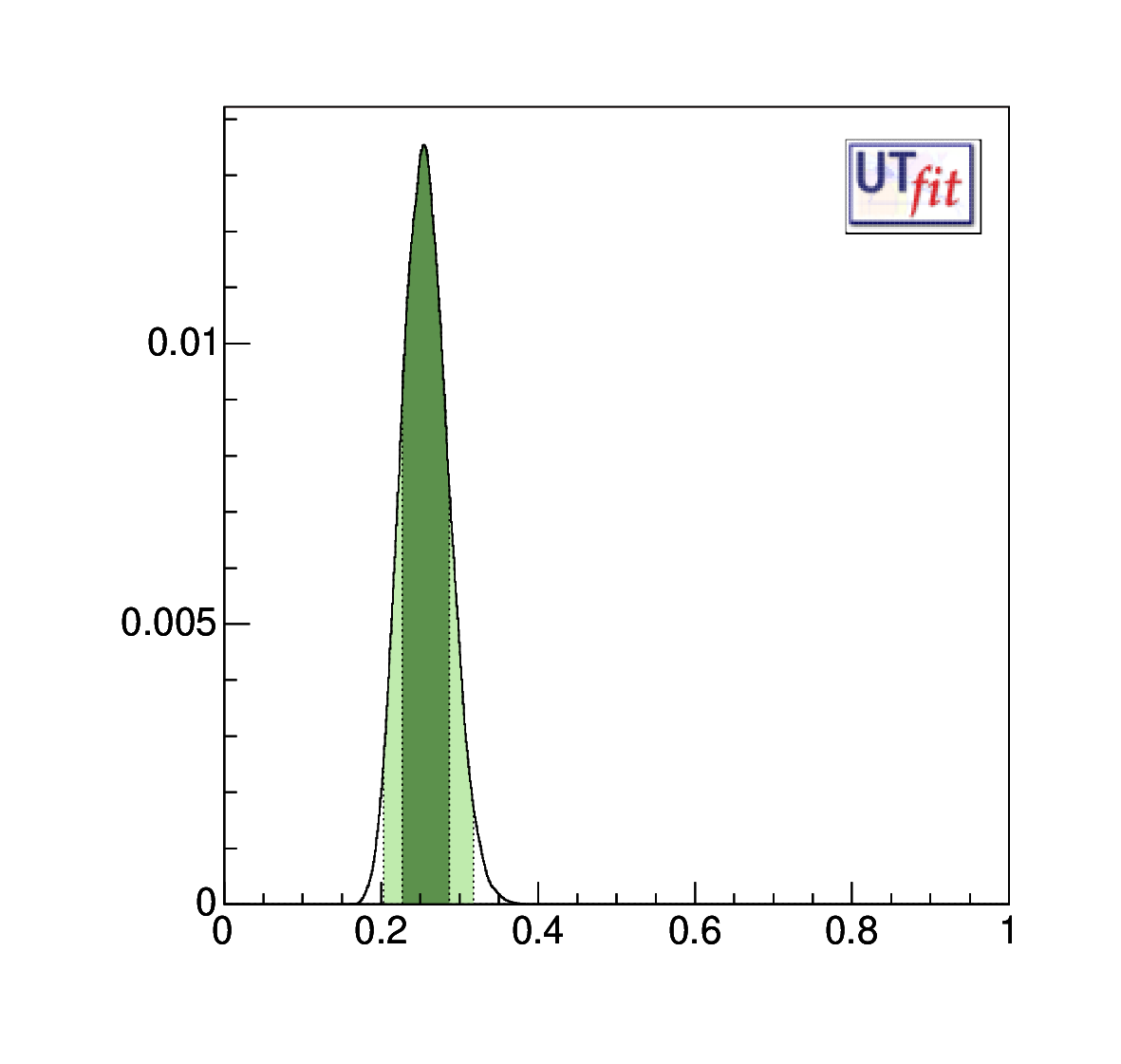}}
\scalebox{0.4}{\includegraphics{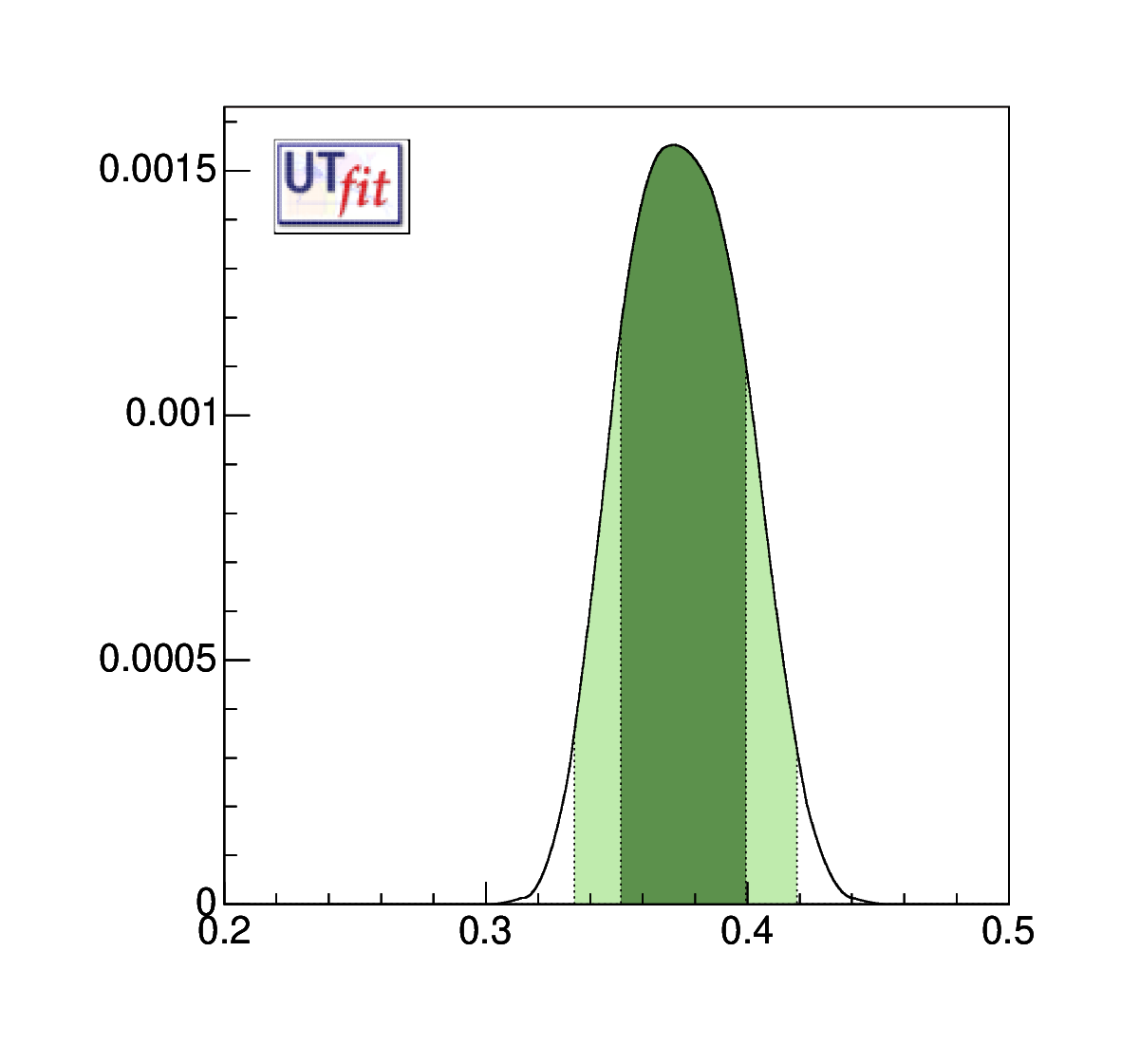}}
\scalebox{0.4}{\includegraphics{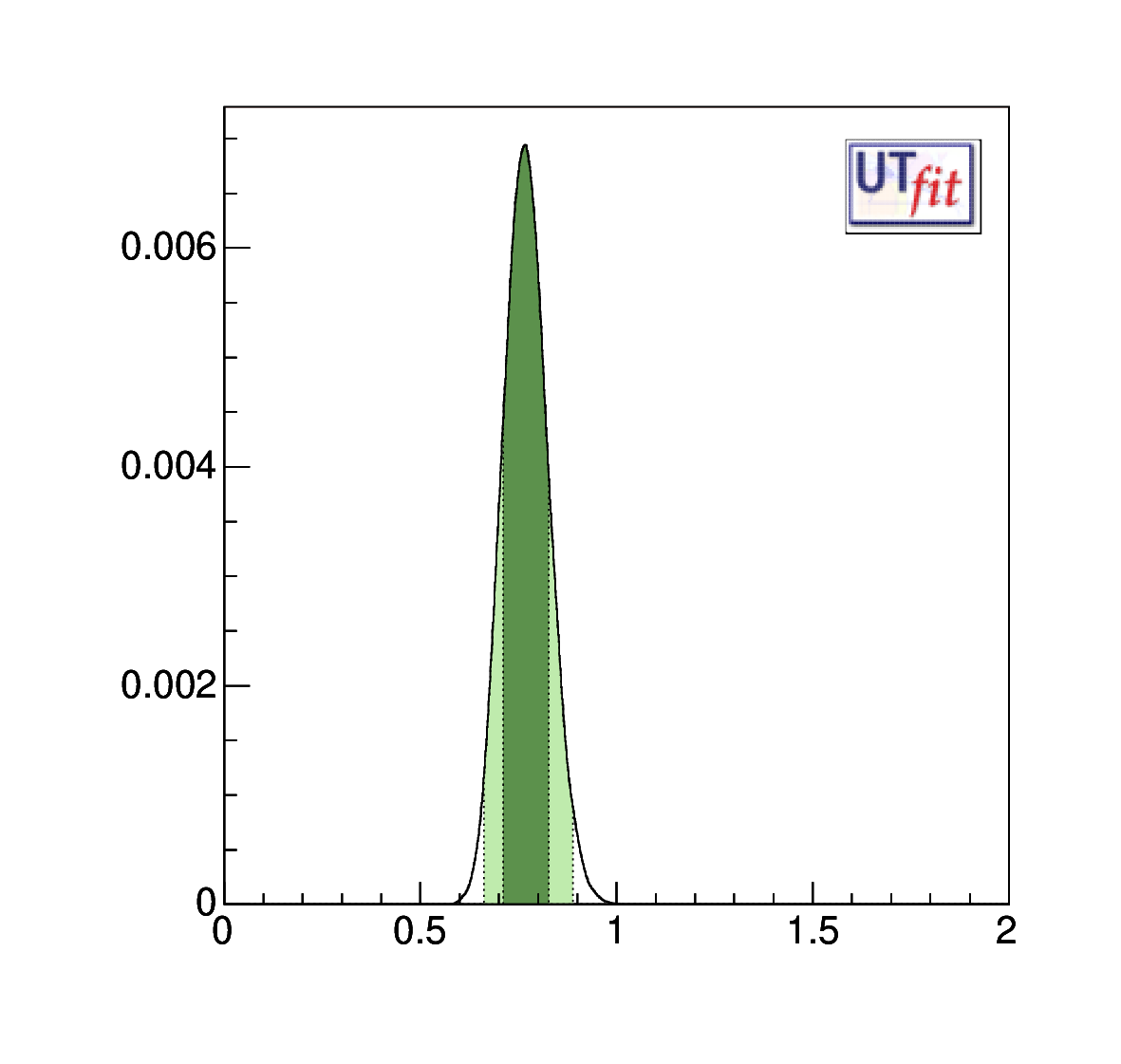}}
\end{center}
\begin{picture}(0,0)%
\setlength{\unitlength}{1pt}%
\put(15,315){{\raisebox{0mm}[0mm][0mm]{%
           \makebox[0mm]{\scalebox{0.9}{\rotatebox{90}{probability density}}}}}}%
\put(125,253){{\raisebox{0mm}[0mm][0mm]{%
           \makebox[0mm]{\scalebox{0.9}{$\BR (\KLtopinunu)~~[10^{-10}]$}}}}}%
\put(243,315){{\raisebox{0mm}[0mm][0mm]{%
           \makebox[0mm]{\scalebox{0.9}{\rotatebox{90}{probability density}}}}}}%
\put(358,253){{\raisebox{0mm}[0mm][0mm]{%
           \makebox[0mm]{\scalebox{0.9}{$\PcX$}}}}}%
\put(131,100){{\raisebox{0mm}[0mm][0mm]{%
           \makebox[0mm]{\scalebox{0.9}{\rotatebox{90}{probability density}}}}}}%
\put(242,38){{\raisebox{0mm}[0mm][0mm]{%
           \makebox[0mm]{\scalebox{0.9}{$\BR (\Ktopinunu)~~[10^{-10}]$}}}}}%
\end{picture}
\caption{Probability density functions of $\BR (\KLtopinunu)$, $\PcX$
and $\BR (\Ktopinunu)$ following from a global CKM analysis performed
by the \utfit\!\!. Dark (light) areas correspond to the $68 \%$ ($95
\%$) probability regions.}           
\label{fig:UTFitpdfs}
\end{figure}

The numerical results for $\BR (\KLtopinunu)$, $\PcX$, and $\BR
(\Ktopinunu)$ obtained by the \ckmfitter and the \utfit are summarized
in \Tab{tab:ckmfittervsutfit}. The corresponding likelihood and
probability density functions are displayed in
\Figsand{fig:CKMFitterpdfs}{fig:UTFitpdfs}. Apart from the CKM 
elements the employed input agrees with the one that has been used to
obtain the numerical values for $\BR (\KLtopinunu)$, $\PcX$, and $\BR
(\Ktopinunu)$ presented earlier in \Eqs{eq:BRKLNLO}, \eq{eq:PcXNNLO}
and \eq{eq:BRNNLO}.  

While the \ckmfitter and the \utfit find comparable errors at $95 \%$
and higher confidence levels (CL), the \utfit obtains significantly
smaller errors at the $68 \%$ CL. This difference is expected, because
both groups treat theoretical errors differently: the \utfit assigns a
probabilistic meaning to them while the \ckmfitter scans for a best fit
value. It can be most easily understood by discussing the value of
$\PcX$ at NNLO. Here the dominant error is the parametric uncertainty in
$\mc(\mc)$, which is treated as a theoretical uncertainty by both
groups. By assigning a flat probability density for $\mc(\mc)$ within
the error range of $1.25 \! \GeV \le \mc (\mc) \le 1.35 \! \GeV$ only
$68 \%$ of this measure is used by the \utfit to compute $\PcX$ at the
$68 \%$ CL. In consequence the error of $\PcX$ found by the \utfit is
smaller by a factor of $0.68$ compared to the error one would obtain by
treating the uncertainty in $\mc (\mc)$ as the 1$\sigma$ range of a
Gaussian distribution.  On the other hand the \ckmfitter uses the whole
parameter range of $\mc(\mc)$ to compute the error on $\PcX$
independently of the CL. For a flat probability density the measure is
proportional to the CL such that at the $95 \%$ CL nearly the whole
range of $\mc (\mc)$ is used by the \utfit\!\!. Correspondingly the
difference in the results for the errors of the \ckmfitter and the
\utfit decreases strongly at the $95 \%$ and higher CL.

A detailed analysis of the individual sources of uncertainty entering
the SM prediction of $\BR (\KLtopinunu)$ and $\BR (\Ktopinunu)$ using
a modified version of the \ckm \hspace{0.4mm} package leads to the
following picture. In both cases residual scale uncertainties are no
longer a dominant source of error as they numerically amount to around
$9 \%$ and $11 \%$ of the total error only. Hence other intrinsic
theoretical errors come to fore. In the case of $\BR (\KLtopinunu)$
the error associated with $r_{K_L}$ is now the main source of
theoretical uncertainty since the error of these isospin breaking
corrections of $\pm 3 \%$ translates into around $15\%$ of the total
uncertainty. In the case of $\BR (\Ktopinunu)$ the uncertainties
associated with the parameters $\dPcu$ and $r_{K^+}$ become
prominent. Numerically the total error introduced by the long-distance
and isospin breaking corrections amounts to about $28 \%$ of the final
uncertainty in $\BR (\Ktopinunu)$. This error is thus slightly larger
than the error due to the charm quark mass which for $\mc (\mc) =
(1.30 \pm 0.05) \! \GeV$ amounts to roughly $20 \%$ of the final
uncertainty. The remaining errors of about $76 \%$ and $41 \%$ for
$\BR (\KLtopinunu)$ and $\BR (\Ktopinunu)$, respectively, are 
due to the uncertainty in the top quark mass, $\as (\MZ)$, and the CKM
elements. The given numbers have been obtained by removing the
individual errors from the fit in the order the have been mentioned in
the text. This study makes clear that if one wants to achieve
predictions of $\BR (\KLtopinunu)$ and $\BR (\Ktopinunu)$ at the level
of $\pm 5 \%$ or below further theoretical improvements concerning the
isospin breaking corrections, long-distance effects, and the
determination of the charm quark mass are indispensable.

\subsection{Final Predictions for {\boldmath $\KLtopinunu$} and
{\boldmath $\Ktopinunu$}}   
\label{subsec:finalpredictions}

Given the sizeable difference of the $68 \%$ CL error intervals
obtained by the \ckmfitter and the \utfit we base our final
predictions on our own analysis of the charm quark contribution and
the branching ratios. To determine our final results for $\BR
(\KLtopinunu)$, $\PcX$, and $\BR (\Ktopinunu)$ we could in principle
add the errors given in \Eqs{eq:BRKLNLO}, \eq{eq:PcXNNLO}, and
\eq{eq:BRNNLO} linearly. These estimates might be too conservative. On
the other hand adding the errors in quadrature would be probably too
optimistic since the uncertainties are not statistically distributed. 
Therefore we quote as the final result the mean of the values obtained
by adding the individual errors once linearly and once in
quadrature. In the case of $\KLtopinunu$ this gives  
\beq \label{eq:KLfinal}
\BR (\KLtopinunu) = \left (2.8 \pm 0.4 \right ) \times 10^{-11} \, , 
\eeq
while in the case of $\Ktopinunu$ one has 
\beq \label{eq:PcXfinal} 
\PcX = 0.38 \pm 0.04 \, , 
\eeq
and 
\beq \label{eq:KPfinal}
\BR (\Ktopinunu) = \left (8.0 \pm 1.1 \right ) \times 10^{-11} \, .
\eeq
The given uncertainties represent the ranges in which we believe that
the true values of $\BR (\KLtopinunu)$, $\PcX$, and $\BR (\Ktopinunu)$
are located with high probability.

\subsection{Impact on the Determination of the CKM Parameters}
\label{subsec:impactonckm}

\begin{figure}[!p]
\begin{center}
\scalebox{0.8}{\includegraphics{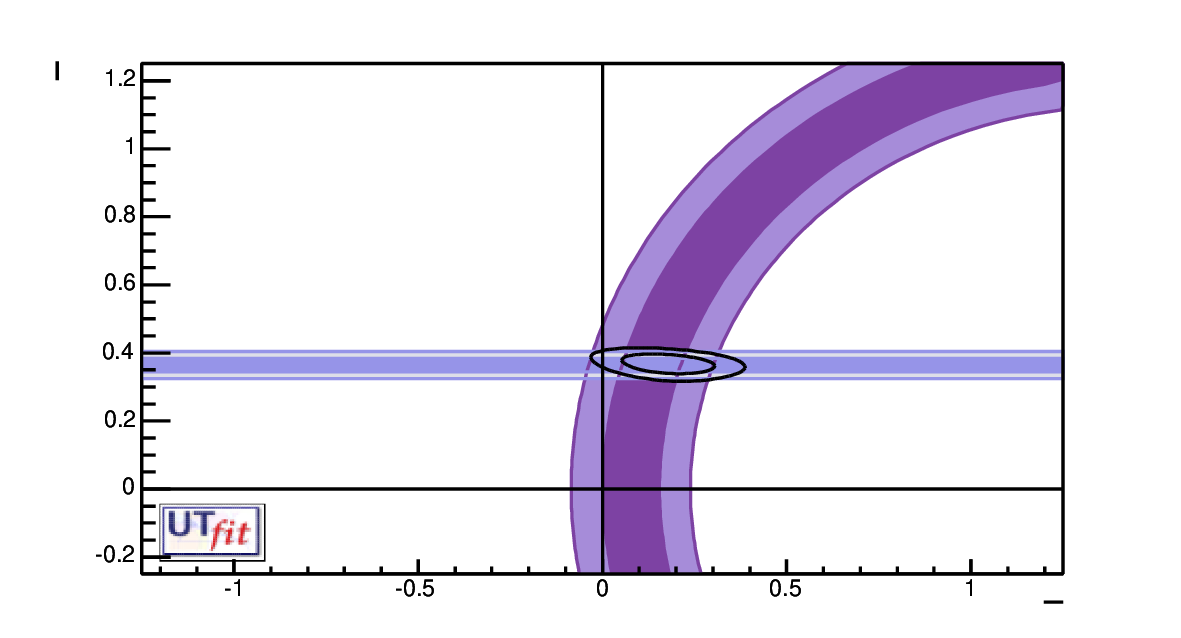}} \vspace{5mm}
\scalebox{0.8}{\includegraphics{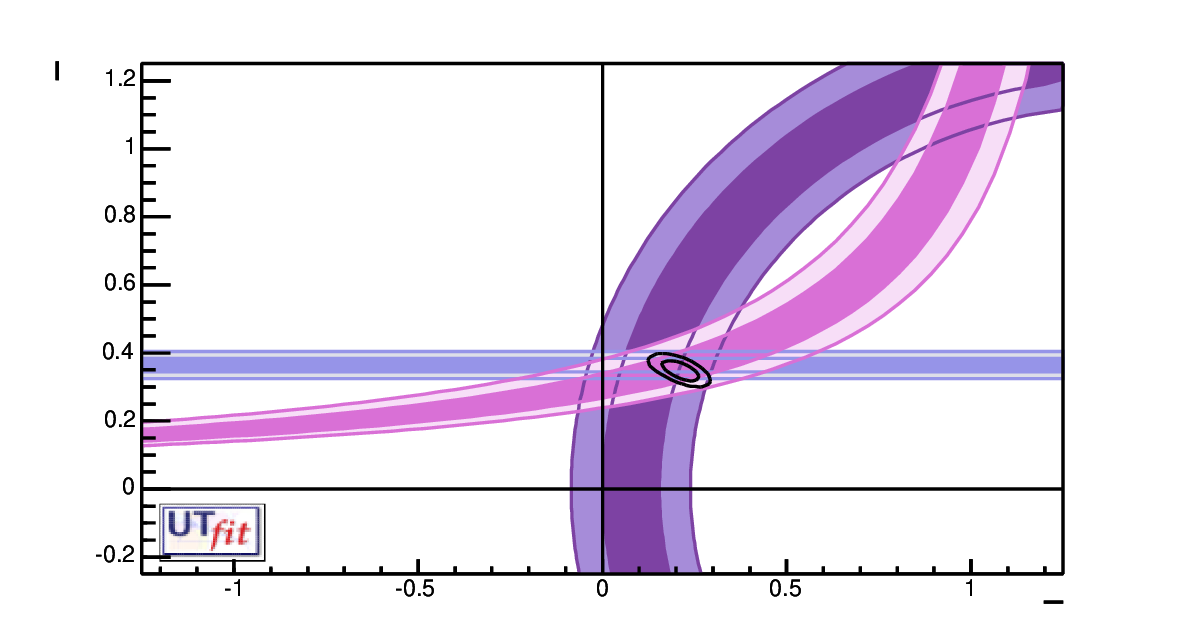}}
\end{center}
\begin{picture}(0,0)%
\setlength{\unitlength}{1pt}%
\put(25,407){{\raisebox{0mm}[0mm][0mm]{%
           \makebox[0mm]{\scalebox{1.15}{\rotatebox{90}{$\bar \eta$}}}}}}%
\put(235,287){{\raisebox{0mm}[0mm][0mm]{%
           \makebox[0mm]{\scalebox{1.15}{$\bar \rho$}}}}}%
\put(140,407){{\raisebox{0mm}[0mm][0mm]{%
           \makebox[0mm]{\scalebox{1.0}{$\BR (\KLtopinunu)$}}}}}%
\put(320,357){{\raisebox{0mm}[0mm][0mm]{%
           \makebox[0mm]{\scalebox{1.0}{$\BR (\Ktopinunu)$}}}}}%
\put(25,160){{\raisebox{0mm}[0mm][0mm]{%
           \makebox[0mm]{\scalebox{1.15}{\rotatebox{90}{$\bar \eta$}}}}}}%
\put(235,40){{\raisebox{0mm}[0mm][0mm]{%
           \makebox[0mm]{\scalebox{1.15}{$\bar \rho$}}}}}%
\put(140,160){{\raisebox{0mm}[0mm][0mm]{%
           \makebox[0mm]{\scalebox{1.0}{$\BR (\KLtopinunu)$}}}}}%
\put(320,110){{\raisebox{0mm}[0mm][0mm]{%
           \makebox[0mm]{\scalebox{1.0}{$\BR (\Ktopinunu)$}}}}}%
\put(150,110){{\raisebox{0mm}[0mm][0mm]{%
           \makebox[0mm]{\scalebox{1.0}{$\eps_K$}}}}}%
\put(410,50){{\raisebox{0mm}[0mm][0mm]{%
           \makebox[0mm]{\scalebox{1.5}{\color{white} $\blacksquare$}}}}}%
\put(410,295){{\raisebox{0mm}[0mm][0mm]{%
           \makebox[0mm]{\scalebox{1.5}{\color{white} $\blacksquare$}}}}}%
\put(25,255){{\raisebox{0mm}[0mm][0mm]{%
           \makebox[0mm]{\scalebox{1.5}{\color{white} $\blacksquare$}}}}}%
\put(25,500){{\raisebox{0mm}[0mm][0mm]{%
           \makebox[0mm]{\scalebox{1.5}{\color{white} $\blacksquare$}}}}}%
\end{picture}
\vspace{-1cm}
\caption{Standard UT from future measurements of $\BR (\KLtopinunu)$
and $\BR (\Ktopinunu)$ with an accuracy of $\pm 10 \%$. Dark (light)
areas correspond to the $68 \%$ ($95 \%$) probability regions. In the
upper (lower) panel the $68 \%$ and $95 \%$ domains following from the
$\Ktopinunus$ constraint (present global CKM fit) as found by the
\utfit are overlaid. The lower plot also shows for comparison the
present constraint coming from $\eps_K$.}           
\label{fig:UTFitUTfit}
\end{figure}

\begin{figure}[!p]
\begin{center}
\vspace{-2.5cm}
\scalebox{0.7}{\includegraphics{ckm_kpinn_all_rhoeta_CKMFitter.eps}}

\vspace{-0.5cm}

\scalebox{0.7}{\includegraphics{ckm_kpinn_all_rhoeta_optimistic_CKMFitter.eps}}
\end{center}
\vspace{-0.5cm}
\caption{Constraints in the $\bar{\rho}$--$\bar{\eta}$ plane arising
from future measurements of $\BR (\KLtopinunu)$ and $\BR (\Ktopinunu)$
with an accuracy of $\pm 10 \%$. The upper (lower) panel shows the
likelihood obtained assuming no (optimistic) improvement of both the 
theoretical and experimental errors of the input parameters. For
comparison the $68 \%$ probability region following from the present
global CKM analysis of the \ckmfitter is also displayed.}           
\label{fig:CKMFitterUTfit}
\end{figure}

As seen in \Eq{eq:ckmerrors} the accuracy of the determination of $|
V_{td} |$, $\sin 2 \beta$, and of the angle $\gamma$ in the UT depends
sensitively on the error in $\PcX$. 

The reduction of the theoretical error in $\PcX$ from $\pm 9.8 \%$
down to $\pm 2.4 \%$ translates into the following uncertainties     
\beq \label{eq:errorcomparison}
\begin{split}
\f{\sigma \left ( | V_{td} | \right)}{| V_{td} |} & = 
\begin{cases}
\pm 4.0 \% \, , & \hspace{1mm} \text{NLO} \, , \\
\pm 1.0 \% \, , & \hspace{1mm} \text{NNLO} \, , 
\end{cases} \\[1mm]
\sigma \left ( \sin 2 \beta \right ) & = 
\begin{cases}
\pm 0.024 \, , & \text{NLO} \, , \\
\pm 0.006 \, , & \text{NNLO} \, , 
\end{cases} \\[1mm]
\sigma \left ( \gamma \right) & = 
\begin{cases}
\pm 4.7^\circ \, , & \hspace{2mm} \text{NLO} \, , \\
\pm 1.2^\circ \, , & \hspace{2mm} \text{NNLO} \, , 
\end{cases} 
\end{split}
\eeq
implying a very significant improvement of the NNLO over the NLO
results. In obtaining these numbers we have used $\sin 2 \beta =
0.724$ and $\gamma = 58.6^\circ$ \cite{Charles:2004jd}, and included
only the theoretical errors quoted in \Eqsand{eq:PcXNLO}{eq:PcXNNLO}.

A comparison of $\sin 2 \beta$ determined from clean $B$-physics
observables with $\sin 2 \beta$ inferred from the $\Ktopinunus$ system
offers a precise and highly non-trivial test of the CKM picture. Both
determinations suffer from small theoretical errors and any
discrepancy between them would signal non-CKM physics. The impact of
future accurate measurements of $\Ktopinunu$ and $\KLtopinunu$ leading
to $\BR (\Ktopinunu) = (8.0 \pm 0.8) \times 10^{-11}$ and $\BR
(\KLtopinunu) = (3.0 \pm 0.3) \times 10^{-11}$ is illustrated in
\Figsand{fig:CKMFitterUTfit}{fig:UTFitUTfit}. As can be seen in
\Fig{fig:UTFitUTfit} the expected precision of the determination of
$(\bar{\rho}, \bar{\eta})$ from the $\Ktopinunus$ system cannot quite
compete with the one from the present global CKM fit performed by the
\utfit\!\!. On the other hand assuming a reduction of the errors in
$\mtpole$, $\mc (\mc)$, $r_{K^+}$, $r_{K_L}$, and $\dPcu$ by a factor
of 3 would put the $\Ktopinunus$ system and the global CKM fit almost
on the same level. The great potential of the $\Ktopinunus$ system 
is clearly visible in the lower panel in \Fig{fig:CKMFitterUTfit}
which shows the constraint in the $\bar{\rho}$--$\bar{\eta}$ plane
obtained by the \ckmfitter adopting this futuristic scenario.  

Obviously, the future of the determination of the standard UT from the
$\Ktopinunus$ system will depend on the uncertainties in the measured
branching ratios, on the value of $\mc (\mc)$ and also on $| V_{cb}
|$. Further theoretical improvement concerning the isospin breaking
and long-distance corrections would be desirable in this respect
too. A corresponding numerical analysis can be found in the updated
version of \cite{Buras:2004uu}, where the NNLO correction to $\PcX$
presented here, will be soon included.   

While the determination of $| V_{td} |$, $\sin 2 \beta$, and $\gamma$
from the $\Ktopinunus$ system is without doubt of interest, with the
slow progress in measuring the relevant branching ratios and much
faster progress in the determination of the angle $\gamma$ from $B_s
\to DK$ system to be expected at LHC, the role of the $\Ktopinunus$
system will shift towards the search for new physics rather than the
determination of the CKM parameters.  

Indeed, determining the CKM parameters from tree diagrams dominated $K$-
and $B$-decays and thus independently of new physics contributions will
allow to find the ``true" values of the CKM parameters and the
so-called reference unitarity triangle \cite{RUT}. Inserting these,
hopefully accurate, values in the formulas for the branching ratios
presented here, will allow to obtain very precise preditions for the
SM rates of both decays. A comparison with future data on these decays
will then give a clear signal of potential new physics contributions
in a theoretical clean enviroment. Even deviations by $20 \%$ from
the SM expectations could be considered as signals of new physics,
whereas it is not possible in most other decays in which the
theoretical uncertainties are at least $10 \%$.   

\section{Summary} 
\label{sec:summary}

In this paper we have calculated the complete NNLO QCD correction of 
the charm quark contribution to the branching ratio for the rare decay
$\Ktopinunu$ in the SM. As the charm quark contribution is essentially
unaffected by new physics, our results are also valid in basically all
extensions of the SM. 
  
The main result of our paper is summarized in the approximate but very
accurate analytic expression for the relevant parameter $\PcX$ as a
function of $\mc (\mc)$ and $\as (\MZ)$ presented in
\Eq{eq:mformula}. The remaining scale uncertainties and the
uncertainty due to higher order terms in the computation of $\as
(\muc)$ from $\as (\MZ)$, that is sizeable at the NLO level as seen in
\Fig{fig:NLONNLOplot}, have been drastically reduced through our
calculation to a level that they can basically be ignored for all
practical purposes. This can be seen in \Tab{tab:pcasmuc} and
\Fig{fig:NLONNLOplot}. Nevertheless, an approximate formula for
$\PcX$ containing the dominant parametric and theoretical errors due
to $\mc (\mc)$, $\as (\MZ)$, $\muc$, and $\mub$ has been given in
\Eq{eq:masterformula}, which should be useful for future
phenomenological analysis of the rare decay $\Ktopinunu$. 

The values of $\PcX$ for different $\mc (\mc)$ and $\as (\MZ)$ are
collected in \Tab{tab:pcasmc}. As $\as (\MZ)$ is already known with an
accuracy of better than $\pm 2 \%$, the main uncertainty in the
evaluation of $\PcX$ resides in the value of $\mc (\mc)$. Our nominal
value for $\PcX = 0.38 \pm 0.04$ used in the NNLO prediction for the
branching ratio of $\Ktopinunu$ in \Eq{eq:KPfinal} corresponds to $\mc
(\mc) = ( 1.30 \pm 0.05 ) \! \GeV$ but the master formulas for $\PcX$
in \Eqsand{eq:mformula}{eq:masterformula} as well as \Tab{tab:pcasmc}
allow one to calculate $\PcX$ and $\BR (\Ktopinunu)$ for other values
of $\mc (\mc)$.       

With the improved recent evaluation \cite{Isidori:2005xm} of the
long-distance contributions to the charm component, that can be
further improved by lattice calculations \cite{Isidori:2005tv}, and
hopefully an increased accuracy on $\mc (\mc)$ and $r_{K^+}$ in the
future, the theoretical computation of the relevant decay rate will
reach an exceptional degree of precision, subject mainly to the
uncertainties in the values of the CKM parameters. As the latter
errors will be reduced to a large extend in the coming years through
the $B$-decay experiments a prediction for $\BR (\Ktopinunu)$ with an
accuracy significantly below $\pm 10 \%$ will be possible before
the end of this decade. Such a precision is unique in the field of
FCNC processes.   

On the other hand, accurate measurements of $\BR (\Ktopinunu)$, in
particular in conjunction with $\BR (\KLtopinunu)$, will provide a
very important extraction of the CKM parameters that compared with the
information from $B$-decays will offer a truly unique test of the CKM
mechanism both in the SM and some of its extensions. The drastic
reduction of the theoretical uncertainty in $\PcX$ achieved by the
computation presented in this paper will play an important role in
these efforts and increases the power of the $\Ktopinunus$ system in
the search for new physics, in particular if $\BR (\Ktopinunu)$ will
not differ much from the SM prediction.

\subsubsection*{Acknowledgments}
\label{subsec:acknowledgments}

We are grateful to X.~Feng  for bringing to our attention the typographical mistakes in 
Eqs.~(\ref{eq:rBtau}), (\ref{eq:CBcoeff}) and (\ref{eq:rhoBtaus}).
We would like to thank Matthias Steinhauser for providing us with an
updated version of {\tt MATAD} \cite{Steinhauser:2000ry} and
A.~H\"ocker, J.~Ocariz, and M.~Pierini for useful communications
concerning the global analysis of the UT. Discussions with
G.~Buchalla, S.~J\"ager, H.~Lacker, A.~Poschenrieder, F.~Schwab,
L.~Silvestrini, D.~St\"ockinger, S.~Uhlig, and A.~Weiler are
acknowledged. Finally we are grateful to the \ckmfitter and the \utfit
for supplying the numbers and plots shown in
\Secsand{subsec:ckmfittervsutfit}{subsec:impactonckm}. This work has
been supported in part by the Bundesministerium f\"ur Bildung and
Forschung under contract 05HT4WOA/3 and the Schweizer
Nationalfonds. U.~H.\ and U.~N.\ have been supported by the DOE under
contract DE-AC02-76CH03000.

\renewcommand{\thesubsection}{A.\arabic{subsection}}
\setcounter{section}{0}  
\renewcommand{\theequation}{A.\arabic{equation}}
\setcounter{equation}{0}  

\section*{Appendix} 

\subsection{Evanescent Operators}
\label{app:evanescent}   

The evanescent operators that arise as counterterms for the one-, two-
and three-loop diagrams with insertions of the current-current
operators
\beq \label{eq:ccpaper} 
\begin{split}
\Qpaper^q_1 & = \left ( \bar s_\sL \gamma_{\mu} T^a q_\sL \right )
\left ( \bar q_\sL \gamma^{\mu} T^a d_\sL \right ) \, , \\[1mm] 
\Qpaper^q_2 & = \left ( \bar s_\sL \gamma_{\mu} q_\sL \right ) \left
( \bar q_\sL \gamma^{\mu} d_\sL \right ) \, , 
\end{split}
\eeq 
can be chosen to be  
\beq \label{eq:ccevanescent}
\begin{split}
E^q_1 & = ( \bar{s}_\sL \gammadown{\mu_1 \mu_2 \mu_3} T^a q_\sL ) (
\bar{q}_\sL \gammaup{\mu_1 \mu_2 \mu_3} T^a d_\sL ) - \left ( 16 - 4
\eps - 4 \eps^2 \right ) Q^q_1 \, , \\[4mm]
E^q_2 & = ( \bar{s}_\sL \gammadown{\mu_1 \mu_2 \mu_3} q_\sL ) (
\bar{q}_\sL \gammaup{\mu_1 \mu_2 \mu_3} d_\sL ) - \left ( 16 - 4 \eps
- 4 \eps^2 \right ) Q^q_2 \, , \\[1mm]    
E^q_3 & = ( \bar{s}_\sL \gammadown{\mu_1 \mu_2 \mu_3 \mu_4 \mu_5} T^a
q_\sL ) ( \bar{q}_\sL \gammaup{\mu_1 \mu_2 \mu_3 \mu_4 \mu_5} T^a
d_\sL ) - \left ( 256 - 224 \eps - \f{5712}{25} \eps^2 \right ) Q^q_1
\, , \\    
E^q_4 & = ( \bar{s}_\sL \gammadown{\mu_1 \mu_2 \mu_3 \mu_4 \mu_5}
q_\sL ) ( \bar{q}_\sL \gammaup{\mu_1 \mu_2 \mu_3 \mu_4 \mu_5} d_\sL )
- \left ( 256 - 224 \eps - \f{10032}{25} \eps^2 \right ) Q^q_2 \, ,
\\[2mm]   
E^q_5 & = ( \bar{s}_\sL \gammadown{\mu_1 \mu_2 \mu_3 \mu_4 \mu_5 \mu_6
\mu_7}  T^a q_\sL ) ( \bar{q}_\sL \gammaup{\mu_1 \mu_2 \mu_3 \mu_4
\mu_5 \mu_6 \mu_7} T^a d_\sL ) - \left ( 4096 - 7680 \eps \right )
Q^q_1 \\[4mm]
E^q_6 & = ( \bar{s}_\sL \gammadown{\mu_1 \mu_2 \mu_3 \mu_4 \mu_5 \mu_6
\mu_7} q_\sL ) ( \bar{q}_\sL \gammaup{\mu_1 \mu_2 \mu_3 \mu_4 \mu_5
\mu_6 \mu_7} d_\sL ) -  \left ( 4096 - 7680 \eps \right ) Q^q_2 \, .  
\end{split}
\eeq
Here the shorthand notation $\gammadown{\mu_1 \cdots \mu_m} =
\gamma_{\mu_1} \cdots \gamma_{\mu_m}$ and $\gammaup{\mu_1 \cdots \mu_m}
= \gamma^{\mu_1} \cdots \gamma^{\mu_m}$ has been used.  The above
operators have been defined such that $i)$ the ADM of the operators
$Q^q_\pm$ introduced in \Eq{eq:Qpm} is diagonal through NNLO, and that
$ii)$ their particular structure differs only by multiples of $\eps^2$
times physical operators from the evanescent operators of the
``traditional'' basis \cite{Buras:1989xd, Ciuchini:1993vr}. The latter
operators can be found by the procedure outlined in \cite{Buras:1992tc}.
Up to three loops they have been given in \cite{Gorbahn:2004my}. Of
course, the above choice is not unique in the sense, that there are many
schemes that would satisfy $i)$ through NNLO. It is however not possible
to define the set of evanescent operators to be invariant under the
interchange of color structures and to achieve $i)$ simultaneously.
Finally, we remark that the $\eps^2$ terms of the above one-loop
evanescent operators $E^q_1$ and $E^q_2$ are unambiguously determined
by condition $i)$. 

In the case of the electroweak box contribution only one evanescent
operator arises as a counterterm for the one-, two-, and three-loop
diagrams considered in this paper. Following \cite{Buchalla:1993wq,
Buchalla:1998ba, Misiak:1999yg} we chose it as     
\beq \label{eq:boxevanescent}
E_\nu = \f{\mc^2}{\gs^2 \mul^{2 \eps}} \sum_{\ell = e, \mu, \tau} (
\bar{s}_\sL \gammadown{\mu_1 \mu_2 \mu_3} d_\sL ) ( \bar
{\nu_\ell}_\sL \gammaup{\mu_1 \mu_2 \mu_3} {\nu_\ell}_\sL  ) - \left (
16 - 4 \eps \right ) Q_\nu \, .  
\eeq   
Here $\mc$ is the charm quark $\MSbar$ mass $\mc (\mul)$ and the
explicit factors $\gs^{-2}$ and $\mu^{-2 \eps}$ follow from the
normalization of $Q_\nu$ in \Eq{eq:Qv}.

\subsection{Change from the ``Standard'' Basis of Current-Current
Operators}     
\label{app:change} 

Beyond LO the anomalous dimensions and the Wilson coefficients depend
on the definition of the operators in $n = 4 - 2 \eps$ dimensions. So
far all of the direct NNLO calculations have been performed in the
operator basis introduced in \cite{Chetyrkin:1997gb}, which we will
call ``standard'' basis from now on. It consists of the following set
of physical and evanescent operators  
\beq \label{eq:qprimeandeprimevector}
\vec{\Qmisiak}^T = ( \Qmisiak^q_1, \Qmisiak^q_2 ) \, , \hspace{1cm}
\vec{\Emisiak}^T = ( \Emisiak^q_1, \Emisiak^q_2, \Emisiak^q_3,
\Emisiak^q_4 ) \, ,      
\eeq
where 
\beq \label{eq:QEmisiak} 
\begin{split}
\Qmisiak^q_1 & = \left ( \bar s_\sL \gamma_{\mu} T^a q_\sL \right )
\left ( \bar q_\sL \gamma^{\mu} T^a d_\sL \right ) \, , \\[1mm] 
\Qmisiak^q_2 & = \left ( \bar s_\sL \gamma_{\mu} q_\sL \right ) \left
( \bar q_\sL \gamma^{\mu} d_\sL \right ) \, , \\[1mm] 
\Emisiak^q_1 & = ( \bar{s}_\sL \gammadown{\mu_1 \mu_2 \mu_3} T^a q_\sL
) ( \bar{q}_\sL \gammaup{\mu_1 \mu_2 \mu_3} T^a d_\sL ) - 16
\Qmisiak^q_1 \\[1mm]  
\Emisiak^q_2 & = ( \bar{s}_\sL \gammadown{\mu_1 \mu_2 \mu_3} q_\sL ) (
\bar{q}_\sL \gammaup{\mu_1 \mu_2 \mu_3} d_\sL ) - 16 \Qmisiak^q_2 \, ,
\\[1mm]  
\Emisiak^q_3 & = ( \bar{s}_\sL \gammadown{\mu_1 \mu_2 \mu_3 \mu_4
\mu_5} T^a q_\sL ) ( \bar{q}_\sL \gammaup{\mu_1 \mu_2 \mu_3 \mu_4
\mu_5} T^a d_\sL ) - 256 \Qmisiak^q_1 - 20 \Emisiak^q_1 \, , \\[1mm]  
\Emisiak^q_4 & = ( \bar{s}_\sL \gammadown{\mu_1 \mu_2 \mu_3 \mu_4
\mu_5} q_\sL ) ( \bar{q}_\sL \gamma^{\mu_1 \mu_2 \mu_3 \mu_4 \mu_5}
d_\sL ) - 256 \Qmisiak^q_2 - 20 \Emisiak^q_2 \, , 
\end{split}
\eeq
while 
\beq \label{eq:qandevector}
\vec{\Qpaper}^T = (\Qpaper^q_+, \Qpaper^q_-) \, , \hspace{1cm}
\vec{\Epaper}^T = ( \Epaper^q_1, \Epaper^q_2, \Epaper^q_3, \Epaper^q_4
) \, ,     
\eeq
denotes the physical and evanescent operators used in this paper. The
operators $\Epaper^q_5$ and $\Epaper^q_6$ and their primed
counterparts have been omitted in the above equations, because they do
not affect the change of scheme up to the order considered here.

In this appendix we demonstrate how the results for the two-loop
initial conditions \cite{Bobeth:1999mk} and the three-loop ADM
\cite{Gorbahn:2004my} of the current-current operators can be 
transformed to our basis. This will serve as a cross-check of our
results for $C^{(k)}_\pm (\muh)$ and $\gamma^{(k)}_\pm$ as given in
\Eqsand{eq:Cpm}{eq:gpm}. In all formulas presented below we set $\ca =
3$ and $\cf = 4/3$, while the number of active quark flavors $\nf$ is
kept arbitrary. 

The transformations relating the primed with the unprimed sets take
the simple form       
\beq \label{eq:transformations}
\vec{\Qpaper} = \hat{R} \hspace{0.4mm} \vec{\Qmisiak} \, ,
\hspace{1cm} \vec{\Epaper} = \hat{M} \left ( \vec{\Emisiak} + \eps \, 
\hat{U} \vec{\Qmisiak} + \eps^2 \, \hat{V} \vec{\Qmisiak} \right ) \, ,
\eeq
where the matrix $\hat{R}$ ($\hat{M}$) describes a rotation of the
physical (evanescent) operators, while the matrix $\hat{U}$
($\hat{V}$) parameterizes a change of basis that consists of adding
multiples of $\eps$ ($\eps^2$) times physical operators to the
evanescent ones. The matrices introduced in \Eq{eq:transformations}
are given by 
{%
\renewcommand{\arraystretch}{1.25}
\beq \label{eq:rmuvmatrices}
\begin{split}
\hat{R} = 
\left (
\begin{array}{cc}
1 & \f{2}{3} \\ 
-1 & \f{1}{3} 
\end{array} 
\right ) , \hspace{1cm} 
\hat{M} = 
\left (
\begin{array}{cccc}
1 & 0 & 0 & 0 \\ 
0 & 1 & 0 & 0 \\
20 & 0 & 1 & 0 \\ 
0 & 20 & 0 & 1 
\end{array} 
\right ) , \\[-2mm] \\[-2mm] 
\hat{U} = 
\left (
\begin{array}{cc}
4 & 0 \\ 
0 & 4\\
144 & 0 \\ 
0 & 144 
\end{array} 
\right ) , \hspace{1cm} 
\hat{V} = 
\left (
\begin{array}{cccc}
4 & 0 \\ 
0 & 4 \\
\f{3712}{25} & 0 \\ 
0 & \f{8032}{75} 
\end{array}
\right ) . 
\end{split}
\eeq
}%

The change of basis in \Eq{eq:transformations} is $\mul$-independent
and leaves, apart from a global rotation, the anomalous dimensions and
the Wilson coefficients invariant. It, however, induces a finite
renormalization of the physical operators. In order to restore the
standard $\MSbar$ renormalization conditions these contributions must
be removed by a change of scheme. Hence a $\eps$-dependent linear
transformation of the operator basis is equivalent to a global
rotation and a change of scheme \cite{Gorbahn:2004my}.

The finite renormalization corresponding to the above change of basis
can be derived with simple algebra. Through $\ord (\as^2)$ we find
\beq \label{eq:Zprimes}
\begin{split}
\hat{Z}^{(1, 0)}_{\QQ} & = -\hat{R} \hspace{0.4mm} \hat{Z}'^{(1,
1)}_{\QE} \hat{U} \hat{R}^{-1} \, , \\[1mm] 
\hat{Z}^{(2, 0)}_{\QQ} & = - \hat{R} \left ( \hat{Z}'^{(2, 1)}_{\QE} 
\hat{U} + \hat{Z}'^{(2, 2)}_{\QE} \hat{V} - \f{1}{2} \hat{Z}'^{(1,
1)}_{\QE} \hat{V} \hat{\gamma}'^{(0)} \right ) \hat{R}^{-1} \, ,
\end{split}
\eeq
where
\beq \label{eq:ZQE22}
\hat{Z}'^{(2, 2)}_{\QE} = \f{1}{2} \left ( \hat{Z}'^{(1, 1)}_{\QE}
\hat{Z}'^{(1, 1)}_{\EE} + \f{1}{2} \hat{\gamma}'^{(0)} \hat{Z}'^{(1,
1)}_{\QE} - \betazero \hat{Z}'^{(1, 1)}_{\QE} \right ) \, .
\eeq
As always, the matrix $\hat{M}$ encoding the rotation of evanescent
operators does not affect the residual finite
renormalization. \Eqsand{eq:Zprimes}{eq:ZQE22} agree with the NLO
formulas of \cite{Herrlich:1994kh, Chetyrkin:1997gb} and generalize
the NNLO formulas of \cite{Gorbahn:2004my}.       

The only feature which has not been discussed in the
literature before is the appearance of the $\eps^2$-dependent
transformation characterized by $\hat{V}$ in 
\Eq{eq:transformations}. It induces a finite $\ord (\as^2)$
renormalization described by the second and third term in the second
line of \Eqs{eq:Zprimes}. The former term is analogous to the first
line of \Eqs{eq:Zprimes}. The latter term stems from a
$\eps$-dependent $\ord (\as)$ renormalization. While the
$\eps$-dependent change of scheme removes the finite terms, it still
leaves the anomalous dimensions and the Wilson coefficients
invariant. Yet it induces a finite $\ord (\as^2)$ renormalization. The
transformation to the standard $\MSbar$ scheme definitions requires
the removal of this finite term, which is achieved by the last term in
the second line of \Eqs{eq:Zprimes}. 

The matrices entering the above equations are found from one- and
two-loop matrix elements of physical and evanescent operators in the
``standard'' basis. The $\ord (\as)$ ADM takes the following form
\cite{Chetyrkin:1997gb, Gorbahn:2004my, Gambino:2003zm} 
\beq \label{eq:misiakADM}
\hat{\gamma}'^{(0)} = \left (
\begin{array}{cc}
-4 & \f{8}{3} \\
12 & 0 
\end{array}
\right ) \, . 
\eeq
For the $\ord (\as)$ mixing into the evanescent operators one obtains
\cite{Chetyrkin:1997gb, Gorbahn:2004my, Gambino:2003zm}      
{%
\renewcommand{\arraystretch}{1.25}
\beq \label{eq:Zoneloop}
\hat{Z}'^{(1, 1)}_{\QE} = 
\left (
\begin{array}{cccc}
\f{5}{12} & \f{2}{9} & 0 & 0 \\ 
1 & 0 & 0 & 0 
\end{array}
\right ) \, , \hspace{1cm} 
\hat{Z}'^{(1, 1)}_{\EE} = 
\left (
\begin{array}{cccc}
-7 & -\f{4}{3} & \f{5}{12} & \f{2}{9} \\ 
-6 & 0 & 1 & 0 \\
0 & 0 & \f{13}{3} & -\f{28}{9} \\ 
0 & 0 & -14 & -\f{64}{3}
\end{array}
\right ) \, .
\eeq
}%
At $\ord (\as^2)$ only the mixing of physical into evanescent
operators is needed. It is given by \cite{Gorbahn:2004my,
Gambino:2003zm}   
{%
\renewcommand{\arraystretch}{1.25}
\beq \label{eq:Ztwoloop}
\hat{Z}^{(2, 1)}_{\QE} = 
\left (
\begin{array}{cccc}
\f{1531}{288} - \f{5}{216} \nf & -\f{1}{72} - \f{1}{81} \nf &  
\f{1}{384} & -\f{35}{864} \\   
\f{119}{16} - \f{1}{18} \nf & \f{8}{9} & -\f{35}{192} & -\f{7}{72} 
\end{array}
\right ) \, . 
\eeq 
}%

The general NNLO formulas relating the initial conditions of the
Wilson coefficients and the ADM in two different schemes have been 
derived recently \cite{Gorbahn:2004my}. They read            
\beq \label{eq:cprime}
\vec{C} (\muh) = \left ( \hat{1} + \f{\as (\muh)}{4 \pi}
\hat{Z}^{(1, 0)}_{\QQ} + \left ( \f{\as (\muh)}{4 \pi} \right )^2
\hat{Z}^{(2, 0)}_{\QQ} \right )^T \! \! ( \hat{R}^{-1} \big )^T
\vec{C}' (\muh) \, ,  
\eeq 
and 
\beq \label{eq:admtransformations}
\begin{split}
\hat{\gamma}^{(0)} & = \hat{R} \hspace{0.2mm} \hat{\gamma}'^{(0)}
\hspace{0.2mm} \hat{R}^{-1} \, , \\[3mm]
\hat{\gamma}^{(1)} & = \hat{R} \hspace{0.2mm} \hat{\gamma}'^{(1)}
\hspace{0.2mm} \hat{R}^{-1} - \left [ \hat{Z}^{(1, 0)}_{\QQ},
\hat{\gamma}^{(0)} \right ] - 2 \betazero \hat{Z}^{(1,0)}_{\QQ} \, ,
\\[2mm]  
\hat{\gamma}^{(2)} & = \hat{R} \hat{\gamma}'^{(2)} \hat{R}^{-1} -
\left [ \hat{Z}^{(2, 0)}_{\QQ}, \hat{\gamma}^{(0)} \right ] -
\left [ \hat{Z}^{(1, 0)}_{\QQ}, \hat{\gamma}^{(1)} \right ] -
\left [ \hat{Z}^{(1, 0)}_{\QQ}, \hat{\gamma}^{(0)} \right ]
\hat{Z}^{(1, 0)}_{\QQ} \\[1mm]
& - 4 \betazero \hat{Z}^{(2, 0)}_{\QQ} - 2 \betaone \hat{Z}^{(1,
0)}_{\QQ} + 2 \betazero \big ( \hat{Z}^{(1, 0)}_{\QQ} \big )^2 \, .    
\end{split}
\eeq
\Eqsto{eq:rmuvmatrices}{eq:admtransformations} allow one to transform
the results for the initial conditions of the Wilson coefficients of
the current-current operators and their ADM in the ``standard'' basis
to the basis used in this paper. This enables us to verify that the
well-established NLO results \cite{Chetyrkin:1997gb, Gorbahn:2004my,
Gambino:2003zm} coincide, after the change of scheme, with the
expressions presented in the second line of
\Eqsand{eq:Cpm}{eq:gpm}. For what concerns the NNLO we confirm both
the two-loop initial conditions of the Wilson coefficients
\cite{Bobeth:1999mk} as well as the corresponding three-loop ADM
\cite{Gorbahn:2004my}.

\pagebreak

\section*{Erratum}

\renewcommand{\theequation}{E.\arabic{equation}}
\setcounter{equation}{0}

\centerline{\bf Abstract}
\vspace*{0.5cm}

\noindent We correct the treatment of anomalous triangle diagrams occuring in
the effective theory in which the heavy top quark is integrated out.  To
this end we determine the initial conditions and anomalous dimensions of the
operator describing the $Z$-mediated coupling of neutrinos to quarks and
further rectify the bilocal renormalization group evolution. Our changes
affect the charm-quark contribution $P_c(X)$ at the next-to-leading and
next-to-next-to-leading orders, but are numerically negligible as they  amount  to relative shifts below a permille.  

\subsection*{\boldmath $Z$  \unboldmath Penguin}

Contrary to the statements in \zsec\ the diagrams in~\figcsmatch~vanish
for $m_t\to \infty$. As a consequence, the Wilson coefficient of the
Chern-Simons operator $Q_{CS}$ 
{also vanishes} and the anomalous diagrams will not drop out in a
next-to-leading (NLO) and next-to-next-to-leading order (NNLO)
calculation of the charm-quark contribution to $K \to \pi^+ \nu \bar
\nu$. There is no more reason to include $Q_{CS}$ in the definition of
$Q_Z$ in \qzequ. Further the Wilson coefficients of $Q_{V}^q $ and
$Q_{A}^q$ are different, so that the definition of $Q_Z$ is not helpful
for the description of the anomalous effects. To maintain the canonical
normalization of the coefficients we define
\beq \label{eq:QZerr}%
Q_{Z}^q = - I^3_q Q_{A}^q \, %
\eeq%
recalling that $I^3_q=\pm 1/2$ denotes the third component of the weak
isospin.  Since complete isospin doublets do not contribute to the
anomalous triangle diagram, it is sufficient to consider diagrams with a
bottom-quark loop stemming from matrix elements involving $Q_{Z}^b$. A
useful operator basis is $(Q_Z^b,Q_Z^{\rm sg}, Q_Z^{\rm nsg})$ with the
flavor-singlet and flavor-non-singlet operators
\begin{equation}
\begin{split}
  Q_Z^{\rm sg} &= Q_Z^s+ Q_Z^d -  Q_Z^c-Q_Z^u \\
        &= 
       \frac12 \sum_{\ell=e,\mu,\tau} \left[ \ov s \gamma_\mu \gamma_5 s + 
                    \ov d \gamma_\mu \gamma_5 d + 
                    \ov c \gamma_\mu \gamma_5 c + 
                    \ov u \gamma_\mu \gamma_5 u  
               \right] \, \ov \nu{}_{\ell L} \gamma^\mu \nu_{\ell L} \,,
        \\[2mm]
  Q_Z^{\rm nsg} &= Q_Z^s+ Q_Z^d +  Q_Z^c + Q_Z^u \\ 
         &= 
       \frac12  \sum_{\ell=e,\mu,\tau} \left[ \ov s \gamma_\mu \gamma_5 s + 
                    \ov d \gamma_\mu \gamma_5 d - 
                    \ov c \gamma_\mu \gamma_5 c - 
                    \ov u \gamma_\mu \gamma_5 u  
               \right] \, \ov \nu{}_{\ell L} \gamma^\mu \nu_{\ell L} \,. 
\end{split}               
\end{equation}
The operator $Q_Z^{\rm nsg}$ does not participate in the renormalization
group (RG) evolution and its coefficient satisfies $C_Z^{\rm
  nsg}(\mu)=1$ at all scales $\muc \leq \mu \leq \muh$.  The
coefficients $C_Z^b$ and $C_Z^{\rm sg}$ have non-trivial initial
conditions at the scale $\muh$, at which the top quark is integrated
out, and further run for $\mu_b \leq \mu \leq \muh$, with their RG
evolution determined by the anomalous dimension $\gamma_A$ calculated
from the renormalization factor $Z_{AA}$ defined in \qacsdefequ.  In our
paper we consider different renormalization schemes for $\gamma_5$: in
$Q_{Z}^b$ we use the 't Hooft-Veltman (HV) scheme, which amounts to the
replacement of $\ov b \gamma^\mu \gamma_5 b$ by $ - (i/3!) \,
\epsilon^{\mu_1 \mu_2 \mu_3 \mu } \, \ov b \gamma_{\mu_1} \gamma_{\mu_2}
\gamma_{\mu_3} b$~[17]. The diagrams for the NLO and NNLO anomalous
dimension describe the mixing of $Q_Z^b$ into $Q_Z^{\rm sg}$ and the
diagrams of the bilocal mixing discussed in \sectwo \ involve $Q_Z^{\rm
  sg}$ with open quark lines only, so that the definition of the naive
dimensional regularization (NDR) scheme is unambigous here.  We perform
these calculation in two schemes, choosing either HV or NDR for the open
quark line. At NNLO, the initial condition of the Wilson coefficient
$C_Z^q$ is the same in HV and NDR:
\begin{equation}
  \label{eq:11}
  C_Z^q(\mu_{\scriptscriptstyle W}) = 1 \, +\, 
       \left (  \frac{\alpha_s(\muh) }{4 \pi} \right )^2  C_Z^{q\,(2)} (\muh)\,,
 \qquad \mbox{with} \quad 
  C_Z^{q\,(2)} (\muh)= 
   2 I_q^3  \,  6 C_F   \ln \frac{m_t^2}{\muh^2}\,,
\end{equation}
and $C_F = 4/3$. It follows that
\begin{eqnarray}
  \label{eq:11cs}
  C_Z^{\rm sg}(\muh) =  
 \left ( \frac{\alpha_s(\muh) }{4 \pi} \right )^2 C_Z^{b\,(2)} (\muh) .
\end{eqnarray}
The results in \Eqsand{eq:11}{eq:11cs} assume that the same scheme 
  for $\gamma_5$ is used in the full standard model (SM) and the
effective five-quark theory. $ C_Z^q(\muh)$ is calculated from the
two-loop diagram obtained by attaching a quark line to the two gluons to
the left diagram in~\figcsmatch\ and a second diagram with reversed
orientation of the quark loop. The finite renormalization constant
$Z_{AA}^{(2,0)} $ defined in \zaaequ~is taken into account here.  The
leading order~(LO) $2\times 2$ anomalous dimension matrix (ADM) $\hat
\gamma_Z^{(0)}$ for $(Q_Z^b,Q_Z^{\rm sg})$ vanishes.  For the NLO and
NNLO ADM we find
\begin{equation}
\label{eq:resgaz}
\hat \gamma_Z^{(n)} = 
  \begin{pmatrix}
   \gamma_A^{(n)} & \gamma_A^{(n)} \\ 
    4 \gamma_A^{(n)} & 4 \gamma_A^{(n)}  
  \end{pmatrix}\,,\qquad n = 1,2 \,,
\end{equation}
{in the HV scheme}, confirming the anomalous dimensions of the 
axial current calculated in [17]: 
\begin{equation} \label{eq:gammaAs}
\gamma_A^{(1)} = 12 C_F \,,
\qquad  \gamma_A^{(2)} =  \frac{284}{3} C_F C_A - \frac{8}{3} C_F f - 
    36 C_F^2 \,. 
\end{equation}
Here $C_A = 3$ and $f$ denotes the number of active quark flavors. A
priori the element $\big (\hat \gamma_Z^{(n)} \big )_{12}$ may depend on
the choice of the scheme for $Q_{Z}^{\rm sg}$: the HV scheme involves a
finite one-loop counterterm [17] to remove a spurious anomaly from
diagrams with an open quark line. This finite counterterm is absent in
the NDR scheme, which instead involves a finite counterterm to the
evanescent operator $E_A^q$ defined in \axevequ. We find that
the analytical results for $\big ( \hat \gamma_Z^{(1)} \big )_{12}$ and
$\big ( \hat \gamma_Z^{(2)} \big )_{12}$ are the same in both
schemes.\footnote{The elements {in the second row of} of $\hat
  \gamma_Z^{(n)}$ involve an insertion of $Q_Z^{\rm sg}$ into a closed
  quark loop and are hence not properly defined in the NDR scheme.  This
  ambiguity is irrelevant for a NNLO analysis of $P_c(X)$.}

The Wilson coefficients $\vec C_Z = (C_Z^b,C_Z^{\rm sg})^T$ for
$\mu_b \leq \mu < \muh$ are obtained in the usual way by solving
\begin{equation}
 \mu \frac{d}{d\mu} \vec C (\mu) = \hat \gamma_Z^T \vec C (\mu). 
\end{equation}
{At the scale $\mu_b$ the bottom quark is integrated out and we have to 
match the five-flavor and four-flavor theories. The corresponding
Hamiltonians are 
\begin{align}
 \Heff^{f=5,Z} &= \f{\pi \aem}{\MW^2 \sws} 
  \left[ C_Z^b Q_Z^b + C_Z^{\rm nsg} Q_Z^{\rm nsg} + C_Z^{5,{\rm sg}}
    Q_Z^{\rm sg} \right], \non\\
 \Heff^{f=4,Z} &= \f{\pi \aem}{\MW^2 \sws} 
  \left[ C_Z^{\rm nsg} Q_Z^{\rm nsg} + C_Z^{4,{\rm sg}}
         Q_Z^{\rm sg} \right] \non
\end{align}
with $C_Z^{f,{\rm nsg}} = 1$ in both theories. 
From the two-loop triangle diagrams with bottom quark
one finds: 
\beq
\label{eq:matchb}
\begin{split}
  C_Z^{4,{\rm sg}} (\mu_b) & = C_Z^{5,{\rm sg}}(\mu_b)
    + C_Z^b (\mu_b) 
  \left (  \frac{\alpha_s(\mu_b) }{4 \pi} \right)^2 6 C_F \ln \frac{m_b^2}{\mu_b^2} \\
& = C_Z^{5,{\rm sg}}(\mu_b)
    + 
      \left ( \frac{\alpha_s(\mu_b) }{4 \pi} \right)^2 6 C_F \ln \frac{m_b^2}{\mu_b^2}
     + {\cal O} \left(\alpha_s ^4 \right) . 
     \end{split}
\eeq
Inserting \Eqsto{eq:11}{eq:resgaz} into the NNLO RG evolution
  equation [39] entails:}
  \beq
\begin{split}
C_Z^{5, {\rm sg}} (\mub) & = \frac{\gamma_A^{(1)}}{2\beta_0} \frac{\alpha_s (\muh) - \alpha_s(\mub)}{4\pi} + \frac{5}{8} \frac{\big (\gamma_A^{(1)} \big)^2}{\beta_0^2} \left ( \frac{\alpha_s (\muh) - \alpha_s(\mub)}{4\pi}\right )^2 \\ & \phantom{xx} - \frac{\beta_0 \gamma_A^{(2)} - \beta_1 \gamma_A^{(1)}}{4\beta_0^2} \frac{\alpha_s^2 (\muh) - \alpha_s^2 (\mub)}{(4\pi)^2} + \left ( \frac{\alpha_s (\muh)}{4 \pi} \right )^2 C_Z^{b \, (2)} (\muh) \,, \\[2mm]
C_Z^{5, b} (\mub) & =1+C_Z^{5, {\rm sg}} (\mub) \,.
\end{split}
\eeq
Numerically, we find
\beq \label{eq:CZsg}
C_Z^{\rm sg} (m_b) = -0.011 \,,
\eeq
with negligible {variation when} $\muh$ is changed between $\MW$ and $\mt$ or if $\mb$, $\mt$, or $\alpha_s (\MZ)$ are
varied within their experimental errors.

\subsection*{Bilocal Evolution}
In order to determine the charm-quark contribution of $K^+\to \pi^+ \nu
\ov \nu$ we need the bilocal mixings of $(C_Z^c, C_\pm)$ and $(C_Z^b,
C_\pm)$ into the coefficient $C_\nu$. The relevant three-loop anomalous
dimension
\begin{equation}
  \gamma^{A \Delta (2)}_{\pm, \, \nu} = 
  6 Z^{\Delta (3, 1)}_{\pm A, \, \nu} - 
  4 Z^{(2, 1)}_{AE} Z^{P (1, 0)}_{\pm E, \, \nu} - 
  4 Z^{\Delta (2, 0)}_{AA} Z^{P (1, 1)}_{\pm A, \, \nu} 
\end{equation}
can be inferred from \zinput. In the NDR scheme we obtain:
\begin{equation}
  \gamma^{A \Delta (2)}_{\pm, \, \nu} = 
  12 C_F (1  \pm 5 C_A) \, .
\end{equation}
{We write the Wilson coefficient for the $Z$-penguin contribution to
$C_\nu $ as $C_\nu^P+ C_\nu^\Delta $. Here $C_\nu^P$ is the piece coming
from the non-anomalous diagrams obtained by solving the
RG equation in \blrgeequ, with the solution quoted
in \pensolequ. $ C_\nu^\Delta$ is the new contribution from the triangle
diagrams. The RG equation read 
\begin{equation}
  \label{eq:fullbilocal}
  \mu \frac{d}{d \mu} 
  \left[
    C_\nu^P(\mu) + C_\nu^\Delta(\mu)
  \right]
  =
  \gamma_\nu 
  \left[
    C_\nu^P(\mu) + C_\nu^\Delta(\mu)
  \right] + \,\,
  4 \!\!\!\!\! \sum_{\substack{i=\pm \\ j = b,{\rm nsg,sg}}} 
  \gamma_{i  j, \, \nu}
  C_i (\mu) C_Z^j (\mu) \, .
\end{equation}
From the definition of $Q_{b,\rm nsg,sg}$ we find
\begin{align}
  \label{eq:bilocal_gammas}
  \gamma_{\pm \, b, \, \nu} = 
  \tfrac12 \gamma^{A \Delta}_{\pm, \, \nu} \, , \qquad\quad
  \gamma_{\pm \, {\rm nsg}, \, \nu} = \gamma^P_{\pm, \, \nu} = 
  -\tfrac12 
  \gamma^A_{\pm, \, \nu} \, , \qquad\quad
  \gamma_{\pm \, {\rm sg}, \, \nu} = 
  \tfrac12 \gamma^A_{\pm, \, \nu}  
  + \tfrac12 \gamma^{A \Delta}_{\pm, \, \nu}  \, .
\end{align} 
The expressions for $\gamma^A_{\pm, \, \nu}$, calculated from the
non-anomalous diagrams, are listed in \gamavequ.  Subtracting the RGE
for $C_\nu^P(\mu)$ in Eq.~\blrgeequ~from \eq{eq:bilocal_gammas} we find:}
\begin{equation}
  \label{eq:1}
    \mu \frac{d}{d \mu} 
    C_\nu^\Delta(\mu) =
    \gamma_\nu C_\nu^\Delta(\mu) + 
    2 \sum_{i=\pm} \left [
      \gamma^{A \Delta}_{i, \, \nu}
      C_i (\mu) C_Z^b (\mu)
      + \left( 
      \gamma^{A \Delta}_{i, \, \nu} 
        + \gamma^A_{i, \, \nu} 
      \right)
      C_i (\mu) C_Z^{\rm sg} (\mu)
    \right ] \,.
\end{equation}
This RG equation splits into two linear differential equations which are
easily solved, as discussed for example in~\cite{Brod:2010mj} for
an analogous case.

\begin{figure}[t!]
\begin{center}
\scalebox{1.0}{\includegraphics{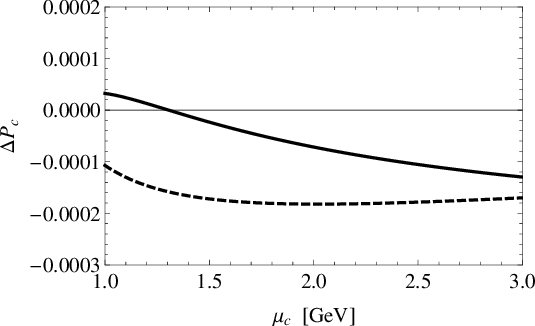}}
\end{center}
\vspace{-4mm}
\caption{$\Delta P_c(X)$ as a function of the renormalization scale
  $\muc$. The dashed (solid) line represents the NLO (NNLO) result.}
\label{fig:thezero}
\end{figure}

{To NNLO} only the one-loop matrix elements involving a double
operator insertion $(Q_\pm^q, Q_Z^{\rm sg})$ gives a non-trivial
contribution to the matching at $\mu_c$. We write the contribution to
the tree-level matrix element $\langle Q_\nu \rangle^{(0)}$ in the
following way
\begin{equation}
  \label{eq:sgmatrixelement}
  \left \langle Q_\pm^{\rm sg}(\muc) \right \rangle = 
  \frac{\as (\muc)}{4 \pi} r^{{\rm sg} \,  (1)}_\pm (\muc) 
  \langle Q_\nu \rangle^{(0)} = 
  - \frac{\as (\muc)}{4 \pi} r^{P \, (1)}_\pm (\muc) 
  \langle Q_\nu \rangle^{(0)} \, ,
\end{equation}
where $r^{P \, (1)}_\pm (\muc)$ codifies the one-loop corrections
detailed in \rmatequ~and $ Q_\pm^{\rm sg}$ is defined
analogously to $Q_\pm^P$ in \bilocinsequ.

We have now collected all ingredients necessary to determine the
  missing contribution $\Delta P_c (X)$ to the charm-quark contribution in
$K^+ \to \pi^+ \nu \bar \nu$, represented by the parameter $P_c(X)$
defined in \pcseriesequ~of the paper. For the NLO and NNLO coefficient
in the perturbative expansion%
\beq%
\Delta P_c (X)= \Delta P_c^{(1)} (X) + \frac{\alpha_s (\muc)}{4 \pi}
\Delta P_c^{(2)} (X) \,, %
\eeq%
we plot the $\mu_c$-dependence in Fig.~\ref{fig:thezero}.  The
charm-quark corrections due to anomalous triangle graphs lead to shifts
in $P_c ( X)$ in the ballpark of $-0.0001$ with the exact value
depending on the renormalization scale.  Recalling that $P_c(X) = 0.369$
($P_c(X) = 0.375$) at NLO (NNLO) the corresponding relative shift
amounts to $-0.05\%$ ($-0.01\%$).  One can understand the smallness of
the NLO contribution to $\Delta P_c(X)$ by inspecting the logarithmic
expansion of our result: the NLO term starts only at 
$\mathcal{O}(\alpha_s^2 \hspace{0.25mm} \ln (\muh^2/\mu_b^2)\hspace{0.25mm} \ln(\muh^2/\mu_c^2) )$, which is of the
same order in $\alpha_s$ as the NNLO correction containing only one
large logarithm. Expanding $\Delta P_c(X)$ in $\alpha_s$ we
find explicitly 
\begin{equation}
\Delta P_c(X) = 
\frac{1}{\lambda^4}
\frac{m_c^2}{\MW^2} \left ( \frac{\alpha_s}{4 \pi} \right)^2 \left\{
   \left(
    \ln \frac{\muh^2}{\mu_b^2}
    -2 \ln \frac{\muh^2}{\mu_c^2} + 1
  \right)  \ln \frac{\muh^2}{\mu_b^2}
\right\} + 
\mathcal{O}
\Big (\alpha_s^2 \ln^0 \frac{\muh^2}{\mu_b^2}, \alpha_s^3\Big) \, ,
\end{equation}
where the $\alpha_s^2$ suppression is accompanied by
 terms with small numerical coefficients of opposite sign.
Notice that the exact result $\Delta P_c(X)$ is smaller by two
orders of magnitude than the total NNLO uncertainty in $P_ c (X)$ of
$\pm 0.015$. In view of these numbers it is clear that our erroneous
treatment of anomalous triangle diagrams has no phenomenological impact.

\subsection*{Acknowledgements} We are grateful to Konstantin Chetyrkin
and Mikolaj Misiak for pointing our mistake out to us. We also thank
Johann K\"uhn for bringing the article \cite{Collins:1978wz}, in
which the large logarithm occuring from the anomalous triangle
diagrams has been discovered, to our attention. We further thank
Mikolaj Misiak and Gerhard Buchalla for a thorough proofreading and
comments on this erratum.


\begin{thebibliography}{99}

\bibitem{UT}
G.~Buchalla and A.~J.~Buras,
Phys.\ Lett.\ B {\bf 333}, 221 (1994) 
[arXiv:hep-ph/9405259] 
and Phys.\ Rev.\ D {\bf 54}, 6782 (1996)
[arXiv:hep-ph/9607447].

\bibitem{Gino}
G.~Isidori,
Annales Henri Poincare {\bf 4}, S97 (2003)
[arXiv:hep-ph/0301159];
in Proceedings of the 2nd Workshop on the CKM Unitarity Triangle,
Durham, England, 2003, eConf {\bf C0304052}, WG304 (2003)
[arXiv:hep-ph/0307014] and references therein. 

\bibitem{Buras:2004uu}
A.~J.~Buras, F.~Schwab and S.~Uhlig,
arXiv:hep-ph/0405132.

\bibitem{Marciano:1996wy}
W.~J.~Marciano and Z.~Parsa,
Phys.\ Rev.\ D {\bf 53}, 1 (1996).

\bibitem{LD}
G.~Ecker, A.~Pich and E.~de Rafael,
Nucl.\ Phys.\ B {\bf 303}, 665 (1988);
D.~Rein and L.~M.~Sehgal,
Phys.\ Rev.\ D {\bf 39}, 3325 (1989);
J.~S.~Hagelin and L.~S.~Littenberg,
Prog.\ Part.\ Nucl.\ Phys.\  {\bf 23}, 1 (1989);
M.~Lu and M.~B.~Wise,
Phys.\ Lett.\ B {\bf 324}, 461 (1994)
[arXiv:hep-ph/9401204];
C.~Q.~Geng, I.~J.~Hsu and Y.~C.~Lin,
Phys.\ Rev.\ D {\bf 50}, 5744 (1994)
[arXiv:hep-ph/9406313] and
Phys.\ Lett.\ B {\bf 355}, 569 (1995)
[arXiv:hep-ph/9506313];
S.~Fajfer,
Nuovo Cim.\ A {\bf 110}, 397 (1997)
[arXiv:hep-ph/9602322];
A.~F.~Falk, A.~Lewandowski and A.~A.~Petrov,
Phys.\ Lett.\ B {\bf 505}, 107 (2001)
[arXiv:hep-ph/0012099].

\bibitem{Isidori:2005xm}
G.~Isidori, F.~Mescia and C.~Smith,
Nucl.\ Phys.\ B {\bf 718}, 319 (2005)
[arXiv:hep-ph/0503107].

\bibitem{Isidori:2005tv}
G.~Isidori, G.~Martinelli and P.~Turchetti,
arXiv:hep-lat/0506026.

\bibitem{CKM}
N.~Cabibbo,
Phys.\ Rev.\ Lett.\  {\bf 10}, 531 (1963); 
M.~Kobayashi and T.~Maskawa,
Prog.\ Theor.\ Phys.\  {\bf 49}, 652 (1973).

\bibitem{Buchalla:1993wq}
G.~Buchalla and A.~J.~Buras,
Nucl.\ Phys.\ B {\bf 412}, 106 (1994)
[arXiv:hep-ph/9308272].

\bibitem{Buchalla:1998ba}
G.~Buchalla and A.~J.~Buras,
Nucl.\ Phys.\ B {\bf 548}, 309 (1999)
[arXiv:hep-ph/9901288].

\bibitem{Bardeen:1978yd}
W.~A.~Bardeen, A.~J.~Buras, D.~W.~Duke and T.~Muta,
Phys.\ Rev.\ D {\bf 18}, 3998 (1978).

\bibitem{X}
G.~Buchalla and A.~J.~Buras,
Nucl.\ Phys.\ B {\bf 398}, 285 (1993)
and B {\bf 400}, 225 (1993).

\bibitem{Misiak:1999yg}
M.~Misiak and J.~Urban,
Phys.\ Lett.\ B {\bf 451}, 161 (1999)
[arXiv:hep-ph/9901278].

\bibitem{LO}
A.~I.~Vainshtein, V.~I.~Zakharov, V.~A.~Novikov and M.~A.~Shifman,
Phys.\ Rev.\ D {\bf 16}, 223 (1977);
J.~R.~Ellis and J.~S.~Hagelin,
Nucl.\ Phys.\ B {\bf 217}, 189 (1983);
C.~Dib, I.~Dunietz and F.~J.~Gilman,
Mod.\ Phys.\ Lett.\ A {\bf 6}, 3573 (1991).

\bibitem{Buras:2005gr}
A.~J.~Buras, M.~Gorbahn, U.~Haisch and U.~Nierste,
Phys.\ Rev.\ Lett.\ {\bf 95}, 261805 (2005)
[arXiv:hep-ph/0508165].

\bibitem{Chern:1974ft}
S.~S.~Chern and J.~Simons,
Annals Math.\  {\bf 99}, 48 (1974).

\bibitem{Larin:1993tq}
S.~A.~Larin,
Phys.\ Lett.\ B {\bf 303}, 113 (1993)
[arXiv:hep-ph/9302240].

\bibitem{ABJ}
S.~L.~Adler,
Phys.\ Rev.\  {\bf 177}, 2426 (1969);
J.~S.~Bell and R.~Jackiw,
Nuovo Cim.\ A {\bf 60}, 47 (1969).

\bibitem{Charles:2004jd}
J.~Charles {\it et al.}  [CKMfitter Group],
Eur.\ Phys.\ J.\ C {\bf 41}, 1 (2005), 
[arXiv:hep-ph/0406184] 
and August 1, 2005 update available at
http://www.slac.stanford.edu/ xorg/ckmfitter/ckm\_results\_summerEPS2005.html.

\bibitem{Ciuchini:2000de}
M.~Ciuchini {\it et al.},
JHEP {\bf 0107}, 013 (2001)
[arXiv:hep-ph/0012308] and September 22, 2005 update available at
http://utfit.roma1.infn.it/. 

\bibitem{Group:2005cc}
J.~F.~Arguin {\it et al.} [The Tevatron Electroweak Working Group],
arXiv:hep-ex/0507091.

\bibitem{Bobeth:2003at}
C.~Bobeth, P.~Gambino, M.~Gorbahn and U.~Haisch,
JHEP {\bf 0404}, 071 (2004)
[arXiv:hep-ph/0312090].

\bibitem{Buchalla:1997kz}
G.~Buchalla and A.~J.~Buras,
Phys.\ Rev.\ D {\bf 57}, 216 (1998)
[arXiv:hep-ph/9707243].

\bibitem{PDG}
S.~Eidelman {\it et al.}  [Particle Data Group],
Phys.\ Lett.\ B {\bf 592} (2004) 1, and 2005 partial update for
edition 2006 available at http://pdg.lbl.gov/.  

\bibitem{x}
T.~Inami and C.~S.~Lim,
Prog.\ Theor.\ Phys.\  {\bf 65}, 297 (1981)
[Erratum-ibid.\  {\bf 65}, 1772 (1981)];
G.~Buchalla, A.~J.~Buras and M.~K.~Harlander,
Nucl.\ Phys.\ B {\bf 349}, 1 (1991).

\bibitem{Melnikov:2000qh}
K.~Melnikov and T.~van~Ritbergen,
Phys.\ Lett.\ B {\bf 482}, 99 (2000) [arXiv:hep-ph/9912391].

\bibitem{Buras:1989xd}
A.~J.~Buras and P.~H.~Weisz,
Nucl.\ Phys.\ B {\bf 333}, 66 (1990).

\bibitem{Dugan:1990df}
M.~J.~Dugan and B.~Grinstein,
Phys.\ Lett.\ B {\bf 256}, 239 (1991).

\bibitem{Herrlich:1994kh}
S.~Herrlich and U.~Nierste,
Nucl.\ Phys.\ B {\bf 455}, 39 (1995)
[arXiv:hep-ph/9412375].

\bibitem{collins}
J.~Collins, Renormalization, Cambridge University Press, New York 1984
and references therein.

\bibitem{IRdimensional}
C.~Greub and T.~Hurth,
Phys.\ Rev.\ D {\bf 56}, 2934 (1997)
[arXiv:hep-ph/9703349];
A.~J.~Buras, A.~Kwiatkowski and N.~Pott,
Nucl.\ Phys.\ B {\bf 517}, 353 (1998)
[arXiv:hep-ph/9710336].

\bibitem{Bobeth:1999mk}
C.~Bobeth, M.~Misiak and J.~Urban,
Nucl.\ Phys.\ B {\bf 574}, 291 (2000)
[arXiv:hep-ph/9910220].

\bibitem{Gambino:2001au}
P.~Gambino and U.~Haisch,
JHEP {\bf 0110}, 020 (2001)
[arXiv:hep-ph/0109058].

\bibitem{Misiak:2004ew}
M.~Misiak and M.~Steinhauser,
Nucl.\ Phys.\ B {\bf 683}, 277 (2004)
[arXiv:hep-ph/0401041].

\bibitem{Chanowitz:1979zu}
M.~S.~Chanowitz, M.~Furman and I.~Hinchliffe,
Nucl.\ Phys.\ B {\bf 159}, 225 (1979).

\bibitem{Ciuchini:1993vr}
M.~Ciuchini, E.~Franco, G.~Martinelli and L.~Reina,
Nucl.\ Phys.\ B {\bf 415}, 403 (1994)
[arXiv:hep-ph/9304257].

\bibitem{Altarelli:1980fi}
G.~Altarelli, G.~Curci, G.~Martinelli and S.~Petrarca,
Nucl.\ Phys.\ B {\bf 187}, 461 (1981).

\bibitem{Chetyrkin:1997gb}
K.~G.~Chetyrkin, M.~Misiak and M.~M\"unz,
Nucl.\ Phys.\ B {\bf 520}, 279 (1998)
[arXiv:hep-ph/9711280].

\bibitem{Gorbahn:2004my}
M.~Gorbahn and U.~Haisch,
Nucl.\ Phys.\ B {\bf 713}, 291 (2005)
[arXiv:hep-ph/0411071].

\bibitem{Misiak:1994zw}
M.~Misiak and M.~M\"unz,
Phys.\ Lett.\ B {\bf 344}, 308 (1995)
[arXiv:hep-ph/9409454].

\bibitem{Chetyrkin:1997fm}
K.~G.~Chetyrkin, M.~Misiak and M.~M\"unz,
Nucl.\ Phys.\ B {\bf 518}, 473 (1998)
[arXiv:hep-ph/9711266].

\bibitem{Gambino:2003zm}
P.~Gambino, M.~Gorbahn and U.~Haisch,
Nucl.\ Phys.\ B {\bf 673}, 238 (2003)
[arXiv:hep-ph/0306079].

\bibitem{Broadhurst:1998rz}
D.~J.~Broadhurst,
Eur.\ Phys.\ J.\ C {\bf 8}, 311 (1999)
[arXiv:hep-th/9803091] 
and references therein. 

\bibitem{olddecoupling}
W.~Wetzel,
Nucl.\ Phys.\ B {\bf 196}, 259 (1982); 
W.~Bernreuther and W.~Wetzel,
Nucl.\ Phys.\ B {\bf 197}, 228 (1982)
[Erratum-ibid.\ B {\bf 513}, 758 (1998)]; 
W.~Bernreuther,
Annals Phys.\  {\bf 151}, 127 (1983) and 
Z.\ Phys.\ C {\bf 20} (1983) 331.

\bibitem{newdecoupling}
S.~A.~Larin, T.~van Ritbergen and J.~A.~M.~Vermaseren,
Nucl.\ Phys.\ B {\bf 438}, 278 (1995)
[arXiv:hep-ph/9411260]; 
K.~G.~Chetyrkin, B.~A.~Kniehl and M.~Steinhauser,
Nucl.\ Phys.\ B {\bf 510}, 61 (1998)
[arXiv:hep-ph/9708255].

\bibitem{Buras:1993dy}
A.~J.~Buras, M.~Jamin and M.~E.~Lautenbacher,
Nucl.\ Phys.\ B {\bf 408}, 209 (1993)
[arXiv:hep-ph/9303284].

\bibitem{Buras:1991jm}
A.~J.~Buras, M.~Jamin, M.~E.~Lautenbacher and P.~H.~Weisz,
Nucl.\ Phys.\ B {\bf 370}, 69 (1992)
[Addendum-ibid.\ B {\bf 375}, 501 (1992)].

\bibitem{IRquarkmass}
A.~J.~Buras, M.~Jamin and P.~H.~Weisz,
Nucl.\ Phys.\ B {\bf 347}, 491 (1990); 
S.~Herrlich and U.~Nierste,
Nucl.\ Phys.\ B {\bf 419}, 292 (1994)
[arXiv:hep-ph/9310311]; 
A.~J.~Buras, P.~Gambino and U.~Haisch,
Nucl.\ Phys.\ B {\bf 570}, 117 (2000)
[arXiv:hep-ph/9911250].

\bibitem{HVscheme}
G.~'t Hooft and M.~J.~G.~Veltman,
Nucl.\ Phys.\ B {\bf 44}, 189 (1972);
P.~Breitenlohner and D.~Maison,
Commun.\ Math.\ Phys.\  {\bf 52}, 11 (1977),
{\bf 52}, 39 (1977) and
{\bf 52}, 55 (1977).

\bibitem{Siegel:1979wq}
W.~Siegel,
Phys.\ Lett.\ B {\bf 84}, 193 (1979).

\bibitem{Stockinger:2005gx}
D.~St\"ockinger,
JHEP {\bf 0503}, 076 (2005)
[arXiv:hep-ph/0503129].

\bibitem{Abbott:1980hw}
L.~F.~Abbott,
Nucl.\ Phys.\ B {\bf 185}, 189 (1981) 
and references therein. 

\bibitem{Adler:1969er}
S.~L.~Adler and W.~A.~Bardeen,
Phys.\ Rev.\  {\bf 182}, 1517 (1969).

\bibitem{Bos:1992nd}
M.~Bos,
Nucl.\ Phys.\ B {\bf 404}, 215 (1993)
[arXiv:hep-ph/9211319] 
and references therein. 

\bibitem{anomalyfree}
D.~J.~Gross and R.~Jackiw,
Phys.\ Rev.\ D {\bf 6}, 477 (1972);
C.~Bouchiat, J.~Iliopoulos and P.~Meyer,
Phys.\ Lett.\ B {\bf 38} (1972) 519;
C.~P.~Korthals Altes and M.~Perrottet,
Phys.\ Lett.\ B {\bf 39}, 546 (1972).

\bibitem{Trueman:1979en}
T.~L.~Trueman,
Phys.\ Lett.\ B {\bf 88}, 331 (1979).

\bibitem{threeloopQCD}
O.~V.~Tarasov,
JINR-P2-82-900;  
O.~V.~Tarasov, A.~A.~Vladimirov and A.~Y.~Zharkov,
Phys.\ Lett.\ B {\bf 93}, 429 (1980);
S.~A.~Larin and J.~A.~M.~Vermaseren,
Phys.\ Lett.\ B {\bf 303}, 334 (1993)
[arXiv:hep-ph/9302208].

\bibitem{Herrlich:1996vf}
S.~Herrlich and U.~Nierste,
Nucl.\ Phys.\ B {\bf 476}, 27 (1996)
[arXiv:hep-ph/9604330].

\bibitem{NNLOmatrixkernel}
M.~Beneke, T.~Feldmann and D.~Seidel,
Nucl.\ Phys.\ B {\bf 612}, 25 (2001) [arXiv:hep-ph/0106067]; 
H.~M.~Asatrian, K.~Bieri, C.~Greub and M.~Walker, 
Phys.\ Rev.\ D {\bf 69}, 074007 (2004) [arXiv:hep-ph/0312063]. 

\bibitem{Chetyrkin:2000yt}
K.~G.~Chetyrkin, J.~H.~K\"uhn and M.~Steinhauser,
Comput.\ Phys.\ Commun.\  {\bf 133}, 43 (2000).

\bibitem{charm}
J.~H.~K\"uhn and M.~Steinhauser,
Nucl.\ Phys.\ B {\bf 619}, 588 (2001)
[Erratum-ibid.\ B {\bf 640}, 415 (2002)]
[arXiv:hep-ph/0109084];
J.~Rolf and S.~Sint  [ALPHA Collaboration],
JHEP {\bf 12}, 007 (2002)
[arXiv:hep-ph/0209255];
A.~H.~Hoang and M.~Jamin,
Phys.\ Lett.\ B {\bf 594}, 127 (2004)
[arXiv:hep-ph/0403083];
O.~Buchm\"uller and H.~Fl\"acher,
arXiv:hep-ph/0507253;
A.~H.~Hoang and A.~V.~Manohar,
arXiv:hep-ph/0509195.

\bibitem{RUT}
T.~Goto, N.~Kitazawa, Y.~Okada and M.~Tanaka,
Phys.\ Rev.\ D {\bf 53}, 6662 (1996)
[arXiv:hep-ph/9506311];
A.~G.~Cohen, D.~B.~Kaplan, F.~Lepeintre and A.~E.~Nelson,
Phys.\ Rev.\ Lett.\  {\bf 78}, 2300 (1997)
[arXiv:hep-ph/9610252];
Y.~Grossman, Y.~Nir and M.~P.~Worah,
Phys.\ Lett.\ B {\bf 407}, 307 (1997)
[arXiv:hep-ph/9704287];
G.~Barenboim, G.~Eyal and Y.~Nir,
Phys.\ Rev.\ Lett.\  {\bf 83}, 4486 (1999)
[arXiv:hep-ph/9905397].

\bibitem{Steinhauser:2000ry}
M.~Steinhauser, 
Comput.\ Phys.\ Commun.\  {\bf 134} (2001) 335 [arXiv:hep-ph/0009029].

\bibitem{Buras:1992tc}
A.~J.~Buras, M.~Jamin, M.~E.~Lautenbacher and P.~H.~Weisz,
Nucl.\ Phys.\ B {\bf 400}, 37 (1993)
[arXiv:hep-ph/9211304].
\end{thebibliography}

\begin{thebibliography}{2}

\bibitem[65]{Brod:2010mj}
  J.~Brod and M.~Gorbahn,
  Phys.\ Rev.\ D {\bf 82} (2010) 094026
  [arXiv:1007.0684 [hep-ph]].

\bibitem[66]{Collins:1978wz}
  J.~C.~Collins, F.~Wilczek and A.~Zee,
  Phys.\ Rev.\ D {\bf 18} (1978) 242.

\end{thebibliography}
\end{document}